\documentclass{aa} 

\usepackage{graphicx}
\usepackage{xspace}
\usepackage{color}
\usepackage{natbib}
\usepackage{longtable}
\usepackage{multirow}
\usepackage{txfonts}
\usepackage[colorlinks=true,citecolor=magenta,linkcolor=blue]{hyperref}
\bibpunct{(}{)}{;}{a}{}{,} 

\newcommand{\mission}[1]{\textit{#1}}
\newcommand{\erasst}{eRASSt~J045650.3$-$203750\xspace}
\newcommand{\jsrc}{J0456$-$20\xspace}
\newcommand{\asko}{ASASSN-14ko\xspace}
\newcommand{\gsn}{GSN~069\xspace}
\newcommand{\gro}{GRO~J1655$-$40\xspace}
\newcommand{\rxj}{RX~J1301.9$+$2747\xspace}
\newcommand{\eroo}{eRO-QPE1\xspace}
\newcommand{\erot}{eRO-QPE2\xspace}
\newcommand{\ictde}{IC~3599\xspace}
\newcommand{\ptde}{\textit{p}TDE\xspace}
\newcommand{\pday}{223\,days\xspace}

\newcommand{\fxfaint}{${P}_\text{X, faint}$\xspace}
\newcommand{\fxdrop}{${P}_\text{X, drop}$\xspace}
\newcommand{\fxrise}{${P}_\text{X, rise}$\xspace}
\newcommand{\fxplat}{${P}_\text{X, plat}$\xspace}

\newcommand{\unitflux}{\text{erg\,cm}^{-2}\,\text{s}^{-1}}
\newcommand{\unitfluxd}{\text{erg\,cm}^{-2}\,\text{s}^{-1}\,\AA^{-1}}
\newcommand{\unitlumi}{\text{erg\,s}^{-1}}
\newcommand{\msun}{\mathrm{M}_\odot}
\newcommand{\rsun}{\mathrm{R}_\odot}
\newcommand{\mbh}{M_\text{BH}}

\newcommand{\ledd}{L_\text{Edd}}

\newcommand{\sflux}{$f_\text{X, soft}$\xspace}
\newcommand{\slumi}{$L_\text{X, soft}$\xspace}
\newcommand{\nxv}{\sigma_\text{rms}^2}

\newcommand{\mul}{$\text{M}_\text{ul}$\xspace}
\newcommand{\mpl}{$\text{M}_\text{pl}$\xspace}
\newcommand{\mpbb}{$\text{M}_\text{pbb}$\xspace}
\newcommand{\mpdb}{$\text{M}_\text{pdb}$\xspace}
\newcommand{\mbkp}{$\text{M}_\text{bkp}$\xspace}
\newcommand{\mcomp}{$\text{M}_\text{comp}$\xspace}
\newcommand{\mmcd}{$\text{M}_\text{mcd}$\xspace}
\newcommand{\xslo}{X2-LO\xspace}
\newcommand{\xshi}{X2-HI\xspace}

\newcommand{\trecur}{223}

\newcommand{\frpfirst}{$\sim6.8\times10^{-12}$}
\newcommand{\frpsecond}{$\sim2.0\times10^{-12}$}

\newcommand{\frdsecond}{30}

\newcommand{\fpffirst}{$\sim1.0\times10^{-11}$}
\newcommand{\fpfsecond}{$\sim2.3\times10^{-12}$}
\newcommand{\fpfthird}{$\sim1.5\times10^{-12}$}

\newcommand{\fpdsecond}{$\sim58$}
\newcommand{\fpdthird}{$\sim54$}
\newcommand{\fpsecondstart}{59454}
\newcommand{\fpsecondend}{59512}
\newcommand{\fpthirdstart}{59645}
\newcommand{\fpthirdend}{59699}

\newcommand{\frsecondstart}{59366}

\newcommand{\chidof}{\chi^2/\text{d.o.f}}

\newcommand*\Circled[1]{\textcircled{{\scriptsize #1}}}

\begin{document}

   \title{Deciphering the extreme X-ray variability of the nuclear transient \erasst}

   \subtitle{A likely repeating partial tidal disruption event}

   \author{Zhu Liu\inst{\ref{ins:mpe}}
           \and A. Malyali\inst{\ref{ins:mpe}}
           \and M. Krumpe\inst{\ref{ins:aip}}
           \and D. Homan\inst{\ref{ins:aip}}
           \and A. J. Goodwin\inst{\ref{ins:curtinu}}
           \and I. Grotova\inst{\ref{ins:mpe}}
           \and A. Kawka\inst{\ref{ins:curtinu}}
           \and A. Rau\inst{\ref{ins:mpe}}
           \and A. Merloni\inst{\ref{ins:mpe}}
           \and G. E. Anderson\inst{\ref{ins:curtinu}}
           \and J. C. A. Miller-Jones\inst{\ref{ins:curtinu}}
           \and A. G. Markowitz\inst{\ref{ins:ncac}, \ref{ins:ucsd}}
           \and S. Ciroi\inst{\ref{ins:padovau}}
           \and F. Di Mille\inst{\ref{ins:carnegie}}
           \and M. Schramm\inst{\ref{ins:saitamau}}
           \and Shenli Tang\inst{\ref{ins:ut}}
           \and D. A. H. Buckley\inst{\ref{ins:saao}, \ref{ins:capeu}, \ref{ins:freeu}}
           \and M. Gromadzki\inst{\ref{ins:warsawu}}
           \and Chichuan Jin\inst{\ref{ins:naoc},\ref{ins:gkd}}
           \and J. Buchner\inst{\ref{ins:mpe}}
          }

   \institute{
              Max-Planck-Institut für extraterrestrische Physik,
                   Gießenbachstraße 1, 85748 Garching, Germany\label{ins:mpe}\\
              \email{liuzhu@mpe.mpg.de}
              \and Leibniz-Institut für Astrophysik Potsdam,
                   An der Sternwarte 16, 14482 Potsdam, Germany\label{ins:aip}
              \and International Centre for Radio Astronomy Research,
                   Curtin University, GPO Box U1987, Perth, WA 6845, Australia\label{ins:curtinu}
              \and Nicolaus Copernicus Astronomical Center, Polish Academy of Sciences, ul.\ Bartycka 18, 00-716 Warszawa, Poland\label{ins:ncac}
              \and University of California, San Diego, Center for Astrophysics and Space Sciences, MC 0424, La Jolla, CA, 92093-0424, USA\label{ins:ucsd}
              \and Dipartimento di Fisica e Astronomia ``G. Galilei'', Università di Padova,
                   Vicolo dell'Osservatorio 3, 35122 Padova, Italy\label{ins:padovau}
              \and Las Campanas Observatory, Carnegie Observatories, Colina El Pino,
                   Casilla 601, La Serena, Chile\label{ins:carnegie}
              \and Graduate school of Science and Engineering, Saitama Univ. 255 Shimo-Okubo,
                   Sakura-ku, Saitama City, Saitama 338-8570, Japan\label{ins:saitamau}
              \and Department of Physics, University of Tokyo, Tokyo 113-0033, Japan\label{ins:ut}
              \and South African Astronomical Observatory, PO Box 9, Observatory Road, Observatory 7935, Cape Town, South Africa\label{ins:saao}
              \and Department of Astronomy, University of Cape Town, Private Bag X3, Rondebosch 7701, South Africa\label{ins:capeu}
              \and Department of Physics, University of the Free State, PO Box 339, Bloemfontein 9300, South Africa\label{ins:freeu}
              \and Astronomical Observatory, University of Warsaw, Al. Ujazdowskie 4, 00-478 Warszawa, Poland\label{ins:warsawu}
              \and National Astronomical Observatories, Chinese Academy of Sciences,
                   20A Datun Road, Beijing 100101, People's Republic of China\label{ins:naoc}
              \and School of Astronomy and Space Sciences, University of Chinese Academy of Sciences,
                   19A Yuquan Road, Beijing 100049, People's Republic of China\label{ins:gkd}
             }

   \date{Received xxx xx, xxxx; accepted xxx xx, xxxx}


  \abstract
  {During its all-sky survey, the extended ROentgen Survey with an Imaging Telescope Array (eROSITA) on board the Spectrum-Roentgen-Gamma (SRG) observatory has uncovered a growing number of X-ray transients associated with the nuclei of quiescent galaxies. Benefitting from its large field of view and excellent sensitivity, the eROSITA window into time-domain X-ray astrophysics yields a valuable sample of X-ray selected nuclear transients. Multi-wavelength follow-up enables us to gain new insights into understanding the nature and emission mechanism of these phenomena.}
   {We present the results of a detailed multi-wavelength analysis of an exceptional repeating X-ray nuclear transient, \erasst (hereafter \jsrc), uncovered by SRG/eROSITA in a quiescent galaxy at a redshift of $z \sim 0.077$. We aim to understand the radiation mechanism at different luminosity states of \jsrc, and provide further evidence that similar accretion processes are at work for black hole accretion systems at different black hole mass scales.}
   {We describe our temporal analysis, which addressed both the long- and short-term variability of \jsrc. A detailed X-ray spectral analysis was performed to investigate the X-ray emission mechanism.}
   {Our main findings are that 1) \jsrc cycles through four distinctive phases defined based on its X-ray variability: an
   \textup{X-ray rising} phase leading to an \textup{X-ray plateau} phase that lasts for $\text{about two}\,$months. This is terminated by a rapid \textup{X-ray flux drop} phase during which the X-ray flux can drop drastically by more than a factor of 100 within one week, followed by an \textup{X-ray faint} state for about two months before the \textup{X-ray rising} phase starts again. 2) The X-ray spectra are generally soft in the rising phase, with a photon index $\gtrsim3.0$, and they become harder as the X-ray flux increases. There is evidence of a multi-colour disk with a temperature of $T_\text{in}\sim70\,\text{eV}$ in the inner region at the beginning of the X-ray rising phase. The high-quality \mission{XMM-Newton} data suggest that a warm and hot corona might cause the X-ray emission through inverse Comptonisation of soft disk seed photons during the plateau phase and at the bright end of the rising phase. 3) \jsrc shows only moderate UV variability and no significant optical variability above the host galaxy level. Optical spectra taken at different X-ray phases are constant in time and consistent with a typical quiescent galaxy with no indication of emission lines. 4) Radio emission is (as yet) only detected in the X-ray plateau phase and rapidly declines on a timescale of two weeks.}
   {\jsrc is likely a repeating nuclear transient with a tentative recurrence time of $\sim$\pday. It is a new member of this rare class. We discuss several possibilities to explain the observational properties of \jsrc. We currently favour a repeating partial tidal disruption event as the most likely scenario. The long-term X-ray evolution is explained as a transition between a thermal disk-dominated soft state and a steep power-law state. This implies that the corona can be formed within a few months and is destroyed within a few weeks.}

   \keywords{X-rays: individuals: \erasst\ -- Accretion, accretion disks -- Galaxies: nuclei -- Black hole physics}
   \titlerunning{\erasst as a repeating partial TDE}
   \authorrunning{Liu et al.}
   \maketitle
%

\section{Introduction}

Supermassive black holes (SMBHs) with masses $\sim10^6-10^9\,\msun$ are thought to be ubiquitous in the centres of all massive galaxies \citep[e.g.][]{kormendy_richstone1995, magorrian_etal1998}. The tight correlation between the mass of the central black hole (BH) and the properties of the host galaxy observed in the local Universe \citep[e.g.][]{ferrarese_merritt2000, gebhardt_etal2000} suggest a coevolution of SMBHs and their host galaxies \citep{kormendy_luis2013}. Our understanding of the growth history and evolution of SMBHs \citep[e.g.][]{merloni_heinz2013} mainly comes from the studies of SMBHs at the centres of active galactic nuclei (AGNs). However, AGNs only comprise a small fraction of the whole galaxy population. Our current understanding is that an AGN represents a particular phase in the evolution of a galaxy, during which the BH grows mainly through radiatively efficient accretion \citep{alexander_hickox2012}. It is thus essential to study the BH demography from a large sample of quiescent galaxies, particularly covering low SMBH masses, and to study SMBHs with extremely low accretion rates, which cannot be explored easily in general AGN studies.

Occasionally, dormant SMBHs will be temporarily fuelled with a sudden influx of gas, which can be caused by ill-fated stars wandering too close to the SMBH. One such scenario comprises tidal disruption events (TDEs; \citealt[][]{rees1988, evans_kochanek1989}), in which a star enters the tidal radius of the SMBH and is torn apart by strong tidal forces. A fraction of the stellar debris falls inwards towards the SMBH. On the other hand, stars might also be on tightly bound orbits around the SMBHs with low eccentricities, resulting in extreme mass-ratio inspirals (EMRIs) undergoing stable Roche-lobe overflow (RLOF) onto the SMBHs. The mass-transfer rate is temporarily enhanced during grazing physical collisions between a pair of RLOFing EMRIs \citep{metzger_stone2017, metzger_etal2022}. During these processes, a fraction of the influx of gas is eventually accreted onto the SMBH, producing energetic nuclear transients. The peak luminosity is in the range of a few percent to close to the Eddington luminosity ($\ledd\equiv1.3\times10^{38}\mbh/\msun\,\unitlumi$), which is comparable to AGN luminosities. Thus nuclear transients provide an effective tool for discovering SMBHs in a large sample of otherwise quiescent galaxies.

Theoretical calculations have suggested that transitions between different accretion modes will take place when the accretion rate reaches certain critical values \citep{meyer_etal2000}. Evidence for transitions of the accretion flow in accreting BH systems, for instance, from a standard thin accretion disk \citep{shakura_sunyaev1973} dominated by thermal emission in the soft state to the advection-dominated accretion flow (ADAF; \citealt{narayan_yi1995, yuan_narayan2014}) in the hard state and vice versa, has been found mainly in stellar mass BH X-ray binaries \citep[BHXRBs; see][for more details on the different accretion states in BHXRBs]{remillard_mcclintock2006}. It is thought that a steady compact jet, which is known to be associated with the hard state of BHXRBs, becomes unstable during the transition from the hard to soft state \citep{fender_etal2004}. On the other hand, powerful transient jets are often observed close to the peak of BHXRB outbursts \citep[see][for a review]{fender_etal2004} during the transition from the hard to the soft state via the steep power-law state. However, strong observational evidence for an accretion-mode transition in individual AGN remains elusive, although it has been invoked to explain the large amplitude X-ray variability in a few AGNs \citep[e.g. NGC 7589;][]{yuan_etal2004, liu_etal2020} and changing-look AGNs \citep[e.g.][]{noda_etal2018}. The Eddington ratio in nuclear transients can change by orders of magnitude on a timescale of hours to years, which is rarely seen in individual AGNs, providing an ideal laboratory for exploring the accretion process across a broad range of Eddington ratios. This might yield new insights into the launch of relativistic jets in BH--accretion systems.

A growing number of nuclear transients have been discovered in the past decades. Among them, TDEs are perhaps the most well known and well studied. The first TDE candidate was discovered in the soft X-ray band using archival \mission{ROSAT} data \citep[][NGC 5905]{komossa_bade1999}. After this, more candidates were found in the X-ray band from archival data \citep[e.g. see][for reviews]{komossa_2015, saxton_etal2021}, and in the UV band using \mission{GALEX} data \citep[e.g.][]{vanvelzen_etal2020}. A few TDEs have also been discovered in the hard X-ray band (e.g. Swift\,1644+57; \citealt{bloom_etal2011, levan_etal2011, zauderer_etal2011}) with the Neil Gehrels Swift Observatory (\mission{Swift}). They can also be luminous in the radio band; the radio emission is thought to arise from relativistic jets launched by the TDEs \citep[e.g.][]{zauderer_etal2011, zauderer_etal2013}. \textcolor{black}{Radio emission from non-jetted TDEs has also been reported recently \citep{vanvelzen_etal2016}, which shed light on the type of outflows associated with both optical/UV and X-ray bright events \citep[see][for a review]{ alexander_etal2020}}. The advance in wide-field high-cadence optical surveys over the past decade, such as the All-Sky Automated Survey for Supernovae (ASAS-SN\footnote{\url{https://www.astronomy.ohio-state.edu/asassn/index.shtml}}), the Palomar Transient Factory (PTF; \citealt{law_etal2009}), the Panoramic Survey Telescope and Rapid Response System (Pan-STARRS; \citealt{chambers_etal2016}), the Zwicky Transient Facility (ZTF; \citealt{bellm_etal2019}), and the Asteroid Terrestrial-impact Last Alert System (ATLAS; \citealt{tonry_etal2018}), has not only greatly enlarged the number of known TDEs \citep[e.g.][]{vanvelzen_etal2020}, but also facilitated the discovery of new classes of nuclear transients that cannot easily be explained by normal TDEs. One example is the new class of extremely energetic transients reported by \citet{kankare_etal2017}. \citet{trakhtenbrot_etal2019} also presented a sample of AGNs that showed extreme flares. \citet{malyali_etal2021} reported a novel nuclear transient discovered by extended ROentgen Survey with an Imaging Telescope Array (eROSITA; \citealt{predehl_etal2021}) in a galaxy without an indication of prior activity. This transient shows double peaks in its optical light curve. TDEs and other unusual nuclear transients are also being discovered in the extremely dust-extincted luminous infrared galaxies in surveys targeting dust-obscured supernovae, which are infrared and radio bright, but faint or undetected at optical and X-ray wavelengths \citep[e.g.][]{mattila18,kool20}.

Recently, a few transients with periodic or repeating flares in X-ray and/or optical/UV have been reported. Quasi-periodic eruptions (QPE) were first detected in the AGNs \gsn and \rxj \citep{miniutti_etal2019, giustini_etal2020}. Recently, \citet{arcodia_etal2021} reported two more events, the \eroo and \erot found in the first eROSITA All Sky Survey (eRASS1), in galaxies without optical signature of AGNs. These sources showed X-ray eruptions with X-ray flux increases of up to several orders of magnitude with a duration $\text{shorter than}~\text{hours}$ and a recurrence time $\text{shorter than } \text{one}$~day. Periodic flares with a longer duration and recurrence time have also been discovered in optical bands. \asko is a periodic nuclear transient discovered by ASAS-SN in the Seyfert\,2 AGN ESO 253-G003 \citep{payne_etal2021}. It flared at regular intervals over the past 7 years. The flares lasted for several months and  recurred every $\sim114$ days. Repeating or periodic transients on much longer timescales (i.e. decades) are more challenging to discover with current surveys. Nevertheless, it has been suggested that the late-time X-ray re-brightening of \ictde, a TDE candidate in a low-luminosity AGN discovered with \mission{ROSAT}, might be caused by a repeating partial tidal disruption event (\ptde) with a recurrence time of 9.5 years \citep[][but see \citealt{grupe_etal2015} for a different interpretation]{campana_etal2015}.

While our understanding of nuclear transients, particularly TDEs, has greatly improved through the enlarged samples, the nature of even the most well-studied nuclear transients can still be controversial \citep[e.g. ASASSN-15lh;][]{zabludoff_etal2021}. Even less is known about the nature of periodic or repeating nuclear transients. Theoretical models involve collisions between a pair of EMRIs orbiting an SMBH (QPEs and \asko; \citealt{metzger_etal2022}) or a \ptde (\asko and IC~3599). On the other hand, the physical processes driving the multi-wavelength emission, which are crucial for understanding the nature of these nuclear transients, are still under hot debate. For instance, the detection of bright optical TDEs came as a surprise, since strong optical emission was not expected in early theoretical TDE work. It has been suggested that the optical emission may be caused by the reprocessing of the UV/X-ray emission by optically thick material \citep[e.g.][]{dai_etal2018}, although others explained the optical emission as shocks generated through collisions of stellar streams \citep{piran_etal2015, lu_etal2020}. While the outflow properties of optical/UV and X-ray selected TDEs can also be probed by radio detections, the source of the radio emission is also debated, including scenarios such as collision-induced spherical outflows \citep[e.g.][]{Goodwin2022} or accretion-driven winds \citep[e.g.][]{Alexander2016,Cendes2021}, or a collimated sub-relativistic jet \citep[e.g.][]{vanvelzen_etal2016,Cendes2022}. However, it is currently unclear whether the X-ray and optical/UV \textcolor{black}{selected TDEs} are populations that are intrinsically different because the sample of X-ray selected nuclear transients is still relatively small. Systematic multi-wavelength studies of an X-ray selected sample can afford us new insights into the nature of these nuclear transients \citep[e.g.][]{sazonov_etal2021}.

\begin{figure*}
    \centering
    \includegraphics[width=\textwidth]{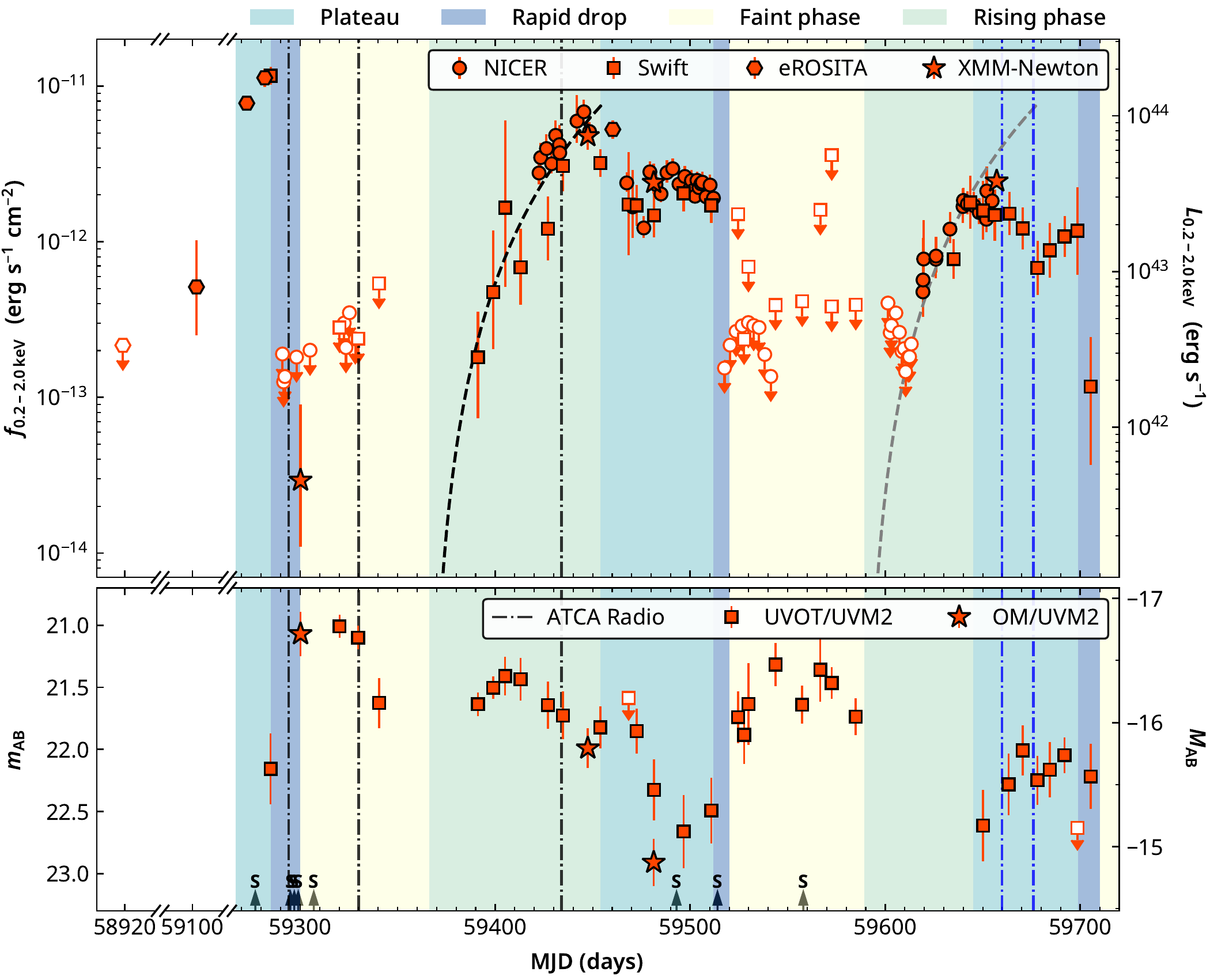}
    \caption{Long-term X-ray and UV light curves for \jsrc. The coloured regions represent the four phases defined in Sect.\,\ref{subsubsec:xray} based on the X-ray light curve: the plateau phase (\fxplat, light cyan), the rapid drop phase (\fxdrop, light blue), the faint phase (\fxfaint, light yellow), and the rising phase (\fxrise, light green). \textup{Upper panel:} Red points are the unabsorbed intrinsic $0.2-2.0\,\text{keV}$ X-ray light curve from eROSITA (hexagons), \mission{Swift}/XRT (squares), \mission{NICER} (circles), and \mission{XMM-Newton} (stars). The error bars indicate 90\% uncertainties. The $3\sigma$ flux and luminosity upper limits are shown with downward arrows. The dashed black line shows the best-fitting power-law model, $f_\text{rs}(t)\propto(t-t_0)^{\beta}$ , where $t_0=59377\pm9$ and $\beta=2.3\pm0.7$ (quoted uncertainties are at $1\sigma$ confidence level), during the rise phase. The dashed grey line shows the same model, but shifted by 223\,days. \textup{Bottom panel:} Long-term UV light curves from \mission{Swift}/UVOT UVM2 (red squares) and \mission{XMM-Newton}/OM UVM2 (red stars). The error bars mark the $1\sigma$ uncertainties. Empty squares with downward arrows indicate 3$\sigma$ upper limits. The vertical dashed-dotted lines mark the dates of the five ATCA radio observations (black: non-detections; blue: detections), and the upward arrows marked with an S indicate the dates when optical spectra were obtained.}
    \label{fig:multi_lc}%
\end{figure*}

In this paper, we present the discovery of a likely repeating nuclear transient, \erasst (hereafter \jsrc), discovered in eRASS2 in a quiescent galaxy at a redshift of $z\sim 0.077$. During its all-sky survey, eROSITA rapidly uncovers transients associated with the nuclei of galaxies that show no apparent signatures of prior AGN activity. \jsrc is one of the most variable X-ray sources in this sample (see Fig.\,\ref{fig:multi_lc}), showing a drastic X-ray flux drop by a factor of 100 within one week. Our follow-up optical spectroscopic observations revealed optical spectra typical of a quiescent galaxy, without signatures of emission lines. More importantly, the long-term X-ray and UV light curves suggest that \jsrc likely is a repeating nuclear transient with a roughly estimated recurrence time of \pday. This adds a member to this rare nuclear transient class. Transient radio emission is also detected in \jsrc, indicating that an outflow or jet may have been launched during its cycle. This paper is structured as follows. In Sect.\,\ref{sec:multi_band} we describe the multi-wavelength data reduction. Our optical spectroscopic analysis, X-ray/UV temporal study, radio data analysis, and X-ray spectral modelling, and short-term variability analysis are presented in Sect.\,\ref{sec:results}. Finally, we discuss and summarise our results in Sect.\,\ref{sec:discussion} and \ref{sec:summary}.

Throughout this paper, we adopt a flat $\Lambda$CDM cosmology with $H_0=67.7\,\text{km\,s}^{-1}\,\text{Mpc}^{-1}$ and $\Omega_m=0.308$ \citep{planck_etal2020}. Therefore, $z=0.077$ corresponds to a luminosity distance of $D_\text{ld}=360\,\text{Mpc}$. All magnitudes are reported in the AB system (not corrected for Galactic extinction).

\section{Observations and data reduction}\label{sec:multi_band}

\jsrc was discovered by {eROSITA} in eRASS2 and was detected in all subsequently performed all-sky scans, eRASS3 and eRASS4. An extensive multi-wavelength follow-up campaign was organised. It is described in this section.

\subsection{{eROSITA}\label{subsec:multi_ero}}
We first discovered \jsrc in its faint X-ray state in a dedicated search for TDE candidates in eRASS2, which consists of ten consecutive eROSITA scans at the location of \jsrc with gaps of $\sim4\,\text{h}$ each between 2020 September 8
and 10. \jsrc was not detected in eRASS1, which covered the location of the source between 2020 March 10 and 11. It was detected again in eRASS3 and eRASS4 with an X-ray flux higher by $\sim25$ and $\sim10$ times than in eRASS2, respectively. The eRASS3
observation was split into two segments because of an orbit correction performed by the Spectrum-Roentgen-Gamma (SRG; \citealt{sunyaev_etal2021}) spacecraft, followed by a test of the cooling system of the CCD cameras. The first segment was performed between 2021 February 27 and 2021 February 28, and the second segment was carried out $\sim8$ days later, between 2021 March 8 and 9. eROSITA observed the source again during eRASS4 between 2021 September 2 and 4. The position of the X-ray transient measured from eRASS3 is $(\text{RA, Dec})=(\text{04:56:49.74}, -20\degr37\arcmin48.44\arcsec)$ with an uncertainty of $1.0\arcsec$. An optical source at an angular distance of $1.0\arcsec$ with coordinates (04:56:49.80, $-20\degr37\arcmin47.99\arcsec$) is identified as the optical counterpart. A false-colour image of the host galaxy, overlaid with positions measured from different instruments, is shown in Fig.\,\ref{fig:fc_img} in Appendix\,\ref{sec:fc_img}.

All eROSITA data were calibrated and cleaned using the pipeline version 946 of the eROSITA Science Analysis Software (eSASS; \citealt{brunner_etal2022}). The photon events from all seven telescopes were merged into one event-list file. The \textsc{srctool} in eSASS (version \texttt{20211004}) was used to extract the X-ray spectra and light curves. A circular region with a radius of $50''$ was chosen as the source region for eRASS3 and eRASS4. However, a smaller source region with a radius of $30''$ was chosen for eRASS2 to increase the signal-to-noise ratio (S/N) as the source is very faint during the eRASS2 observation. \jsrc is not detected in eRASS1, so for that epoch we used the source region defined for eRASS2. A source-free annular region with an inner radius of $90''$ and outer radius of $120''$ was selected as the background region for all the eROSITA observations. Assuming a model consisted of an absorbed power-law (\texttt{TBabs*zashift*cflux*powerlaw} in \textsc{Xspec}, hereafter \mul) with a photon index fixed at 3.0 and the Galactic absorption ($N_\text{H, Gal}$) fixed at $3.3\times 10^{20}~\mathrm{cm^{-2}}$ \citep{willingale_etal2013}, we estimated a $3\sigma$ upper limit of $2.2\times10^{-13}\,\unitflux$ in the $0.2-2.0~\text{keV}$ energy band for eRASS1 (see Appendix\,\ref{sec:cal_ul}). For all the other eRASS observations, the X-ray flux was estimated via X-ray spectral modelling. More details about the eROSITA observations are listed in Table\,\ref{tab:logxray} in Appendix\,\ref{sec:obs_log}.

\subsection{\mission{XMM-Newton}\label{subsec:multi_xmm}}
Pre-approved \mission{XMM-Newton} target of opportunity (ToO) observations were performed on 2021 March 27 (Obs-ID: 0862770201, hereafter X1; PI: Krumpe) and 2022 March 19 (Obs-ID: 0884960601, hereafter X4; PI: Liu). In addition, we requested two longer \mission{XMM-Newton} Director's Discretionary Time (DDT) observations performed on 2021 August 21 and September 21 (PI: Liu; Obs-IDs: 0891801101 and 0891801701, hereafter X2 and X3, respectively) to investigate the short-term X-ray variability. The details of the \mission{XMM-Newton} observations are listed in Table\,\ref{tab:logxray}.

All \mission{XMM-Newton} data were reduced using the latest calibration files. The observation data files (ODFs), retrieved from the \mission{XMM-Newton} Science Archive, were reduced using the \mission{XMM-Newton} Science Analysis System software (\textsc{SAS}, version 19.1; \citealt{gabriel_etal2004}). For each observation, the \textsc{SAS} tasks \textsc{emchain} and \textsc{epchain} were used to generate the event lists for the European Photon Imaging Camera (EPIC) MOS \citep{turner_etal2001} and pn \citep{struder_etal2001} detectors, respectively. High-background flaring periods were identified and filtered from the event lists. {\jsrc is detected in all the \mission{XMM-Newton} observations except for X1 in the standard source detection pipeline using \texttt{edetect\_chain}. However, we detected \jsrc in X1 with \textsc{ml\_det} of 11.8 in a customised pipeline, during which the source detection \texttt{edetect\_chain} task was run on the combined MOS1 and MOS2 data over the $0.2-1.0\,$keV band. To increase the S/N, we selected a circular region with a radius of $15''$ as the source region for the MOS1 and MOS2 images for X1. A larger circular region with a radius of $45''$} was chosen for the MOS1, MOS2, and pn images for all the other observations. A source-free circular region with a radius of $100''$ was chosen as the background region for the MOS cameras. The background region for the pn camera was selected from a circular region with a radius of $60''$ centred at the same CCD read-out column as the source position. X-ray events with pattern $\leq12$ for MOS and $\leq4$ for pn were selected to extract the X-ray spectra. We used the \texttt{SAS} task \texttt{RMFGEN} and \texttt{ARFGEN} to generate the response matrix and ancillary files, respectively. We only extracted the X-ray spectra from the MOS1 and MOS2 data for X1. They were combined into one MOS spectrum using the \textsc{addascaspec} task to increase the S/N, and were then rebinned to have at least one count in each energy bin. For all the other observations, the X-ray spectra (MOS1, MOS2, and pn) were rebinned to have at least 20 counts for each background-subtracted channel and to avoid oversampling the intrinsic energy resolution by a factor larger than 3.

The \mission{XMM-Newton} Optical Monitor (OM) observations were taken in image+fast mode using the UVM2 filter. The OM imaging- and fast-mode data were analysed using the \texttt{omichain} and \texttt{omfchain} tasks, respectively. The fast-mode data are dominated by background because the source is faint in the UV. Only the image-mode data were therefore used. \jsrc is detected in the OM/UVM2 data in all four observations, except for X4. We only report the OM/UVM2 results for the three observations in which \jsrc is detected. We found no significant short-term UVM2 variability in any of the three observations. We thus report the OM/UVM2 photometry using the combined OM data set of the three observations. The measured AB magnitudes and flux densities for each observation are listed in Table\,\ref{tab:sw_uvot}.

\subsection{\mission{Swift} observations\label{subsec:multi_sw}}
An extensive X-ray and UV follow-up campaign  of \jsrc was performed with \mission{Swift} (PIs: Krumpe \& Liu). Three \mission{Swift} observations (obsids: 00014135030, 00014135031, 00014135032) were observed within 2 days. We therefore stacked these three observations, and assign this stacked observation the surrogate ObsID 00014135999 for the purpose of brevity in Table\,\ref{tab:sw_uvot} and \ref{tab:logxray}. We used the XRT (X-Ray Telescope) online data analysis tool\footnote{\url{http://www.swift.ac.uk/user_objects}} \citep{evans_etal2009} to check whether the source was detected for each individual observation \citep[for more details about the XRT source detection, see][]{evans_etal2020}. For observations in which the source was detected, we used the same online tool to generate the X-ray spectra. The X-ray flux was then calculated through X-ray spectral fitting. For observations in which the source was not detected, we obtained the total and background photon counts as well as the $3\sigma$ count rate upper limits using the XRT online tool. The \mul model was then used to convert the count rate upper limits into the $0.2-2.0\,\text{keV}$ flux upper limits.

The \mission{Swift} Ultraviolet/Optical Telescope (UVOT) data were reduced using the UVOT analysis pipeline provided in \textsc{Heasoft} with the UVOT calibration version \texttt{20201215}. Source counts were extracted from a circular region with a radius of $5''$ centred at the source position, and a $20''$ radius circle was chosen as the background region centred at a source-free region close to the position of \jsrc. The task \textsc{uvotsource} was used to perform photometric measurements. We confirmed the overall accuracy of the photometric measurements with a nearby star. The standard deviation derived from the star is about 0.1\,mag, which is smaller than the typical UVOT uncertainty for \jsrc. We further verified that small-scale sensitivity (SSS) does not affect the UVOT data at the position of \jsrc. A summary of the \mission{Swift}/UVOT observations is listed in Table\,\ref{tab:sw_uvot}.

\subsection{\mission{NICER} observations\label{subsec:multi_ni}}
The monitoring of \jsrc with the Neutron star Interior Composition Explorer (\textit{NICER}; \citealt{gendreau_etal2016}) was triggered shortly after the eRASS3 detection (PIs: Liu \& Krumpe). The \mission{NICER} data were analysed using the \texttt{heasoft} software (version 6.29c) with the latest calibration files (version 250210707). Data were downloaded from the \mission{NICER} data archive on HEASARC. Noisy detectors were excluded for each observation by running a $5\sigma$ clipping on all 52 detectors. The merged level 2 event-list file for the remaining detectors were regenerated following the standard \texttt{nicerl2} pipeline\footnote{\url{https://heasarc.gsfc.nasa.gov/docs/nicer/analysis_threads/nicerl2}}. The background and total X-ray spectra were then generated using the \texttt{BACKGEN3C50} pipeline\footnote{\url{https://hera.gsfc.nasa.gov/docs/nicer/tools/nicer_bkg_est_tools.html}}. Following the suggestion from \citet{remillard_etal2022}, additional selection criteria were applied to exclude periods during which $S0_\text{net}$ is higher than 2 and $\text{hbg}_\text{net}$ is higher than 0.05. The \texttt{nicerarf} and \texttt{nicerrmf} tasks were used to generate the response matrix and ancillary file for each observation\footnote{\jsrc was very faint (close to the sensitivity limit) during \mission{NICER} observation 4595020126, which consists of seven snapshots. \jsrc was detected in three snapshots, but it was not detected in the other four snapshots, likely because of the short exposure time and/or background variability. In this work, we report the results from the three detection snapshots for this observation.}, respectively.

The \mission{NICER} background spectra were generated using an empirical background model \citep{remillard_etal2022} with a Gaussian uncertainty rather than a Poisson uncertainty. The \mission{NICER} data are also often dominated by systematic uncertainties. We therfore calculated the $3\sigma$ photon count upper limits for \mission{NICER} using the following formula: $N_\mathrm{3\sigma} = 3\sqrt{N_\mathrm{bkg}+(N_\mathrm{bkg}\times \sigma_\mathrm{sys})^2}$, where $N_\mathrm{bkg}$ is the total background photon count in a given energy range, and $\sigma_\mathrm{sys}$ is the systematic uncertainty. In this work, we estimated the $3\sigma$ photon count upper limits for \mission{NICER} non-detections over the $0.4-1.0\,\text{keV}$ band\footnote{This energy range was chosen to avoid potential contamination from optical loading in the energy band below 0.4\,keV} with a systematic uncertainty of $\sigma_\mathrm{sys}=0.05$. Observations with $N_\mathrm{tot,  0.4-1.0\,\text{keV}}>N_\mathrm{bkg}+N_\mathrm{3\sigma}$ were considered as detected. For observations in which \jsrc was detected, the \mission{NICER} X-ray spectra were then rebinned using the \texttt{ftgrouppha} tool with the optimal binning scheme from \citet{kaastra_bleeker2016} and a minimum photon count of 20 per bin. The X-ray flux was then estimated through X-ray spectral modelling (Sect.\,\ref{subsubsec:xray_spec_long}). For non-detections, the \mul model was again used to convert the $3\sigma$ count rate upper limits ($N_\mathrm{3\sigma}/t_\text{exposure}$) into the $0.2-2.0\,\text{keV}$ flux upper limits.

\subsection{Optical spectroscopic observations}
We performed several optical spectroscopic follow-up observations of the optical counterpart of \jsrc. The observations were made from early March 2021 until the end of 2021 (Table~\ref{tab:opt_spec}) and are marked in Fig.~\ref{fig:multi_lc}.

{SALT:} Two pairs of optical spectra, both 200\,s exposure, were obtained using the RSS instrument \citep{burgh_etal2003} on the Southern African Large Telescope \citep[SALT;][]{buckley_etal2006}, on the night of 2021 March 4. We used the PG0900 grating and a slit width of $1.5$\arcsec. The first pair of spectra covered the wavelength range $3636-6699$\,\AA,\xspace  and the second pair of redder spectra covered $6031-8981$\,\AA, both at a resolution of 5.5~\AA. Sky conditions were clear, with a seeing of $\sim1.2$\arcsec. To reduce the spectra, we used the PyRAF-based PySALT package\footnote{\url{https://astronomers.salt.ac.za/software/pysalt-documentation}} \citep{crawford_etal2010}, which includes corrections for gain and cross-talk and performs bias subtraction. We extracted the science spectrum using standard IRAF\footnote{\url{https://iraf.net}} tasks, including wavelength calibration (neon and argon calibration lamp exposures were taken, one immediately before and one immediately after the science spectra, respectively), background subtraction, and one-dimensional spectrum extraction. The pupil (i.e. the view of the mirror from the tracker) moves during all SALT observations, causing the effective area of the telescope to change during exposures. Therefore, no absolute flux calibration is possible. However, by observing spectrophotometric standards during twilight, we were able to obtain relative flux calibration, allowing the recovery of the correct spectral shape and relative line strengths.

{NTT}: The source was observed with the ESO Faint Object Spectrograph and Camera v.2 (EFOSC2; \citealt{buzzoni_etal1984}) mounted on the ESO New Technology Telescope (NTT) on 2021 March 26 (proposal ID 106.21RU.001, PI: Malyali). We used grism 13 and the $1.2\arcsec$ slit oriented 11 degrees off the parallactic angle. We obtained two consecutive exposures of 1800\,s each with a seeing of about $1$\arcsec during the observation. The spectrum has a wavelength range of $3685-9315\,\AA$ with a dispersion of 2.77\,\AA/pixel. The data were reduced and calibrated using the \texttt{esoreflex} pipeline \citep[][v2.11.5]{freudling_etal2013}.
The He+Ar arcs were used to obtain the wavelength calibration,  and the standard star LTT3864 was used for flux calibration, which was observed with the same grism and the same slit oriented along the parallactic angle.

{FORS2:} We observed \jsrc with the FORS2 instrument \citep{appenzeller_etal1998} on UT1 of the Very Large Telescope Array on 2021 April 3 (proposal ID 106.21RU, PI: Krumpe). We used grisms 300V and 300I (+OG590 filter), with combined exposure times of 1000~s per grism. We made use of a 1.3\arcsec\xspace slit under clear observing conditions. All reductions and calibrations of the data were performed using the \texttt{esoreflex} pipeline (v2.11.3). We made use of He+HgCd+Ar (300V) and He+Ar (300I) arcs for the wavelength calibration and used observations of the flux standard LTT4816, which was observed using the same instrumental set-up.

{Gemini:} This target was observed with the GMOS-S Hamamatsu detectors in two seasons in 2021 March and 2021 October. The data were reduced using the Python spectroscopic data reduction pipeline (PypIt; \citealt{Prochaska2020}) tool. The CuAr arc lamp was used for the wavelength calibration, and the standard star EG274 was taken for flux calibration.

{Magellan:} A long-slit spectrum was obtained on 2021 October 26 with LDSS3-C mounted at Clay. The VPH-All grism was used in combination with a $1\arcsec$ slit to cover a wavelength range of $3700-10000\,\AA$ with a dispersion of about 2\,\AA/pixel and a resolution of about 8.5\,\AA. The slit was oriented along the parallactic angle. 3 $\times$ 600\,s consecutive exposures were acquired to effectively remove cosmic rays. The seeing during the observation was around $0.5\arcsec$. The spectra were reduced with IRAF following the usual procedure of overscan subtraction, flat-field correction, and wavelength calibration by means of a He-Ne-Ar lamp. The flux calibration was performed through the observation of the standard star LTT 9239. The same aperture slit was used, again oriented along the parallactic angle. Finally, the three exposures were sky subtracted and averaged to obtain the spectrum of the source.

{WiFeS:} We obtained a spectrum of \jsrc with the Wide Field Spectrograph \citep[WiFeS;][]{dopita_etal2010} mounted on the ANU 2.3\,m telescope at Siding Spring Observatory on 2021 December 10 (proposal ID 4210177, PI: Miller-Jones). We used the R3000 and B3000 gratings and obtained a NeAr arc lamp exposure after the target exposure. The total spectral range is from 3500 to 9000 \AA. The data were reduced using the PyWiFeS reduction pipeline \citep{childress_etal2014}. The pipeline produces three-dimensional
sets consisting of spatially resolved, bias subtracted, flat-fielded, wavelength- and flux-calibrated spectra for each slitlet. We then extracted background-subtracted spectra from the slitlets that provided the most significant flux using the task \texttt{apall} in IRAF. We used the white dwarf EG21 as the flux standard.

\begin{table}[]
\caption{\textbf{Spectroscopic observations of \jsrc.}}
\centering
\label{tab:opt_spec}
\begin{tabular}{llllll}
\hline\hline
UT date    & Tel.     & Inst.      & Exp     & Slit       & Airmass\\
           &          &            & (ks)    & ($\arcsec$)&        \\\hline
2021-03-04 & SALT     & RSS        & 0.2     & 1.5        & 1.32   \\
2021-03-22 & Gemini   & GMOS       & 0.3     & 1.0        & 1.43   \\
2021-03-22 & Gemini   & GMOS       & 0.3     & 1.0        & 1.35   \\
2021-03-27 & NTT      & EFOSC2     & 3.6     & 1.2        & 1.33   \\
2021-04-03 & VLT      & FORS2      & 1.0     & 1.3        & 1.30   \\
2021-10-06 & Gemini   & GMOS       & 0.3     & 1.0        & 1.30   \\
2021-10-27 & Magellan & LDSS3-C    & 1.8     & 1.0        & 1.10   \\
2021-12-10 & ANU      & WiFeS      & 2.4     & ---        & 1.02   \\
\hline
\end{tabular}
\end{table}

\begin{figure*}
  \centering
  \includegraphics[width=1.0\columnwidth]{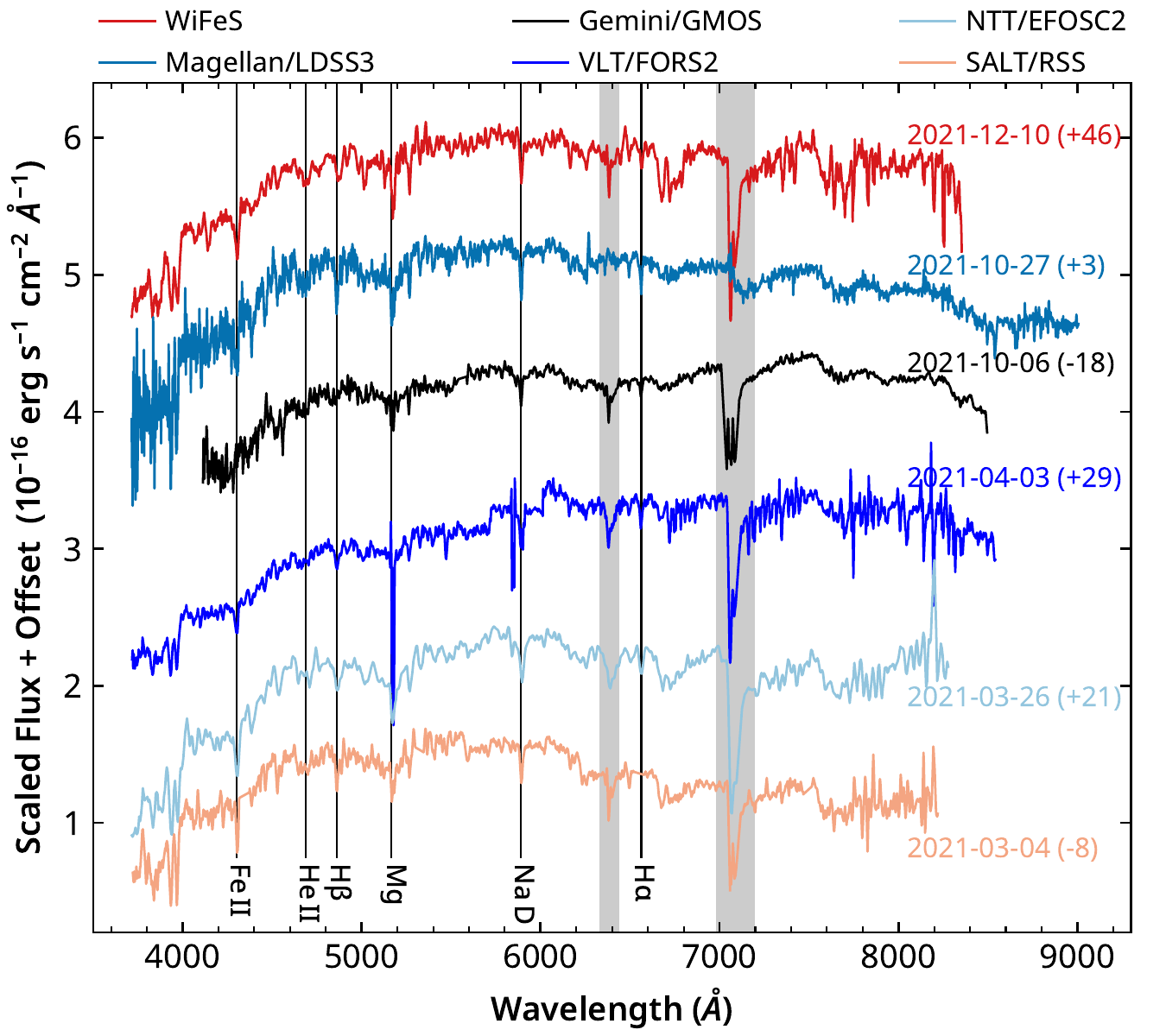}
  \includegraphics[width=1.0\columnwidth]{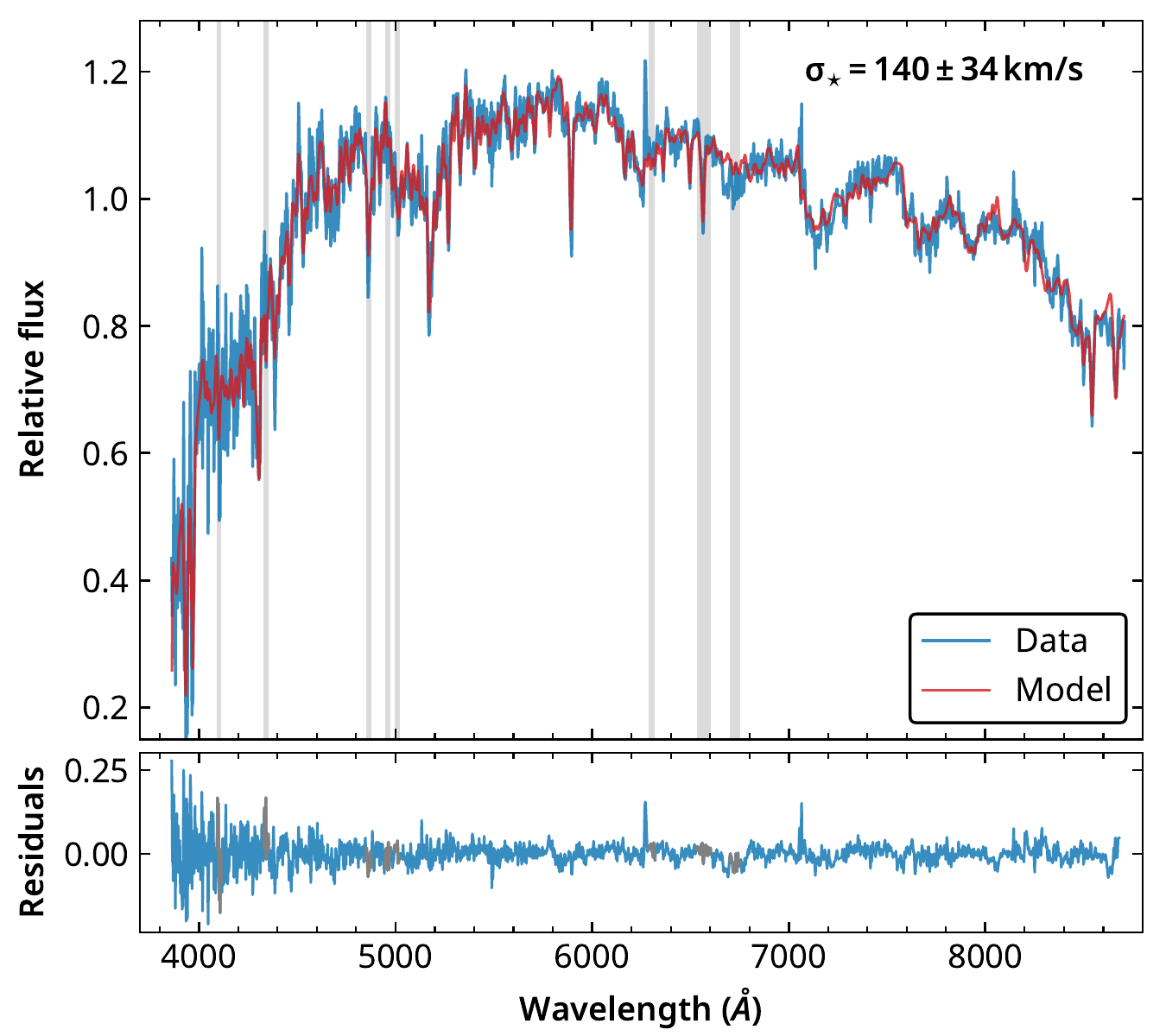}
  \caption{Optical spectroscopic observations of \jsrc and stellar velocity measurement. \textup{Left panel:} The optical spectra are rescaled and shifted for illustration purposes. The values quoted in parentheses are the offset between the date of the optical observation and the date of the closest rapid X-ray flux drop, in units of days. The Magellan LDSS3-C spectrum is corrected for telluric absorption. \textup{Right panel:} Magellan LDSS3-C rest-frame spectrum is shown with a solid dark blue line, and the solid red line shows the best-fitting model using the \texttt{pPXF} package. The blue lines at the bottom are the fit residuals, and the grey points and shadows mark the regions that are excluded from the fitting. We measured an intrinsic velocity dispersion of $140\pm34\,\text{km}\,\text{s}^{-1}$ from the Magellan data. We also measured the intrinsic velocity dispersion using the WiFeS data, which have a higher resolution than the Magellan LDSS3-C data. The intrinsic velocity dispersion obtained from the WiFeS data is $120\pm20\,\text{km\,s}^{-1}$, which is consistent with that from the Magellan data (see Sect.~\ref{subsubsec:msigma}).}
  \label{fig:m_sigma}
\end{figure*}

Overall, we found that the optical spectra are consistent with a typical quiescent galaxy with no indication of strong evolution or emission lines (see the left panel of Fig.\,\ref{fig:m_sigma}). We also extracted the optical light curves at the position of \jsrc from the ATLAS and ZTF database. No significant optical variability was found. We measured a redshift of $z=0.077$ in all spectra for \jsrc.

\subsection{ATCA radio observations\label{subsec:atca}}
We observed the position of \jsrc with the Australia Telescope Compact Array (ATCA) radio telescope on five occasions between 2021 March and 2022 April at 2--21\,GHz (project code C3334; PI: Anderson). All data were reduced in the Common Astronomy Software Application \citep[CASA v5.6.3,][]{McMullin2007} using standard procedures including flux and band-pass calibration with PKS 1934--638 and phase calibration with PKS 0454--234. Images of the target field of view were created using the \texttt{CASA} task \texttt{tclean,} and the flux density of the source was extracted in the image plane by fitting an elliptical Gaussian fixed to the size of the synthesised beam using the \texttt{CASA} task \texttt{imfit}.

No radio emission was detected at the location of \jsrc in three observations during 2021 at either 5.5 or 9\,GHz. However, on 2022 March 22, we detected a point source at both 5.5 and 9\,GHz. The coordinates of the radio source are $\text{(RA, Dec)}=\text{04h56m49.7s}, -20\degr37\arcmin47.4\arcsec$, measured from the 9\,GHz data. The positional uncertainty is characterised by an ellipse with a semi-minor axis of $0.1\arcsec$, a semi-major axis of $0.87\arcsec$, and a rotation angle of $155\degr$. We subsequently triggered a follow-up observation at 2.1, 5.5, 9, 17, and 21\,GHz on 2022 April 6 to characterise the radio spectrum of the transient (see Fig.\,\ref{fig:atca_obs}). A point source around the location of \jsrc is detected at 5.5, 9, and 17\,GHz. Unfortunately, the 2.1\,GHz field of view contains a bright AGN just outside of the primary beam, which hindered the deconvolution process and led to high image noise, even after self-calibration of the target field. The ATCA radio observations are summarised in Table~\ref{tab:ATCA}.

To assess the possibility of short-time variability of the radio emission, we extracted the 5.5 and 9\,GHz flux densities of \jsrc when the source was brightest, on 2021 March 22, for individual 10-minute scans. There is no statistically significant variability between the  six scans we analysed. The probability of the flux density at each frequency is constant at $P=0.95$ (5.5\,GHz) and $P=0.6$ (9\,GHz).

\begin{table}
\caption{ATCA radio observations of \jsrc}
\centering
\label{tab:ATCA}
\begin{tabular}{lccc}
\hline\hline
Date  & Frequency  & Array   & Flux Density  \\
    & (GHz) & config.  & ($\mu$Jy)\\
\hline

 2021-03-21   & 5.5  &  6D  & $<78$ \\

   &  9 &   & $<66$ \\

    \hline

 2021-04-26   &  5.5  &  6D  & $<36$ \\

    & 9  &    & $<47$ \\

    \hline

 2021-08-07   & 5.5  &  EW367 & $<125$ \\

   &  9 &   & $<57$ \\

    \hline

 2022-03-22   & 5  & 6A   & $257\pm11$ \\

  & 6  &   & $278\pm11$ \\

  &  9 &   & $311\pm12$ \\

    \hline

 2022-04-06   & 2.1  & 6A  &  $<738$ \\

& 5.5  &   & $61\pm18$ \\

  & 9  &   & $103\pm23$ \\

 & 17  &   & $117\pm22$ \\

 & 21  &   & $<198$ \\

\hline

\hline

\end{tabular}
\tablefoot{Upper limits are reported at 3$\sigma$.}
\end{table}

\section{Data analysis and results}\label{sec:results}

\subsection{Long-term multi-wavelength light curves}\label{subsubsec:xray}
\begin{figure}
\centering
\includegraphics[width=\columnwidth]{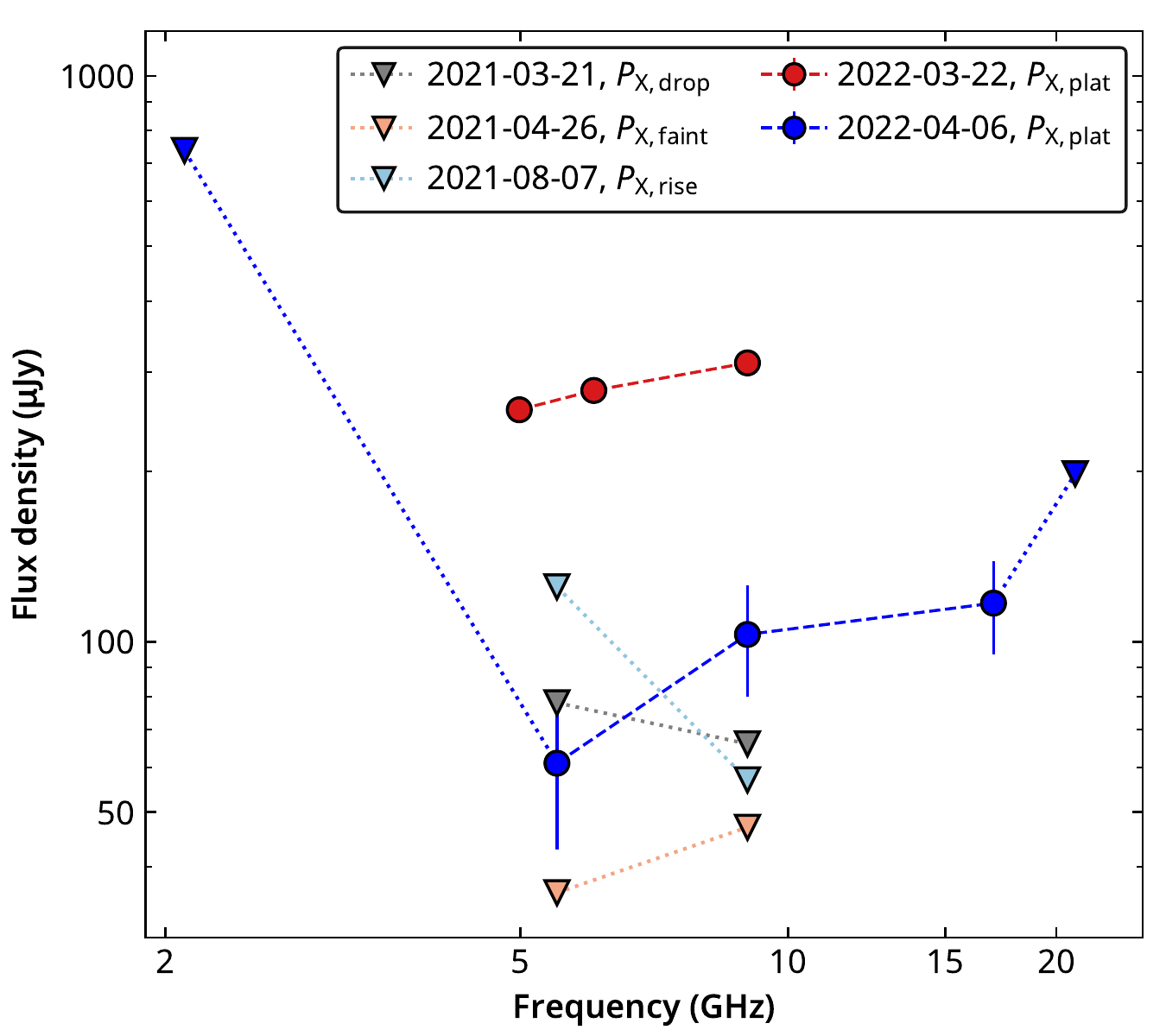}
\caption{ATCA radio observations of \jsrc. Filled circles with error bars ($1\sigma$ uncertainties) represent the observed radio flux densities at different frequencies. Downward triangles indicate the $3\sigma$ upper limits. \jsrc is detected only in the \fxplat phase (blue and red).}\label{fig:atca_obs}
\end{figure}

We show the long-term X-ray/UV light curves of \jsrc in Fig.\,\ref{fig:multi_lc}. The unabsorbed rest-frame $0.2-2.0\,\text{keV}$ X-ray flux (\sflux) was calculated from X-ray spectral modelling (see Sect.\,\ref{subsec:xray_spec}). We also show the time evolution of the radio spectra of \jsrc in Fig.\,\ref{fig:atca_obs}. \jsrc is detected only in the X-ray plateau phase in the latest two radio observations.

It is clear from Fig.\,\ref{fig:multi_lc} that \jsrc showed a drastic drop of the X-ray flux by more than a factor of $\gtrsim100$ ($\sim300$) within $ \text{about one}$ week (two weeks). By contrast, the UV only showed moderate variability, with $\Delta m\approx 1.5  (0.8)\,\text{mag}$ on a much longer timescale of several months (three weeks). Subsequent X-ray/UV follow-up observations revealed that the source showed repeating and complicated long-term X-ray variability.
We divided the long-term temporal evolution of \jsrc into four phases based on the X-ray light curve: the rapid flux drop phase (\fxdrop), the faint phase (\fxfaint), the rising phase (\fxrise), and the plateau phase (\fxplat). The phases are represented with different background colours in Fig.\ref{fig:multi_lc}. In Fig.\,\ref{fig:multi_lc} we also indicate the times of the ATCA observations. We summarise the multi-wavelength characteristics of each phase below.

\begin{figure*}
  \centering
  \includegraphics[width=\textwidth]{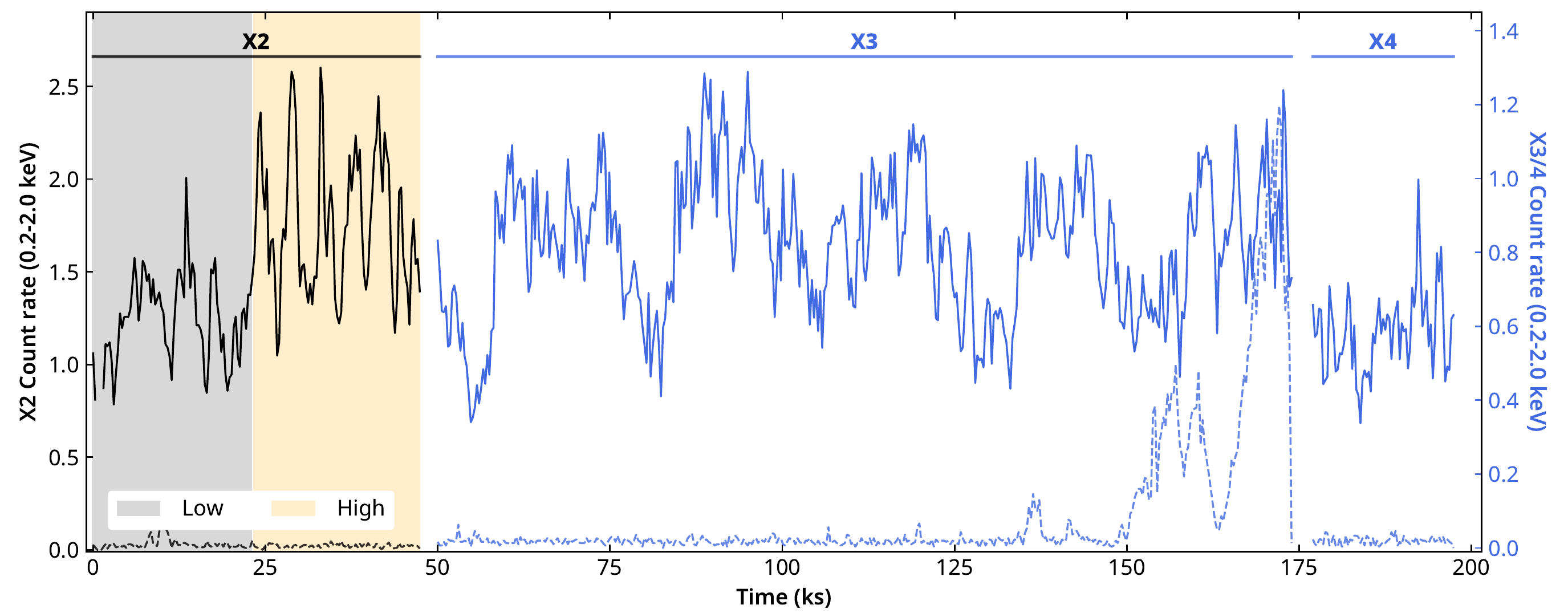}
  \caption{\mission{XMM-Newton} pn light curve in the $0.2-2.0\,\text{keV}$ energy range, with a bin size of 300\,s. The solid and dotted lines show the source and background light curves, respectively. The light curves for X2 are shown in black, and the blue lines show the light curves for X3 and X4. The high-background flare in the last 27\,ks of X3 is clearly visible. This high-background period is excluded from our temporal and spectral analysis. The time intervals we used to extract the X-ray spectra in the high- and low-flux state of X2 are marked in orange and grey, respectively.}
  \label{fig:xray_pn_lc}
\end{figure*}

\begin{table*}
\centering
\caption{Characteristics of the four different X-ray phase.}
\begin{tabular}{llcrcr}
\hline\hline
\multicolumn{2}{l}{\multirow{1}{*}{X-ray phase}} & \multicolumn{1}{l}{\multirow{1}{*}{Duration}} & \multicolumn{1}{c}{\sflux$^a$}                                     & \multicolumn{1}{c}{UV}                                                           & \multicolumn{1}{r}{\multirow{1}{*}{Radio}}  \\
\hline
\multirow{3}{*}{\fxdrop}  & 1st & $\lesssim2$~weeks  & $2.9^{+6.0}_{-1.8}\times10^{-14}$    & Sudden rise   & Non-detection \\
                          & 2nd & $\lesssim1$~weeks  & $<1.5\times10^{-13}$     & Sudden rise   & …             \\
                          & 3rd & $\lesssim1$~weeks  & $1.2_{-0.8}^{+1.3}\times10^{-13}$     & Sudden rise   & …             \\\hline
\multirow{2}{*}{\fxfaint} & 1st & $\lesssim3$~months & $<1.0\times10^{-13}$   & Bright        & Non-detection \\
                          & 2nd & $\lesssim3$~months & $<0.6\times10^{-14}$   & Bright        & …             \\\hline
\multirow{2}{*}{\fxrise}  & 1st & $\gtrsim2$~months  & \frpfirst              & Rising, then decline           & Non-detection \\
                          & 2nd & $\gtrsim1$~months  & \frpsecond             & …                            & …             \\\hline
\multirow{3}{*}{\fxplat}  & 1st & …                  & \fpffirst              & …                            & …             \\
                          & 2nd & $\sim2$~months     & \fpfsecond             & Decline                        & …             \\
                          & 3rd & $\sim2$~months     & \fpfthird              & Plateau                        & Rapid decline \\\hline
\end{tabular}
\tablefoot{Ellipsis dots mean that no observations or measurements are available.\\
\tablefoottext{a}{\sflux is the intrinsic $0.2-2.0\,$keV flux in units of $\unitflux$. For the \fxplat phase, \sflux denotes the average flux in the \fxplat phase.}}

\end{table*}

\subsubsection{Rapid X-ray flux drop phase \fxdrop}
The most prominent temporal feature of \jsrc is the drastic X-ray flux drop that occurred shortly after the eRASS3 and first \mission{Swift} (hereafter Swift1) observations around MJD~59290, during which \sflux dropped from $\sim1.2\times10^{-11}\,\unitflux$ to a $3\sigma$ upper limit of $\sim1.3\times10^{-13}\,\unitflux$ (measured from \mission{NICER} observations) in 7\,days. A faint source is detected in X1, which was taken $\sim15\,$ days after Swift1, with \sflux $=3^{+6}_{-2}\times10^{-14}\,\unitflux$ (see Section~\ref{subsubsec:xray_spec_long}). A second \fxdrop phase occurred $\sim\trecur$\,days later around MJD~59514, when the \sflux dropped by a factor of $>10$ in 6\,days. A third \fxdrop likely occurred at MJD~59698, about 408\,days after Swift1, in which \sflux decreased from $1.2^{+1.1}_{-0.6}\times10^{-12}\,\unitflux$ to $1.2^{+1.3}_{-0.8}\times10^{-13}\,\unitflux$ in 7\,days. Confirmation from subsequent X-ray observations was not possible because \jsrc was blocked by the Sun.

The UV variability, however, is very different from the X-ray during the \fxdrop phase. Although we only have sparse UV coverage during the \fxdrop, it is likely that the observed three \fxdrop phases all began at the time when the UV was faint, with UVOT/UVM2 magnitudes of $22.2\pm0.3$, $22.5\pm0.3$, and $<22.6$\,mag at the start of the first, second, and third \fxdrop, respectively. In contrast to the X-ray band, the UV brightness increased during the \fxdrop phase.

We observed \jsrc in the radio bands with ATCA soon after the X-ray flux drop (e.g. $\sim10$~days after Swift1) during the first \fxdrop phase. No significant radio emission was detected at the position of \jsrc with $3\sigma$ upper limits of 78 and 66~$\mu\text{Jy}$ in the 5.5 and 9~$\text{GHz}$ bands, respectively.

\subsubsection{X-ray faint phase \fxfaint}
Immediately after the \fxdrop phase, \jsrc was not detected in the X-ray band for $\text{about two}$\,months (hereafter the \fxfaint phase), with a $3\sigma$ flux upper limit as low as $\sim1.0\times10^{-13}\,\unitflux$ in the \mission{NICER} and \mission{Swift} monitoring programs. Combining all the \mission{Swift}/XRT non-detections during the first and second \fxfaint phases, we obtained $3\sigma$ upper limits of $\sim1.0$ and $\sim0.6\times10^{-13}\,\unitflux$ with total exposures of 7998 and 17393 seconds, respectively. We obtained a slightly lower $3\sigma$ \sflux upper limit of $\sim4.2\times10^{-14}\,\unitflux$ for the \fxfaint phase by combining all the \mission{Swift}/XRT non-detections. Using the current data set, we estimate that the \fxfaint phase lasted between 60 and 90~days.

Whilst the X-ray became considerably fainter, the UV rose to its brightest magnitude, with $m_\text{UV}=21.0\pm0.1$ and $21.3\pm0.2$ observed during the first and second \fxfaint phase, respectively. The UV brightness shows an initial rise in the second \fxfaint phase, followed by an overall decline.

We observed \jsrc with ATCA in the 5.5 and 9~$\text{GHz}$ bands during the first \fxfaint phase. The radio observation was carried out quasi-simultaneously with one of the \mission{Swift} observations during which the UV was fairly bright, with $m_\text{UV}=21.1\pm0.1$. We did not detect significant radio emission at the position of \jsrc with $3\sigma$ upper limits of 36 and 47~$\mu\text{Jy}$ in the 5.5 and 9~$\text{GHz}$ bands, respectively.

\subsubsection{X-ray rising phase \fxrise}
The two \fxrise phases, both following a \fxfaint phase, were well captured by our \mission{Swift} and \mission{NICER} monitoring with an observed faintest \sflux of $1.8^{+1.7}_{-1.1}\times10^{-13}\,\unitflux$. To characterise the profile of the X-ray light curve during the \fxrise phase, we modelled the first \fxrise phase X-ray light curve (MJD 59390--59460) with a power-law function $f_\text{rs}(t)=A*(t-t_0)^{\beta}$. The Python package \texttt{emcee} was used to estimate the uncertainties. We obtained a best-fitting index of $\sim2.3\pm0.7$ and $t_0=\mathrm{MJD}\,59377\pm9$. Interestingly, this best-fitting power-law function can match the light curve profile of the second \fxrise phase very well by simply shifting $t_0$ by $\sim\trecur$~days (see Fig.\,\ref{fig:multi_lc}). In contrast, the peak \sflux in the second \fxrise phase (\frpsecond~$\unitflux$) is fainter than that in the first phase (\frpfirst~$\unitflux$). We also note that significant short-term X-ray variability is clearly detected in the \mission{XMM-Newton} observation (Fig.\,\ref{fig:xray_pn_lc}) at a time close to the peak X-ray flux in the \fxrise phase. The second \fxrise phase ($\gtrsim\frdsecond$~days) is $\sim30$\,days shorter than the first phase.

The UV light curve is well sampled only in the first \fxrise phase. \jsrc is generally relatively bright in the UV band during the \fxrise phase. The UV was initially in a rising phase, which is similar to the X-ray, followed by a decline, while the X-ray flux continued to increase. However, this cannot be confirmed in the second \fxrise phase because we lack UV observations.

A radio observation by ATCA was carried out quasi-simultaneously with the X-ray/UV observation during the first \fxrise phase. We again did not detect significant radio emission at the position of \jsrc, with $3\sigma$ upper limits of 125 and 57~$\mu\text{Jy}$ in the 5.5 and 9~$\text{GHz}$ bands, respectively. We note that \jsrc was in a fairly X-ray bright state with \sflux of $\sim3.0\times10^{-12}\,\unitflux$ at the time of the ATCA observation.

\subsubsection{X-ray plateau phase \fxplat}
Following the first and second \fxrise phases, the X-ray flux of \jsrc was stable within a factor of 2 for $\text{about two}$\,months (the \fxplat phase) during MJD~\fpsecondstart--\fpsecondend\xspace and \fpthirdstart--\fpthirdend\xspace (Fig.~\ref{fig:multi_lc}) before it transited to the \fxdrop phase. During the \fxplat phases, significant short-term variability by a factor of 2 on a timescale of a few hours (Fig.\ref{fig:xray_pn_lc}) was seen in the \mission{XMM-Newton} observations. The \sflux observed in the two eRASS3 segments ($7.8\pm0.8\times10^{-12}$ and $1.1^{+0.2}_{-0.1}\times10^{-11}\,\unitflux$) and the Swift1
observation ($1.2^{+0.2}_{-0.1}\times10^{-11}\,\unitflux$) are very similar. In addition, their X-ray spectral properties (discussed below) are also consistent with the properties observed in the \fxplat phase. We thus concluded that \jsrc was in the \fxplat phase during the eRASS3 and Swift1 observations (MJD\,59272--59285, the first observed \fxplat phase). The average $0.2-2.0\,\text{keV}$ flux in each \fxplat phase showed a monotonic decrease. Compared to the first \fxplat, the average \sflux decreased by a factor of $\sim4$ and 7 in the following \fxplat phases, respectively. The duration of the \fxplat is estimated to be \fpdsecond~days for the second \fxplat and \fpdthird~days for the third \fxplat, implying that the duration of the \fxplat phase does not change significantly. A dip for $\text{less than about two}$ weeks can also be seen in the X-ray band in both the second and third \fxplat phases. The \sflux dropped by a factor of $\sim2$ during the two X-ray dips.

The UV showed very different variability during the second and third \fxplat phases. In the second \fxplat phase, a decline in the UV brightness is clearly seen in the \mission{Swift}/UVM2 light curves, with an indication of a short UV plateau phase at the beginning of the \fxplat. In contrast, the overall UV brightness does not vary significantly during the third \fxplat phase, with evidence of short-term variability on a timescale of $\text{about one}$~week.

We carried out two radio observations during the third \fxplat phase. \jsrc was detected in both radio observations. The flux densities of \jsrc during the first observation measured at 5, 6, and 9~GHz are $257\pm11$, $278\pm11$, and $311\pm12~\mu\text{Jy}$ (Fig.\,\ref{fig:atca_obs}), respectively. For the second radio observation, which was performed $\sim16$~days after the first, the source was detected in the 5.5, 9, and 17~GHz with flux density of $61\pm18$, $103\pm23$, and $117\pm22~\mu\text{Jy}$ (Fig.\,\ref{fig:atca_obs}), respectively. We measured a $3\sigma$ upper limit of $738$($198$)~$\mu$Jy at the 2.1(22)~GHz band. Our results suggest that the radio emission during the two detections declined rapidly; the flux density decreased by a factor of $\sim3(4)$ in $\text{about two}$~weeks in the 9(5.5)~GHz band (see Fig.~\ref{fig:atca_obs}). Overall, the radio flux density at 9(5.5)~GHz increased by factors of $\gtrsim5(3)$, 6(7), and 5(2) compared to that in the \fxdrop, \fxfaint, and \fxrise phase, respectively.

\begin{figure*}
  \centering
  \includegraphics[width=1.0\columnwidth]{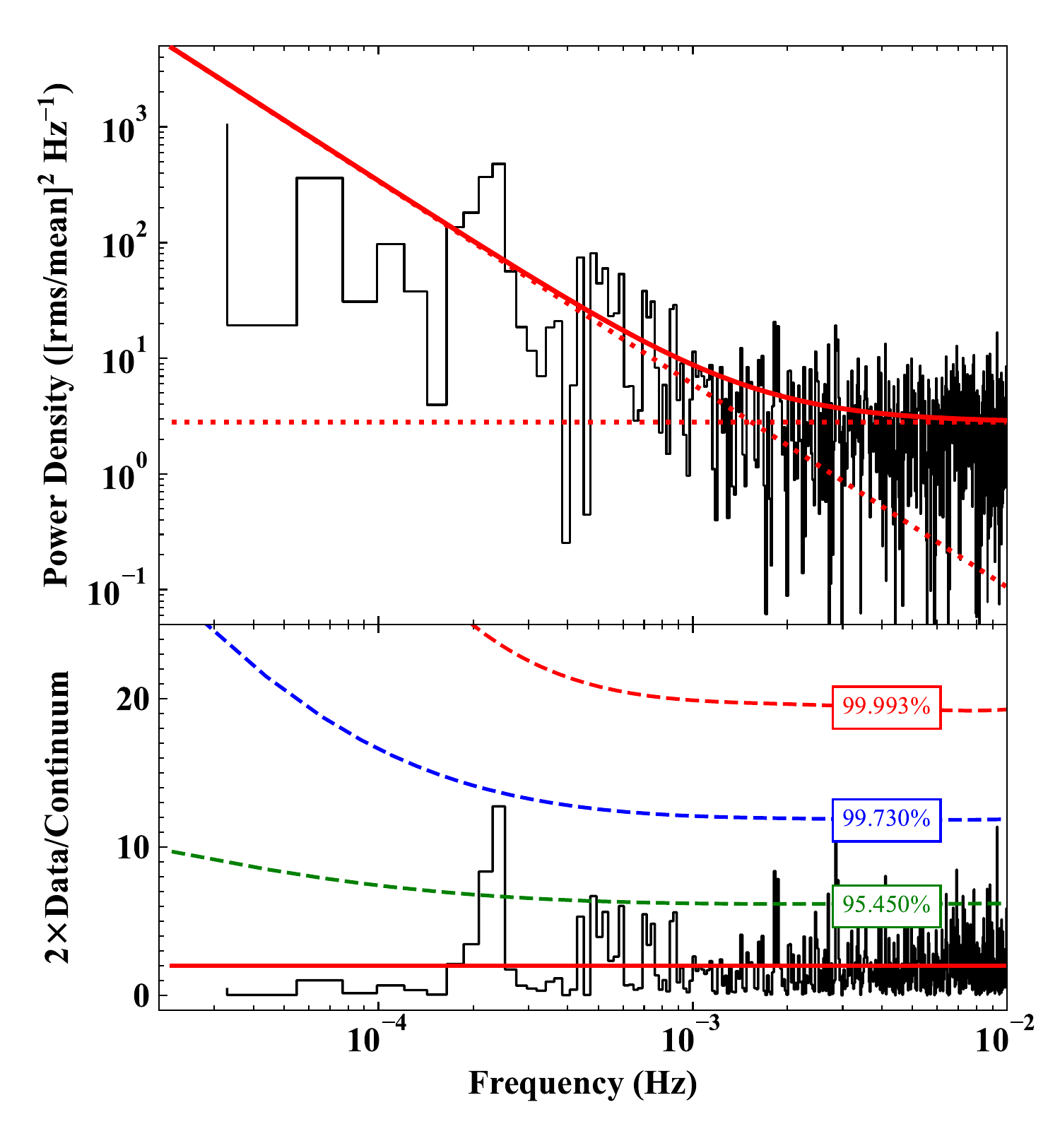}
  \includegraphics[width=1.0\columnwidth]{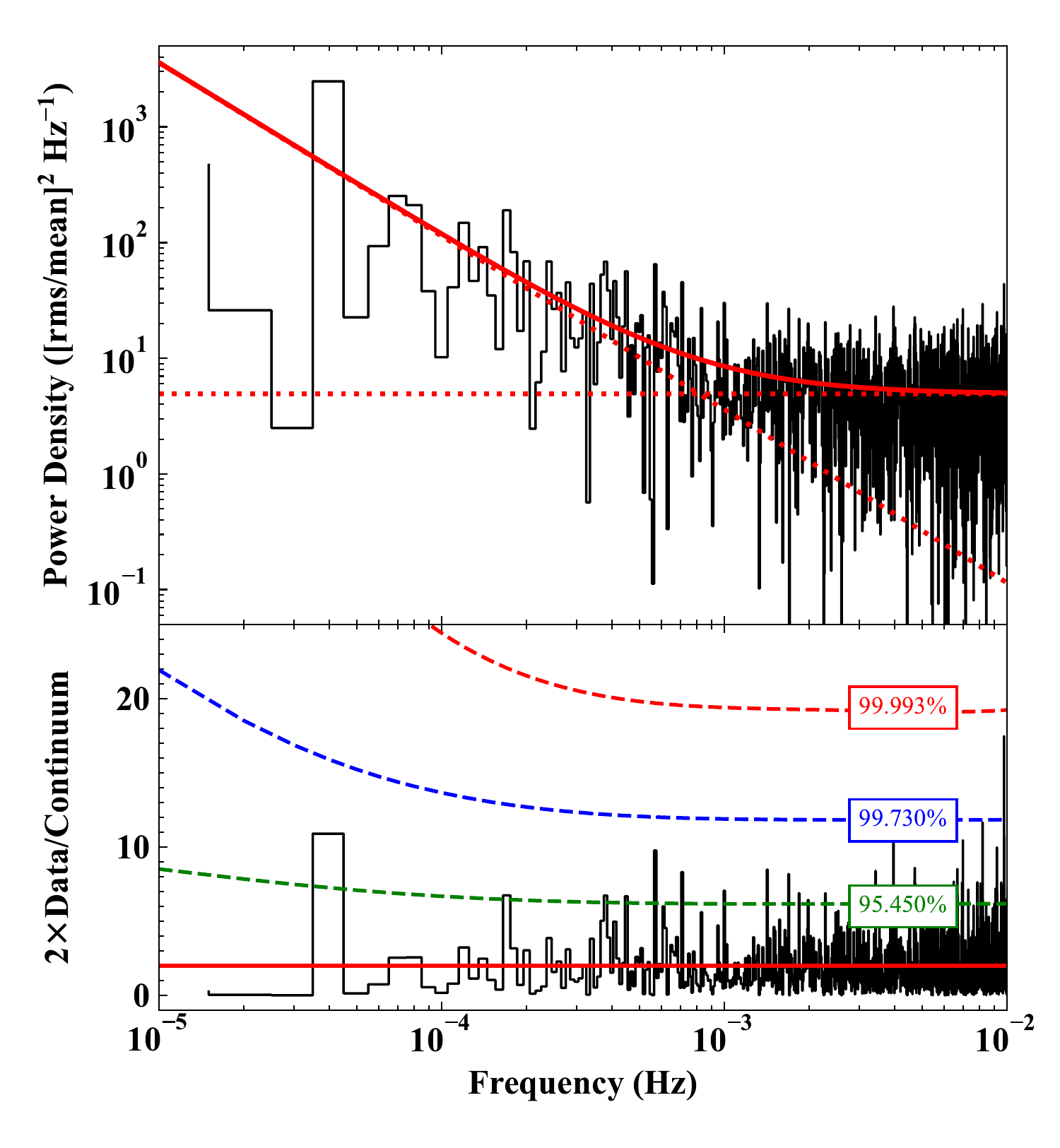}
  \caption{PSDs for X2 and X3. \textup{Left panel:} PSD calculated using the pn light curve of X2 in the $0.3-0.8\,\text{keV}$ energy range with time bin-size of 50\,s (upper panel). The solid red line indicates the best-fitting power-law plus Poisson constant model. The dotted red line shows the power-law component. The bottom panel shows the ratio of the data to the model multiplied by 2.  This ratio roughly indicates the significance of the deviation of the observed power from the model continuum at a given frequency. The 2, 3, and $4\sigma$ confidence levels are marked with dashed green, blue, and red lines, respectively. \textup{Right panel:} Same as the left panel, but for X3. A peak at a frequency of $2.5\times10^{-4}\,\text{Hz}$ is seen in the PSD of X2 with a significance of $\sim2.9\sigma$.}
  \label{fig:xray_psd}
\end{figure*}

\subsection{Short-term X-ray temporal variability}
The \mission{XMM-Newton} pn $0.2-2.0\,\text{keV}$ light curves for X2, X3, and X4 (Fig.\,\ref{fig:xray_pn_lc}) exhibit significant short-term variability with an amplitude larger than a factor of 2 on a timescale of hours during all three \mission{XMM-Newton} observations.  We performed a timing analysis in the frequency domain in order to better characterise the broadband variability properties, characterise the putative repeating signal implied by the light curve (henceforth referred to as a quasi-periodic oscillation, or QPO), and to deconvolve a putative QPO from stochastic, broadband red-noise variability. We calculated the power spectral densities (PSDs) for X2 and X3 in the very soft $0.3-0.8\,\text{keV}$ energy band. The first 1750\,s data from X2 were excluded because of the large gap in the light curve. Only the first 100~ks data of X3 were used as the remaining 27~ks were severely affected by high-background flaring.

The calculated PSDs from X2 and X3 are shown in the left and right panels of Fig.~\ref{fig:xray_psd}, respectively. The PSDs were normalised so that the integrated PSD represents the fractional rms variability \citep[e.g.][]{belloni_hasinger1990}. We found an indication of a tentative QPO signal in the PSDs of X2.

However, stochastic red-noise processes will generate power in the PSD across a wide band of frequencies. These processes can spuriously mimic a few-cycle QPO, particularly in the low-frequency regime of the periodogram, where data sampling can be relatively sparse \citep{vaughan_etal2016}. We thus emphasise that because of the low temporal frequency of the QPO and the very small number of cycles sampled, any claim of a QPO (any claim that we have successfully deconvolved red noise from a QPO signal) is tentative at best. To assess the significance of the QPO, we first modelled the PSDs and then ran a Markov chain Monte Carlo (MCMC) code to estimate the confidence level of the QPO. The PSDs were fit with a power-law continuum model, adding a constant component to account for the Poisson white noise. The maximum likelihood estimate (MLE) method \citep{vaughan2010} was used to estimate the model parameters. We found a best-fitting power-law index of $1.75\pm0.22$ ($1.50\pm0.18$) for X2 (X3). A QPO-like signal at frequency of $2.5\times10^{-4}\,\text{Hz}$ is shown in the ratio of the data to model times 2 ($2R_\text{PSD}$, bottom panel in Fig.\,\ref{fig:xray_psd}). This ratio roughly indicates the significance of the deviation of the observed power from the model continuum at a given frequency \citep{vaughan2010}.

To calculate the significance of the signal, we first simulated the continuum model using the initial values of the MLE parameters, assuming a uniform prior probability density function \citep{vaughan2010}. The \textsc{PYTHON} \texttt{emcee} package was used to perform MCMC sampling in order to draw from the posterior of model parameters. We generated $10^{5}$ posterior predictive periodograms and simulated light curves for each of them using the method presented in \citet{timmer_koenig1995}. The MLE method  was then applied to fit the simulated light curves using the same power-law plus white-noise model, and the $2R_\text{PSD}$ was calculated for each of the $10^{5}$ simulated light curves. The 2, 3, and $4\sigma$ confidence levels were then estimated by calculating the 95.45, 99.73, and 99.993 percentiles of the $2R_\text{PSD}$ at each frequency bin (see the dashed lines in the bottom panel of Fig.\,\ref{fig:xray_psd}). Our results suggest that a signal at a frequency of $2.5\times10^{-4}\,\text{Hz}$ is tentatively detected at a $2.9\sigma$ level in X2, but no significant signal is found in X3.

\subsection{X-ray spectral analysis\label{subsec:xray_spec}}

\begin{table*}
\caption{Results of the X-ray spectral fitting. $N_\text{H, Gal}$ is the column density of the Galactic absorption; $E_\text{Br}$ is the break energy in the \texttt{bknpower} component; $\Gamma_1$ and $\Gamma_2$ are the photon indices at lower and higher energies compared to $E_\text{Br}$, respectively; $kT$ is the temperature in the \texttt{bbody} model; $T_\text{in}$ is the temperature of the inner region of the accretion disk in the \texttt{diskbb} model; $\Gamma$ is the photon index in the \texttt{powerlaw} model; $\Gamma_\text{hot}$ and $\text{CF}_\text{hot}$ are the power-law photon index and covering fraction for the hot corona, respectively, modelled with the \texttt{thcomp} model; $\Gamma_\text{warm}$ and $\text{CF}_\text{warm}$ are for the warm corona; $kT_\text{e}$ is the electron temperature of the warm corona; $f_\text{soft, diskbb}$, $f_\text{soft, bb}$, $f_\text{soft, po}$ are the intrinsic flux calculated over the rest-frame $0.2-2.0$~keV band for the \texttt{diskbb}, \texttt{bbody}, and \texttt{powerlaw} model, respectively; $f_\text{soft}$ is the total intrinsic $0.2-2.0$~keV flux.}
\label{tab:xray_spec}
\begin{tabular}{@{}lccccccc@{}}
\multicolumn{7}{c}{\mbkp \texttt{const*TBabs*(cflux*bknpower)}}\\[0.5mm]
\hline\hline
Parameters & $N_\text{H, Gal}$         & $\Gamma_1$             & $E_\text{Br}$          & $\Gamma_2$             & $\log\,f_\text{soft}$   & $\chidof$    \\
           & ($10^{20}\,\mathrm{cm}^{-2}$)  &  & (keV)&  & ($\unitflux$)  &  &      \\\hline\hline
X2       & $4.4\pm{0.2}$        & $3.43\pm{0.03}$        & $1.86_{-0.14}^{+0.18}$ & $2.54_{-0.13}^{+0.11}$ & $-11.21\pm{0.02}$ & $291.3/263$  \\[0.5mm]
X3       & $5.1\pm{0.3}$        & $3.31\pm{0.03}$        & $1.93_{-0.13}^{+0.16}$ & $2.46_{-0.12}^{+0.10}$ & $-11.49\pm{0.02}$ & $287.6/287$  \\[0.5mm]
X4       & $5.4_{-0.6}^{+0.8}$  & $3.33_{-0.07}^{+0.10}$ & $1.72_{-0.15}^{+0.37}$ & $2.61_{-0.18}^{+0.20}$ & $-11.61\pm{0.04}$ & $141.7/145$  \\[0.5mm]
\xslo     & \multirow{2}{*}{$4.6\pm{0.2}$}        & $3.35_{-0.03}^{+0.04}$ & $2.23_{-0.27}^{+0.20}$ & $2.30_{-0.20}^{+0.22}$ & $-11.30\pm{0.02}$ & \multirow{2}{*}{$444.3/412$}  \\[0.5mm]
\xshi     &                                       & $3.52\pm{0.04}$        & $1.72_{-0.11}^{+0.12}$ & $2.60_{-0.11}^{+0.10}$ & $-11.12\pm{0.02}$ &     \\[0.5mm]
\hline
\smallskip
\end{tabular}

\begin{tabular}{@{}lccccccc@{}}
\multicolumn{8}{c}{\mpbb \texttt{const*TBabs*zashift*(cflux*powerlaw+cflux*bbody)}}\\[0.5mm]
\hline\hline
Parameters & $N_\text{H, Gal}$         & $kT$                  & $\Gamma$                & $\log\,f_\text{soft, bb}$ & $\log\,f_\text{soft, po}$ & $\log\,f_\text{soft}$   & $\chidof$    \\
           & ($10^{20}\,\mathrm{cm}^{-2}$) & (eV) &  &  ($\unitflux$)   & ($\unitflux$) &  ($\unitflux$)  &      \\\hline
X2       & $2.3\pm{0.3}$        & $108\pm{4}$           & $2.86\pm{0.05}$         & $-12.03\pm{0.04}$         & $-11.54\pm{0.03}$   & $-11.42\pm0.03$   & $367.8/263$  \\[0.5mm]
X3       & $2.5\pm{0.3}$        & $119\pm{4}$           & $2.75\pm{0.05}$         & $-12.35\pm{0.04}$         & $-11.82\pm{0.03}$   & $-11.71\pm0.03$   & $367.0/287$  \\[0.5mm]
X4       & $2.3\pm{0.8}$        & $128_{-10}^{+9}$      & $2.74\pm{0.13}$         & $-12.50_{-0.12}^{+0.08}$  & $-11.97\pm{0.09}$   & $-11.86\pm0.07$   & $145.1/145$  \\[0.5mm]
\xslo     & \multirow{2}{*}{$2.3\pm{0.3}$}        & $117\pm{5}$           & $2.84\pm{0.07}$         & $-12.18_{-0.08}^{+0.06}$  & $-11.62\pm{0.04}$   & $-11.51\pm0.04$   & \multirow{2}{*}{$519.9/412$}  \\[0.5mm]
\xshi     &         & $105\pm{4}$           & $2.86\pm{0.06}$         & $-11.91_{-0.05}^{+0.04}$  & $-11.50\pm{0.04}$   & $-11.35\pm0.03$   &   \\[0.5mm]
\hline
\smallskip
\end{tabular}

\begin{tabular}{@{}lccccccc@{}}
\multicolumn{8}{c}{\mpdb \texttt{const*TBabs*zashift*(cflux*powerlaw+cflux*diskbb)}}\\[0.5mm]
\hline\hline
Parameters & $N_\text{H, Gal}$         & $T_\text{in}$          & $\Gamma$               & $\log\,f_\text{soft, diskbb}$   & $\log\,f_\text{soft, po}$ & $\log\,f_\text{soft}$   & $\chidof$    \\
           & ($10^{20}\,\mathrm{cm}^{-2}$) & (eV) &  &  ($\unitflux$)   & ($\unitflux$) &  ($\unitflux$)  &      \\\hline
X2       & $2.4\pm{0.3}$        & $144\pm_{-6}^{+5}$     & $2.76\pm{0.06}$        & $-11.85_{-0.04}^{+0.03}$  & $-11.61\pm{0.04}$   & $-11.41\pm{0.03}$ & $338.3/263$ \\[0.5mm]
X3       & $2.6\pm{0.3}$        & $160\pm{6}$            & $2.64\pm{0.06}$        & $-12.15_{-0.04}^{+0.03}$  & $-11.89\pm{0.03}$   & $-11.70\pm{0.03}$ & $333.3/287$ \\[0.5mm]
X4       & $2.5_{-0.7}^{+0.8}$  & $171_{-15}^{+14}$      & $2.64\pm{0.16}$        & $-12.30_{-0.11}^{+0.07}$  & $-12.05\pm{0.11}$   & $-11.86\pm{0.08}$ & $144.7/145$  \\[0.5mm]
\xslo     & \multirow{2}{*}{$2.4\pm{0.3}$}        & $154_{-8}^{+7}$        & $2.75\pm{0.09}$        & $-11.98_{-0.07}^{+0.06}$  & $-11.68\pm{0.06}$   & $-11.51\pm{0.04}$ & \multirow{2}{*}{$491.8/412$} \\[0.5mm]
\xshi     &        & $139\pm{6}$            & $2.77\pm{0.07}$        & $-11.74\pm{0.04}$         & $-11.57\pm{0.05}$   & $-11.34\pm{0.03}$ &  \\[0.5mm]
\hline
\smallskip
\end{tabular}

\begin{tabular}{@{}lccccccccc@{}}
\multicolumn{10}{c}{\mcomp \texttt{const*TBabs*zashift*(cflux*thcomp*thcomp*diskbb)}}\\
\hline\hline
Parameters & $N_\text{H, Gal}$  & $T_\text{in}$ & $\Gamma_\text{hot}$ & $\text{CF}_\text{hot}$ & $\Gamma_\text{warm}$ & $kT_\text{e}$ & $\text{CF}_\text{warm}$ & $\log\,f_\text{soft}$ & $\chidof$  \\
           & ($10^{20}\,\mathrm{cm}^{-2}$) & (eV) &  &    & & (keV) & & ($\unitflux$)  &      \\\hline
X2 & $3.5\pm0.4$         & $64_{-10}^{+13}$ & $2.40_{-0.18}^{+0.14}$ & $0.14_{-0.05}^{+0.06}$ & $3.04_{-0.85}^{+0.36}$ & $0.25_{-0.07}^{+0.10}$ &  $\geq0.23$ & $-11.32\pm0.09$          & 283.5/261 \\[0.5mm]
X3 & $3.8_{-0.4}^{+0.5}$ & $61_{-7}^{+21}$  & $2.29_{-0.14}^{+0.13}$ & $0.14_{-0.05}^{+0.06}$ & $2.97_{-0.65}^{+0.15}$ & $0.27_{-0.07}^{+0.06}$ &  $\geq0.30$ & $-11.62_{-0.07}^{+0.12}$ & 280.0/285 \\[0.5mm]\hline
\end{tabular}
\tablefoot{X2 was taken at the end of the \fxrise phase. X3 and X4 were taken during the \fxplat phase.}
\end{table*}

The \textsc{Xspec} software (version \texttt{12.12.0a}; \citealt{arnaud1996}) was used to fit all X-ray spectra. The \mission{Swift}/XRT X-ray spectra have low photon counts, thus the Cash statistic (\citealt{cash1979}, Cstat in \textsc{Xspec}) is used. The $\chi^2$ statistic is used for \mission{XMM-Newton} and \mission{NICER} spectral fitting. The \textsc{Xspec} models \texttt{TBabs} and \texttt{zTBabs} (\citealt{wilms_etal2000}, abundances were set to \texttt{wilms} in \textsc{Xspec}) were used to model the Galactic and host galaxy absorption, respectively.

\subsubsection{Modelling time-averaged \mission{XMM-Newton} data}\label{subsubsec:xmm_ave_spec}
\begin{figure}
  \centering
        \includegraphics[width=\columnwidth]{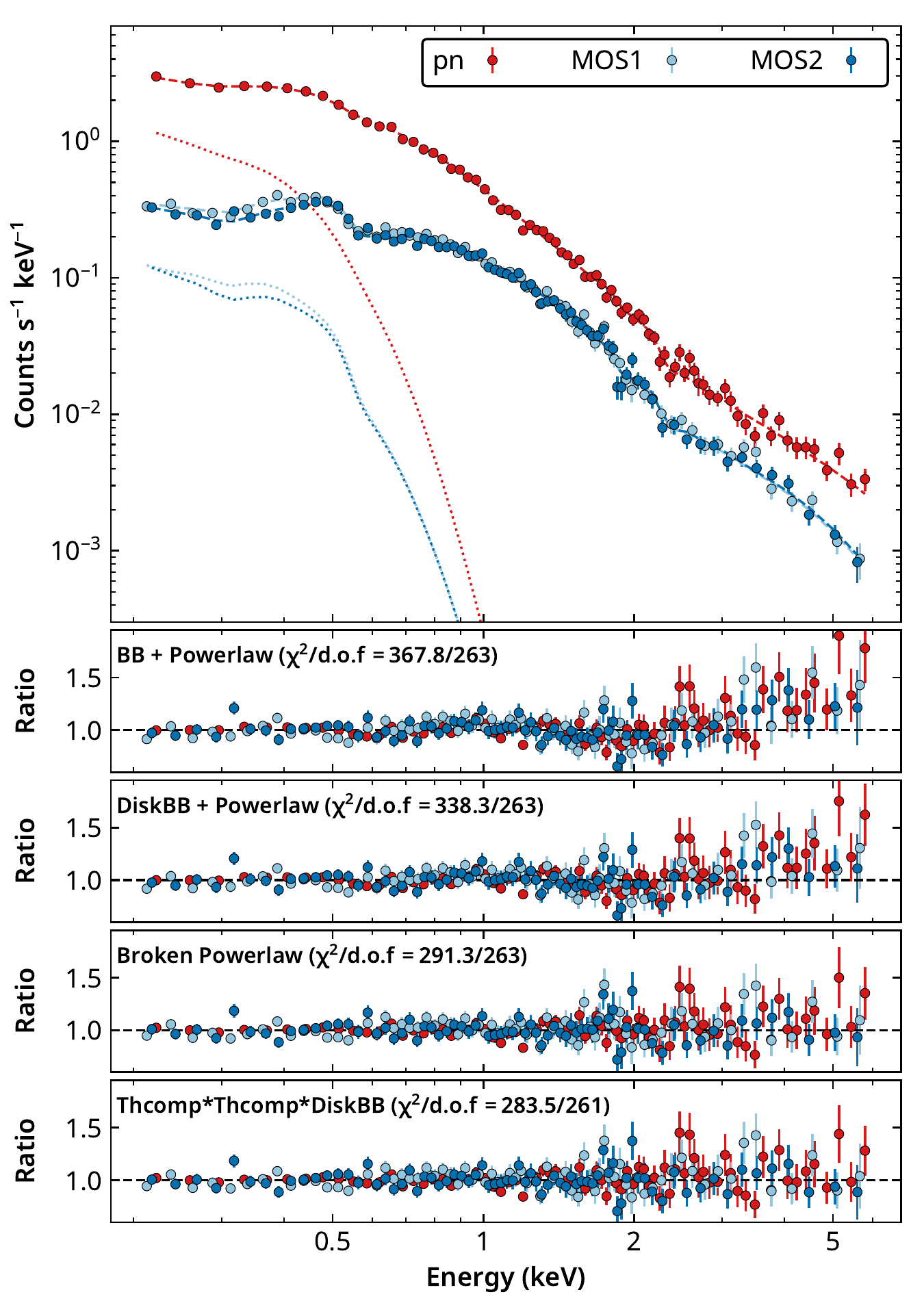}
  \caption{X-ray spectral modelling. \textup{Upper panel:} X-ray spectra of \jsrc using the data from X2, with EPIC/pn shown in red, EPIC/MOS1 in dark blue, and EPIC/MOS2 in light blue. The best-fitting Comptonisation model (\mcomp) is shown as dashed lines, and the dotted lines represent the \texttt{diskbb} component in the \mcomp model. \textup{Bottom panels:} Data/model ratio for the \mpbb, \mpdb, \mbkp, and \mcomp models (from top to bottom).}
  \label{fig:spec_fit}
\end{figure}

The three \mission{XMM-Newton} observations (X2, X3, and X4) provide the best S/N for X-ray spectral modelling. The data above $6.0~\text{keV}$ are dominated by background; we thus fitted the data from the three EPIC cameras (pn, MOS1, and MOS2) simultaneously over the $0.2-6.0~\text{keV}$ energy range. A constant component was added to account for the flux calibration difference between the three cameras.

We first fitted each of the time-averaged data of the three \mission{XMM-Newton} observations (i.e. X2, X3, and X4) with a simple power-law model modified by both Galactic and host galaxy absorption  (i.e. \texttt{const*TBabs*zTBabs*cflux*powerlaw} in \textsc{Xspec}). No strong evidence for significant host galaxy absorption is found in each data set. We thus removed the host galaxy absorption component from our spectral modelling (the \mpl model, i.e. \texttt{const*TBabs*cflux*powerlaw}). The simple power-law model cannot fit the X-ray spectra of X2 well with $\chidof=569.8/265$. The fit can be significantly improved by adding a blackbody component, that is, \texttt{const*TBabs*zashift*(cflux*powerlaw+cflux*bbody)} (hereafter \mpbb), which gives a $\chidof$ of $367.8/263$. The spectral fitting can be further improved by replacing the blackbody component with a multi-colour disk component, that is, \texttt{const*TBabs*zashift*(cflux*powerlaw+cflux*diskbb)} (hereafter \mpdb), which gives a $\chidof$ of $338.3/263$.

The same models were also used to fit the X3 and X4 data. We again found that the \mpl model cannot fit the X3 and X4 data well with $\chidof$ of $577.3/289$ and $177.9/147$, respectively. We obtained acceptable fits for X4 with both the \mpbb ($\chidof=145.1/145$) and \mpdb ($\chidof=144.7/145$) models. Similar to X2, the \mpdb model, which gives a $\chidof$ of $333.3/287$, can fit the X3 spectra better than the \mpbb model ($\chidof=367.0/287$). However, systematical features at energies above $3~\text{keV}$ are clearly seen in the residuals for both X2 and X3 (see Fig.\,\ref{fig:spec_fit}). The details of the spectral fitting are listed in Table\,\ref{tab:xray_spec}.

The excesses at higher energy for X2 and X3 suggests the need for an additional spectral component. We then tried to fit the data with a broken power-law model (\texttt{const*TBabs*(cflux*bknpower)}, hereafter \mbkp). The \mbkp model can fit both the X2 and X3 spectra well with $\chidof$ of $291.3/263$ and $287.6/287$, respectively. The best-fitting break energy for X2 is $1.86_{-0.14}^{+0.18}\,\text{keV}$ with $\Gamma_1=3.43\pm{0.03}$ and $\Gamma_2=2.54_{-0.13}^{+0.11}$. We obtained a shallower $\Gamma_1$ ($3.31\pm{0.03}$) for X3, while the best-fitting $E_\text{Br}=1.93_{-0.13}^{+0.16}$ and $\Gamma_2=2.46_{-0.12}^{+0.10}$ are consistent within their uncertainties with X2. The \mbkp model can also fit the X4 data well. The best-fitting parameters for X4 are consistent with those from X2 and X3. Overall, the \mbkp model is preferred over the \mpbb and \mpdb models for both X2 and X3, while all the three models give acceptable fits for X4. In Table\,\ref{tab:xray_spec} we list the best-fitting parameters of the \mbkp model for the three observations.

The broken power-law fitting may be an indication that there are two coronae in \jsrc. The power-law component in the soft X-ray band could be due to inverse Comptonisation of the soft accretion disk photons by a warm corona with an electron temperature of a few keV or lower, and a hot corona may contribute to the hard-band power law. We then fit the X2 and X3 data with a model consisting of two inverse Comptonisation components, \texttt{const*TBabs*zashift*(cflux*thcomp*thcomp*diskbb)} (\mcomp). This model was used to fit the X-ray spectra of bright AGNs \citep[e.g.][]{jin_etal2021}. To better constrain the temperature of the multi-colour component, which may well peak in the UV band, we fitted the $0.2-6.0\,\text{keV}$ X-ray data along with the OM/UVM2 data. The Galactic dust reddening is estimated to be $E_{\text{B-V}}=0.03$\,mag using the dust map \citep{schlafly_finkbeiner2011}. The electron temperature of the hot corona cannot be constrained and was therefore fixed at $150\,\text{keV}$. Overall, the best-fitting parameters for the two time-averaged spectra are consistent with each other. We found that a low electron temperature is required for the warm corona components for both X2 ($kT_\text{e}=0.25^{+0.10}_{-0.07}\,\text{keV}$) and X3 ($kT_\text{e}=0.27^{+0.06}_{-0.07}\,\text{keV}$). The covering factor of the hot corona is small ($\text{CF}_\text{hot}=0.14_{-0.05}^{+0.06}$) in both cases, and only a lower limit can be given for the warm corona ($\text{CF}_\text{warm}$ $>0.23$ for X2, and $>0.32$ for X3). The inner temperature of the multi-colour disk is constrained to be $\approx60\,\text{eV}$ for both data with $T_\text{in}=64_{-10}^{+13}\,\text{eV}$ for X2 and $61_{-7}^{+21}\,\text{eV}$ for X3. The details of the best-fitting parameters and the uncertainties are listed in Table\,\ref{tab:xray_spec}. The results suggest that the time-averaged X-ray spectral properties do not evolve significantly over a timescale of several months.

\subsubsection{Short-term X-ray spectral variability}\label{sec:xray_spec_short}
\begin{figure}
  \centering
  \includegraphics[width=\columnwidth]{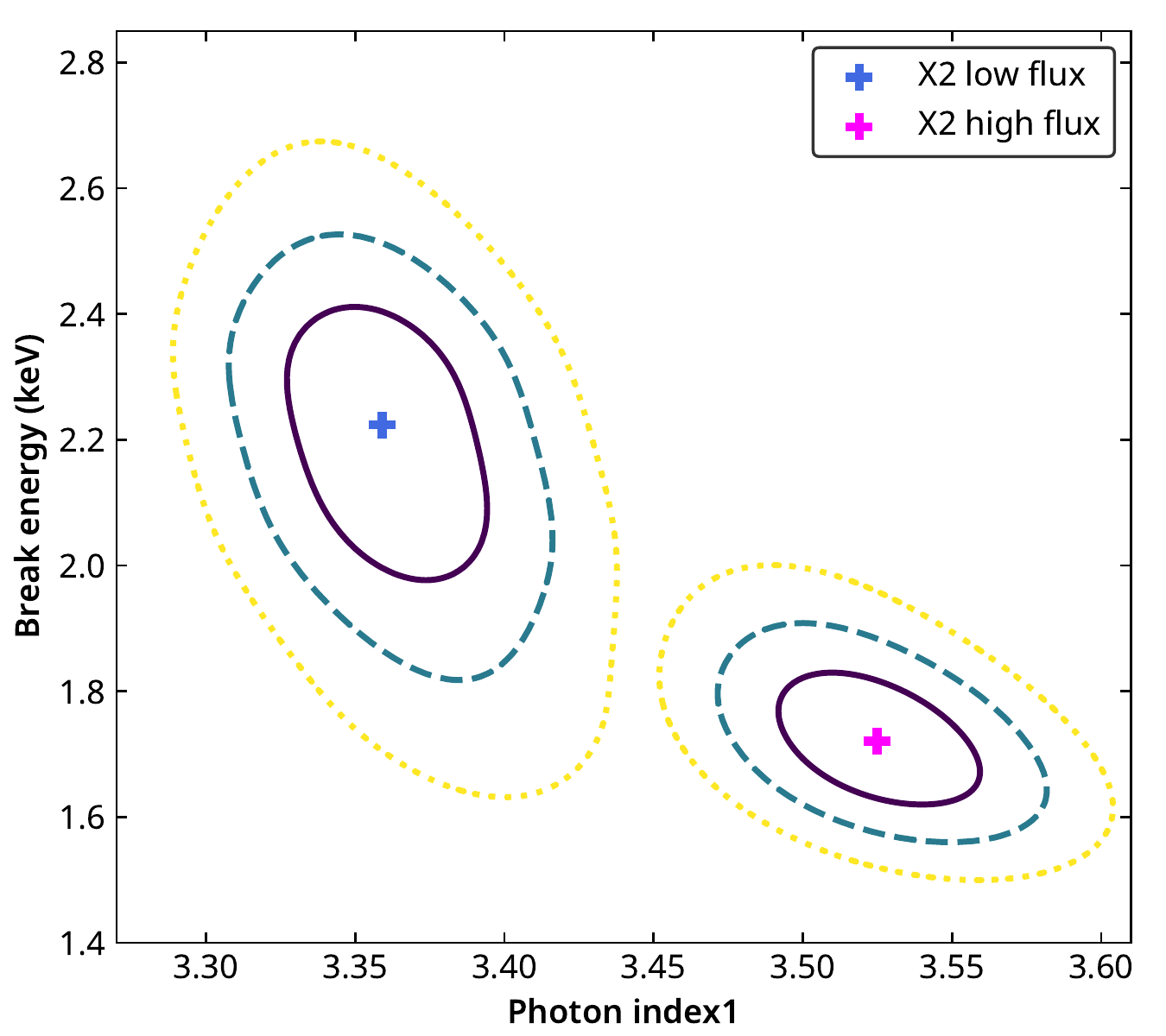}
  \caption{Contours of the photon index vs break energy for \xslo and \xshi; solid, dashed, and dotted lines denote 1 ($\Delta\chi^2=2.3$), 2 ($\Delta\chi^2=6.18$), and 3$\sigma$ ($\Delta\chi^2=11.83$) confidence contours, respectively.}
  \label{fig:phoind_brE}
\end{figure}

Fig.\,\ref{fig:xray_pn_lc} shows significant X-ray variability in   \jsrc on timescales as short as a few hours during the X2 and X3 observations. To test whether the X-ray spectral properties evolved on short timescales, we split the X2 event lists into two segments. As shown in Fig.\,\ref{fig:xray_pn_lc}, the source was in a relatively low X-ray flux state in the first segment (\xslo), while the overall X-ray flux is higher in the second segment (\xshi). The same models were used to fit the X-ray spectra extracted from the two segments. We found that the temperature of the blackbody (multi-colour disk) is slightly lower in the high flux state when it is fitted with the \mpbb (\mpdb) model (see Table\,\ref{tab:xray_spec}). The photon index in the soft X-ray band changed significantly in the two flux states when fitted with the \mbkp model. This is clearly illustrated in Fig.\,\ref{fig:phoind_brE} , in which we show the $\Delta\chi^2$ contours for the photon index in the soft X-ray band and the break energy in the \mbkp model. Our spectral analysis suggests that \jsrc shows rapid X-ray spectral evolution on a timescale of a few hours.

\subsubsection{Long-term X-ray spectral evolution}\label{subsubsec:xray_spec_long}
\begin{figure}
  \centering
  \includegraphics[width=1.0\columnwidth]{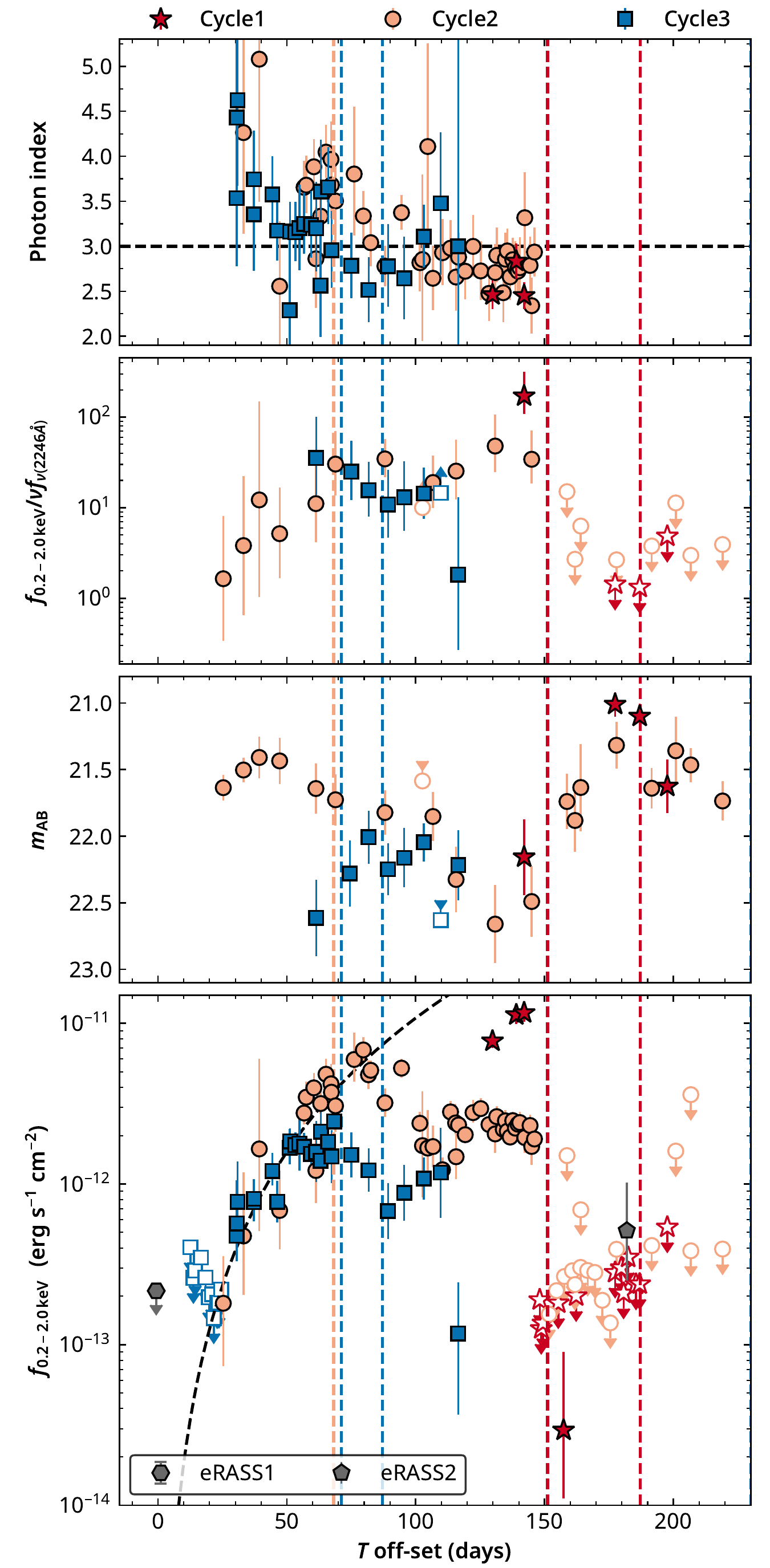}
  \caption{From top to bottom: Evolution of the photon index (measured using the \mpl model), ratio of the $f_{\text{0.2-2.0\,keV}}$ to the 2246\AA\ monochromatic flux, apparent UVM2 AB magnitude, and $\log(f_{\text{0.2-2.0\,keV}})$ against the time offset. The time off-set is calculated relative to the inferred $t_0$ assuming a period of \trecur~days. The whole data set was split into three cycles that were determined by the time when the rapid X-ray flux drop occurred. The vertical dashed lines represent the time (relative to $t_0$ of each cycle) of the ATCA radio observations. Radio emission was detected only in cycle 3 (dashed blue lines).}
  \label{fig:xuv_char}
\end{figure}

\jsrc showed dramatic long-term X-ray variability with four distinctive phases. It is interesting to probe whether a spectral evolution occurred when the source transitioned between different flux levels. The S/N of most of the available X-ray data is low. Thus, we fitted the \mission{Swift}/XRT (except for the one at around MJD 59391, hereafter Swift5), \mission{NICER}, and eROSITA eRASS3/4 spectra with the simple power-law model (\mpl). We also fixed $N_\text{H, Gal}$ at $3.3\times10^{20}~\mathrm{cm^{-2}}$ \citep{willingale_etal2013}. The distribution of the best-fitting photon index is shown in the upper panel of Fig.\,\ref{fig:xuv_char} (the definition of the three cycles is given in Section~\ref{sec:discussion}). It is evident that the spectra are very soft at low fluxes  (or at the \fxrise phase, $\Gamma\gtrsim 3.0$) and become harder as the X-ray flux increases ($\Gamma\lesssim 3.0$), although the uncertainties are large.

The X-ray spectra during the faintest observed X-ray flux (i.e. in eRASS2, X1, and Swift5) are very soft and essentially lack source photons above 0.6\,keV in the Swift5 XRT data and 1.0\,keV in the eRASS2 and X1 data. We thus fitted each of the eRASS2, X1, and Swift5 spectra with a multi-colour disk model modified by Galactic absorption (\texttt{TBabs*zashift*cflux*diskbb}, hereafter \mmcd) and fixed the Galactic column density at $3.3\times10^{20}~\mathrm{cm^{-2}}$. We obtained a best-fitting inner disk temperature $T_\text{in}$ of $64^{+29}_{-18}\,\text{eV}$ with \sflux of $5.1^{+5.1}_{-2.6}\times10^{-13}\,\unitflux$ for eRASS2. The value of $T_\text{in}$ cannot be constrained by the \mission{Swift}/XRT data with only a lower limit of $29\,$eV. We thus fixed it at its best-fitting value of $78\,\text{eV}$. This resulted in a \sflux of $1.8^{+1.8}_{-1.1}\times10^{-13}\,\unitflux$. For X1, we obtained a best-fitting inner disk temperature of $99^{+120}_{-40}~\text{eV}$ and \sflux of $2.9^{+6.0}_{-1.8}\times10^{-14}\,\unitflux$.

\subsection{Black hole mass estimation}\label{subsec:est_mbh}
We used two independent observed empirical relations, the $M_\text{BH}-\sigma_{\star}$ relation \citep{ferrarese_merritt2000, gebhardt_etal2000} and the $M_\text{BH}-\sigma_\text{rms}^2$ relation (\citealt{ponti_etal2012}; $\sigma_\text{rms}^2$ is the normalised excess variance, see below for more details), to estimate the mass of the central BH in \jsrc.

\subsubsection{$\mbh$ measured from the $M_\text{BH}-\sigma_{\star}$ relation}\label{subsubsec:msigma}
The Magellan LDSS3-C spectrum was used as it provides the best balance between wavelength coverage and S/N. Using the spectral lamp, we measured an instrument  FWHM of $4.6$ pixels across the LDSS3-C band coverage. When the instrumental dispersion ($2\,\AA$ per pixel) is taken into account, this resulted in an instrumental FWHM of $\sim9\,\AA$, corresponding to an instrumental velocity resolution of $\sim208$, 181, and 125\,$\text{km\,s}^{-1}$ at observed wavelengths of 5500, 6340, and 9200$\,\AA$, respectively. We adopted the \texttt{pPXF} package, a Python implementation of the penalised pixel-fitting method \citep{cappellari2017}, to measure the stellar velocity dispersion $\sigma_{\star}$. The Indo-US Library of Coudé Feed Stellar Spectra Library, consisting of 1237 stars covering the region of $3460-9464\,\AA$ \citep{valdes_etal2004}, was used to construct stellar templates to fit the Magellan spectrum. We created 1000 spectral realisations by resampling the Magellan data within the uncertainties, and performed \texttt{pPXF} fitting over the rest-frame $3900-8700\,\AA$ range for each of the realisations. The mean ($140\,\text{km\,s}^{-1}$) and standard deviation ($34\,\text{km\,s}^{-1}$) of the best-fitting values of $\sigma_{\star}$ are reported here as the intrinsic velocity dispersion and the $1\sigma$ uncertainty, respectively. This results in a BH mass of $\log(\mbh/\msun)=7.4\pm0.5$ using the relation reported in \citet{gultekin_etal2009}. The WiFeS data have a better spectral resolution ($\sim1.7\,\AA$, $\sigma_\text{inst}\sim50\,\text{km\,s}^{-1}$) than the Magellan data, although their S/N is lower. We again performed \texttt{pPXF} fitting over the blue band ($3900-5900\,\AA$) of 1000 spectral realisations generated by resampling the WiFeS data. We measured a velocity dispersion of $120\pm20\,\text{km\,s}^{-1}$, which leads to a BH mass of $\log(\mbh/\msun)=7.0\pm0.4$. This value is consistent with the value measured from Magellan data.

\subsubsection{$\mbh$ measured from $\sigma_\text{rms}^2-M_\text{BH}$ relation}
A strong correlation between the $\mbh$ and the normalised excess variance $\sigma_\text{rms}^2$ has been reported for AGNs \citep{ponti_etal2012, pan_etal2015}. We also estimated the $\mbh$ of \jsrc using this empirical correlation, assuming that its short-term X-ray variability is similar to that of AGNs. The $\sigma_\text{rms}^2$ was calculated using the following formula \citep[e.g.][]{nandra_etal1997}:
$$ \sigma_\text{rms}^2 = \frac{1}{N\mu^2} \sum_{i=1}^{N}[(X_i-\mu)^2-\sigma_i^2]\text{,}$$
where $N$ is the number of bins in the X-ray light curve, and $\mu$ is the unweighted arithmetic mean of the photon count rates of the light curve. $X_i$ and $\sigma_i$ are the count rates and uncertainties in each time bin, respectively. The $\sigma_\text{rms}^2$, however, depends on the time length, energy region, and time bin-size of the light curves. In this work, the $\nxv$ was calculated using the \mission{XMM-Newton} EPIC/pn light curves over the $0.3-0.7\,\text{keV}$ energy range with time bin-size of $250\,\text{s}$. To better estimate the uncertainty of the $\nxv$, we calculated $\nxv$ over 14 segments with $10\,\mathrm{ks}$ observation data of each using the X2 and X3 observations. We then estimated the $\log(\nxv)$ and its $1\sigma$ uncertainty using the mean and standard deviation of the $\log(\nxv)$ values calculated from the 14 segments, which give a $\log(\nxv)=-1.5\pm0.3$. We estimated a BH mass of $\log(\mbh/\msun)=6.4\pm0.2$ using the best-fitting model for a $10\,\text{ks}$ segment presented in \citet{ponti_etal2012}.

The value of $\mbh$ measured from the $\nxv-\mbh$ relation is  lower than that from the $M_\text{BH}-\sigma_{\star}$ relation. We note that the relation between $\nxv$ and $\mbh$ was derived from a sample of persistently accreting AGNs with overall steady accretion flow. Thus, it does not necessarily apply to \jsrc, which may have a non-steady accretion disk. For instance, the $\nxv$ of the jetted TDE Swift 1644+57 shows a long-term evolution and can change by a factor of $\sim5$ in different flux states \citep{jin2021}. In this work, we assumed a BH mass of $10^7\,\msun$ for \jsrc. We note that  the main conclusions of this paper do not change when a lower BH mass (e.g. $\mbh=3\times10^6\,\msun$, the mass measured from the $\sigma_\text{rms}^2-M_\text{BH}$ relation) is adopted.

\section{Discussion\label{sec:discussion}}
The long-term X-ray light curve of \jsrc is characterised by four distinctive phases (\fxdrop, \fxfaint, \fxrise, and \fxplat). We define an X-ray variability cycle for \jsrc as the time between the start of two consecutive \fxrise phases (at the starting point of the dashed black and grey lines in the upper panel Fig.\,\ref{fig:multi_lc}). We therefore observed three cycles (MJD before 59377 for cycle1, MJD after 59598 for cycle3, and the MJD in between for cycle2), and each of the four phases was observed at least twice (see Fig.\,\ref{fig:multi_lc}). We derived a start time of MJD~\frsecondstart\xspace for cycle2 from the best-fitting $t_0$ of the second \fxrise (see Sect\,\ref{subsubsec:xray}). The start times for cycle1 and cycle3 were then estimated by assuming a recurrence time of \trecur~days.
Fig.\,\ref{fig:xuv_char} shows the X-ray flux against the time offset for the 3 cycles, where the time offsets were calculated with respect to the start time for each cycle. The profile of the first \fxrise phase is well described by a power-law function (Fig.\,\ref{fig:multi_lc} and bottom panel of Fig.\,\ref{fig:xuv_char}). Remarkably, the same function can also match the profile of the second \fxrise phase well. The characteristic of the long-term X-ray light curve strongly suggests that \jsrc is a repeating, or even periodic, X-ray nuclear transient with a roughly estimated recurrence time of \trecur\,days. This adds a member to this rare class of nuclear transients. We further note that, unlike \asko and \ictde, no strong broad or narrow emission lines (e.g. the [\ion{O}{III}]$\lambda4959/5007$ or the Balmer lines) are detected from our low- and medium-resolution optical spectroscopic data, indicating that the host galaxy of \jsrc exhibited no signs of prior AGN activity.

\begin{figure}
  \centering
  \includegraphics[width=\columnwidth]{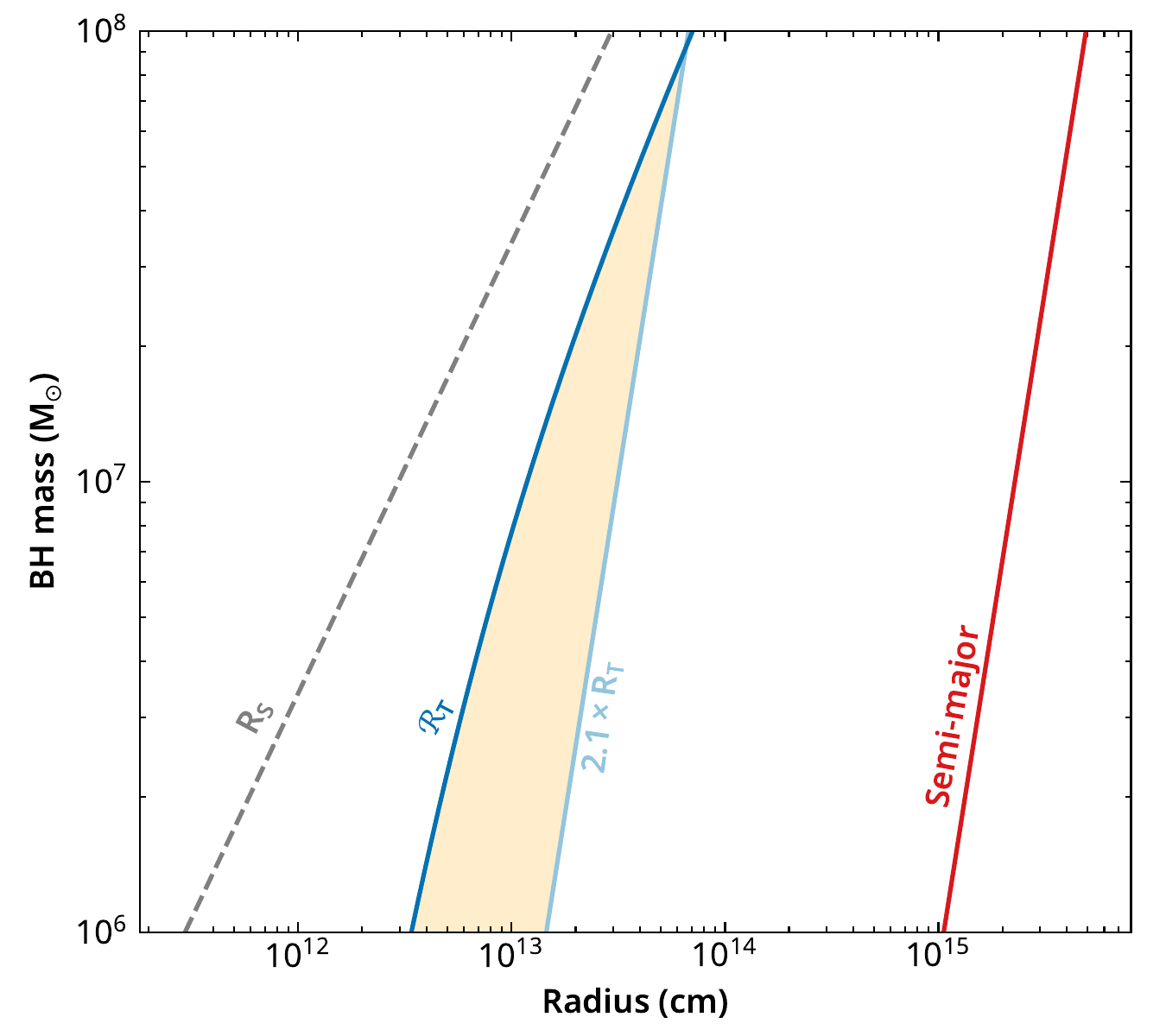}
  \caption{Different radii as a function of BH mass. The Schwarzschild radius ($R_{\text{S}}$) is shown by the dashed grey line. The radius ($\mathcal{R}_{\text{T}}$) at which a full disruption of a solar-like main-sequence star would happen is marked by the solid blue line. The solid light blue line shows the tidal radius ($R_{\text{T}}$) multiplied by 2.1, which is roughly the radius at which a partial tidal disruption event would happen. The $2.1\times R_{\text{T}}$ line also indicates the radius at which RLOF could happen. We show the semi-major axis for a period of \trecur~days in red.}
  \label{fig:comp_r}
\end{figure}

\subsection{Nature of \jsrc}
Using the best-fitting \mcomp models from X2 and X3 (see Sect.\,\ref{subsec:xray_spec} and Table\,\ref{tab:xray_spec}), which consist of a multi-colour disk and two Componization components, we can roughly estimate the bolometric luminosity from \slumi using a correction factor, $\kappa=L_\text{bol}/L_\text{X, soft}$, of $\sim3$. The peak values of \slumi are $\sim1.6$, 1.0, and $0.3\times10^{44}\,\unitlumi$ (see Fig.\ref{fig:multi_lc}) for cycle1, cycle2, and cycle3, respectively. Thus, when we assume a BH mass of $10^7\,\msun$, the observed peak Eddington ratios $\lambda_\text{Edd}=L_\text{bol}/L_\text{Edd}$ in cycle1, cycle2, and cycle3, are $\sim0.37$, 0.23, and 0.07, respectively.

The total energy released in each cycle can be estimated by
$$E_\text{tot}=L_\text{fa} \Delta t_\text{fa}+L_\text{pl} \Delta t_\text{pl}+\int_0^{\Delta t_\text{rs}}L_\text{rs}(t)~\text{d}t\text{,}$$
where $\Delta t_\text{fa}$, $\Delta t_\text{pl}$, and $\Delta t_\text{rs}$ are the duration of the \fxfaint, \fxplat, and \fxrise phases, respectively. $L_\text{fa}$ and $L_\text{pl}$ are the average bolometric luminosity during the \fxfaint and \fxplat phases, respectively, and we adopted the best-fitting power-law function of $f_\text{rs}(t)$ found in cycle2 to describe the profile of $L_\text{rs}(t)=\kappa_\text{rs} 4\pi D_\text{ld}^2 f_\text{rs}(t)$ for the \fxrise phase for each of the three cycles. The bolometric correction factor, $\kappa_\text{rs}$, could change from $\sim20$ at faint X-ray flux level to $\sim3$ at bright X-ray flux level (see Sect.\,\ref{subsec:mech_uvxray}) during the \fxrise phase. We conservatively assumed an average value of 15 for $\kappa_\text{rs}$. Using a bolometric correction factor of $\kappa=3$,  $L_\text{pl}$ is estimated to be 4.8, 1.1, and $0.7\times10^{44}\,\unitlumi$ for cycle1, cycle2, and cycle3, respectively. The values of $L_\text{fa}$ are poorly constrained. We thus adopted a value of $9.0\times10^{43}\,\unitlumi$, derived from Swift5 (see Sect.\,\ref{subsec:mech_uvxray}), for $L_\text{fa}$ for each of the three cycles. The values of $\Delta t_\text{fa}$, $\Delta t_\text{rs}$ and $\Delta t_\text{pl}$ are well constrained for cycle2 and cycle3 with $\Delta t_\text{pl}=2\,$ months and $\Delta t_\text{fa}=3\,$ months for both cycles, and $\Delta t_\text{rs}=3\,$months for cycle2 and 2\,months for cycle3. For cycle1, we assumed the values are 3, 4, and 2\,months for $\Delta t_\text{fa}$, $\Delta t_\text{rs}$, and $\Delta t_\text{pl}$, respectively. We then calculated a total released energy of $E_\text{tot}=1.5$, 0.5, and $0.2\times10^{52}\,\text{erg}$ for cycle1, cycle2, and cycle3, respectively. When we assume that half of the tidally disrupted debris returns to the SMBH and that the radiation efficiency is 0.1, the total mass of the disrupted debris is about $0.17\,\msun$ for cycle1, $0.06\,\msun$ for cycle2, and $0.02\,\msun$ for cycle3.

\subsubsection{Partial tidal disruption event}
The very soft X-ray spectra in eRASS2 and at the beginning of the rising phase as well as the observed peak $0.2-2.0\,\text{keV}$ luminosity of $\sim1.6\times10^{44}\,\unitlumi$ are consistent with the expectation from a TDE. A \ptde is then required to produce the repeating X-ray flares seen in \jsrc. The estimated Eddington ratios are much lower than the Eddington accretion rate, but are reasonable for a \ptde \citep[e.g.][]{guillochon_ramirez2013}. The required mass loss for each cycle is also low enough for a repeating \ptde.

Fig.\,\ref{fig:comp_r} shows the characteristic radii for a main-sequence star with mass of $1\,\msun$ and radius of $1\,\rsun$ as a function of $\mbh$. The radius ($\mathcal{R}_{\text{T}}$) at which a full TDE will occur was calculated using the equation presented in \citet{ryu_etal2020a}. The radius at which a \ptde can be expected to happen is approximately 2.1 times the tidal radius ${R}_{\text{T}}$ \citep{ryu_etal2020b}. The orange shadowed region marks the parameter space where a \ptde can occur. Clearly, a \ptde is possible for \jsrc even for a solar-type main-sequence star. The semi-major axis of the orbit of the star, assuming a period of \trecur~days, is also shown in Fig.\,\ref{fig:comp_r}. The semi-major axis is always at least ten times larger than $2.1\mathcal{R}_{\text{T}}$. A repeating \ptde with period of \trecur~days therefore remains a possible scenario for the origin \jsrc.

The orbital period of a star, which was fed to the SMBH through the traditional two-body scattering, can be estimated by $P\simeq\pi R_\star^{3/2}(\mbh/M_\star)/\sqrt{2GM_\star}$ \citep{cufari_etal2022}, where $R_\star$ and $M_\star$ are the radius and mass of the star, respectively. For a solar-like star and $\mbh$ of $10^7\msun$, the orbital period should be around $10^3\,$yr, which is much longer than the recurrence time of \trecur~days for \jsrc. Recently, \citet[][see also \citealt{amaro-seoane_etal2012}]{cufari_etal2022} has proposed that the Hills mechanism, according to which a star in a binary system can be captured on a tightly bound orbit around the SMBH after the binary system is destroyed by the tides of the SMBH, can produce the short period ($\sim114\,$days) observed in \asko. With a semi-major axis $a_\star$ of the binary system of about $0.007\,$au\footnote{$P\simeq\pi a_\star^{3/2}(\mbh/M_\star)^{1/2}/\sqrt{2GM_\star}$, \citet{cufari_etal2022}}, this mechanism can also generate the short period observed in \jsrc. The tidal radius at which the Hills mechanism will take place can be calculated by $r_\text{t}=a_\star(\mbh/M_\star)^{1/3}$ \citep{hills1988}, which is about $15R_\text{g}$ ($\sim2.3\times10^{13}\,\text{cm}$, $R_\text{g}=G\mbh/c^2$ is the gravitational radius) for \jsrc assuming $a_\star=0.007\,$au. This also agrees with the radius required for a \ptde (Fig.\,\ref{fig:comp_r}).

The \fxdrop phase, however, is rarely seen in normal TDEs, which often show a decline in their X-ray light curve that is broadly consistent with the predicted $f\propto t^{-\beta}$ law with $\beta=5/3$ \citep[e.g.][]{komossa2015, saxton_etal2021}. It has been suggested that the fall-back rate of a \ptde can noticeably deviate from this relation. For instance, the slope calculated from hydrodynamic simulations in \citet{guillochon_ramirez2013} can be as steep as $\beta>3$. \citet{coughlin_nixon2019} advocated that the asymptotic late-time fallback rate in a \ptde scales with $t^{-9/4}$. In the recent simulations by \citet{ryu_etal2020c}, the decline in the fallback rate was argued to depend on the mass of the remnant, for instance, $\beta=2$ for strong \ptde while $\beta=5$ for a weak \ptde in a $10^{6}\msun$ BH. The slope of the X-ray light curve during the \fxdrop phase strongly depends on the time of the peak X-ray flux ($t_\text{X, peak}$), which is unknown. We nevertheless estimated that $\beta$ is larger than 8 using the Swift1 detection and the first \mission{NICER} upper limit in the first \fxdrop phase, assuming that Swift1 was taken at least 10~days\footnote{The $t_\text{X, peak}$ is very likely before the first eRASS3 observation, which was taken $\sim11$~days before Swift1.} after the $t_\text{X, peak}$. It is thus unlikely that the rapid X-ray drop reflects a dramatic decrease in the fall-back rate. Rather, we interpret the X-ray flux drop as a transition of the accretion state (see Sect.\,\ref{subsec:acc_trans} for more details).

Further evidence to support \jsrc as a \ptde is the gradual decrease of the observed peak X-ray flux from cycle1 to cycle3. The same trend is also observed in the UV light curve, but it is less significant. This is consistent with the expectation from a \ptde, as less material is accreted onto the BH if the surviving star becomes more compact after each passage. However, we caution that, as pointed out by \citet{ryu_etal2020c}, the stellar remnant could be expanded by additional heat and rotation, therefore a more severe \ptde is not impossible.

\subsubsection{Other potential scenarios}
Alternative scenarios, such as a superluminous supernova near the centre of the galaxy nucleus, can be ruled out as the X-ray luminosity evolution as well as the X-ray spectral properties of \jsrc are very different from these events.
Periodic flares can also be produced by collisions between consecutive EMRI that are in the process of stable mass-transfer via RLOF onto an SMBH \citep{metzger_stone2017}. However, the recurrence time for this model is normally decades or even hundreds of years, which is much longer than the observed timescale for \jsrc. The predicted timescale can be significantly reduced, as short as hours, if the two stars reside in a common orbital plane \citep{metzger_etal2022}. This model has been proposed to explain the QPEs and the periodic flares observed in \asko. The BH mass estimated for \jsrc is around $10^7\msun$, which is one order of magnitude lower than that in \asko. It is still possible to produce the observed recurrence timescale for \jsrc as long as the stars have a much lower mean density (e.g. $\rho<0.1\rho_\odot$). However, we note that the typical mass loss per collision from this model is in the order of $2\times10^{-4}\msun$ for $\mbh$ of $10^7\msun$ and period of \trecur\,days (equation 15 in \citealt{metzger_etal2022}), which is much lower than the estimated mass loss for \jsrc. We thus disfavour a pair of EMRIs as the cause for the repeating flares in this object.

Disk instabilities, including the radiation pressure instability, have been proposed to explain the repeating rapid soft X-ray variability in the BHXRB GRS~1915+105 \citep{belloni_etal1997} and also the soft X-ray flares in the galaxy \ictde \citep{grupe_etal2015} and the AGN NGC~3599 \citep{saxton_etal2015}. In this scenario, in the X-ray faint state, the accretion disk (either from a pre-existing AGN or a newly formed disk after a TDE) is truncated at a certain radius $R_\text{tr}$. This truncated disk will be slowly refilled at a steady accretion rate from the outer disk. The time needed to refill the inner region (with inner radius of $R_0$) of the accretion disk is governed by the viscous timescale and can be estimated \citep{saxton_etal2015} by
$$\tau_\text{fill} \sim 0.33\alpha^{-8/10}M_6^{6/5}m_\text{Edd}^{-3/10}\left[\left(\frac{R_\text{tr}}{R_\text{g}}\right)^{5/4}-\left(\frac{R_0}{R_\text{g}}\right)^{5/4}\right]~\text{months,}$$
where $\alpha$ is the viscosity parameter. $M_6$ is the BH mass in units of $10^6\msun$ and $R_\text{g}$ is the gravitational radius. $m_\text{Edd}$ is the mass accretion rate in units of Eddington-limited accretion rate $\dot{M}_\text{Edd}=1.4\times10^{18}(\mbh/\msun)~\text{g~s}^{-1}$. Assuming a BH mass of $10^7\msun$ with Eddington-limited accretion rate $m_\text{Edd}=1$ and inner radius $R_0=3R_\text{g}$, we estimate the value of $\tau_\text{fill}$  to be $\gtrsim56~$months even for a small truncation radius of $4~R_\text{g}$ with $\alpha=0.1$, which is much longer than the observed duration ($\lesssim3~$months) of the faint state for \jsrc. This timescale can be significantly shortened as in a toy model developed by \citet{sniegowska_etal2020} (see also \citealt{pan_etal2021}), in which the radiation pressure instability only operates in a narrow ring between a thin outer disk and an optically thin hot inner ADAF. However, our X-ray spectral analysis suggests that the temperature of the inner-region of the multi-colour disk does not change at different X-ray flux states. This is inconsistent with the radiation instability, which predicts a hotter (cooler) disk at high (low) accretion rates. We thus also disfavour the disk instability as the cause for the repeating X-ray flares in \jsrc.

\subsection{Mechanism for the X-ray and UV emission}\label{subsec:mech_uvxray}
\begin{figure}
    \centering
    \includegraphics[width=\columnwidth]{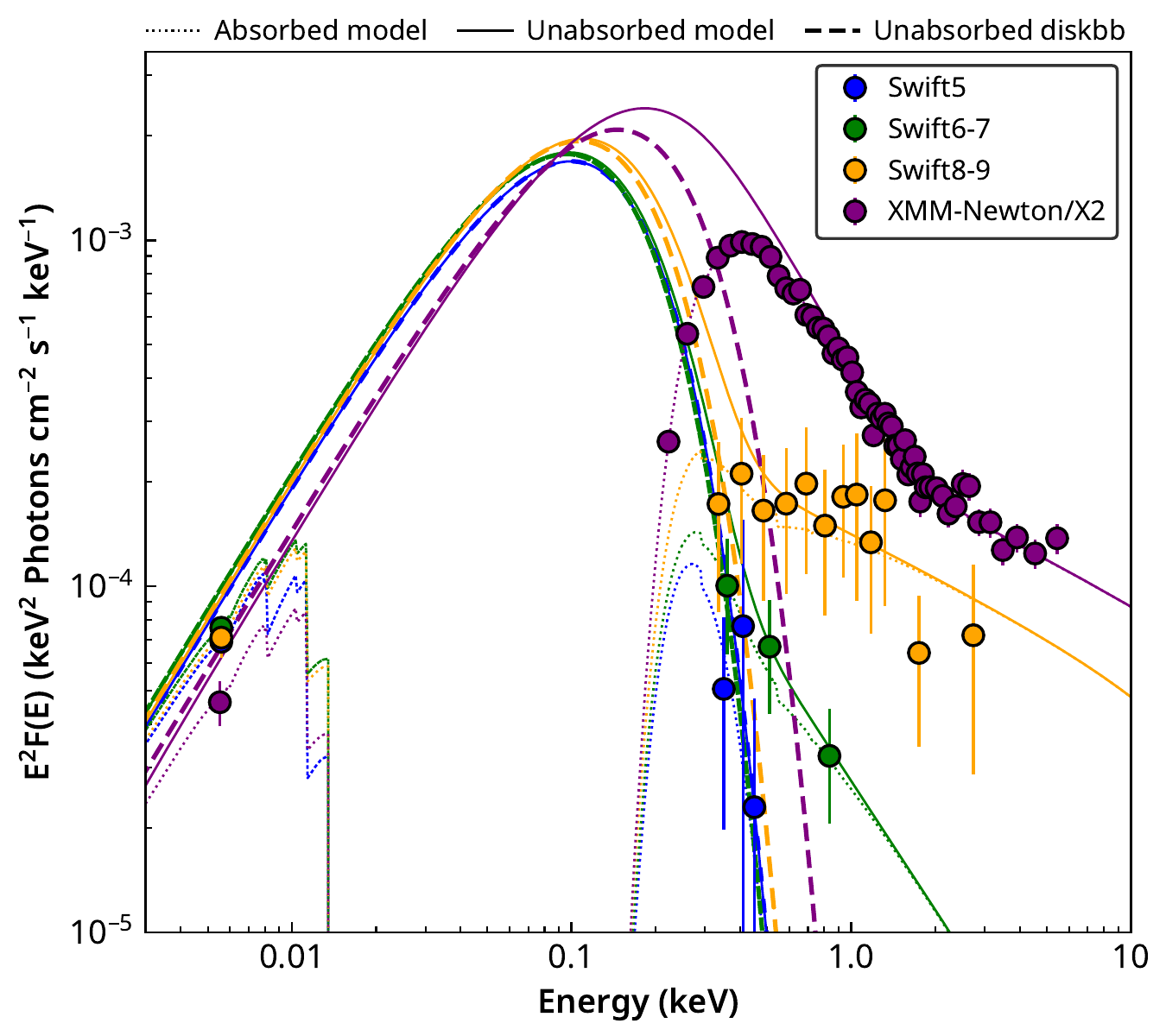}
    \caption{Evolution of the UV/X-ray SED during the first \fxrise phase. The solid circles are the unfolded spectra for Swift5 (blue), Swift6-7 (green), Swift8-9 (orange), and XMM-Newton/X2 (purple). The dotted lines are the absorbed models. The solid and dashed lines represent the unabsorbed intrinsic model and the diskbb component, respectively.}
    \label{fig:uvxray_sed}
\end{figure}

\begin{figure}
  \centering
  \hspace{-0.2cm}\includegraphics[width=1.0\columnwidth]{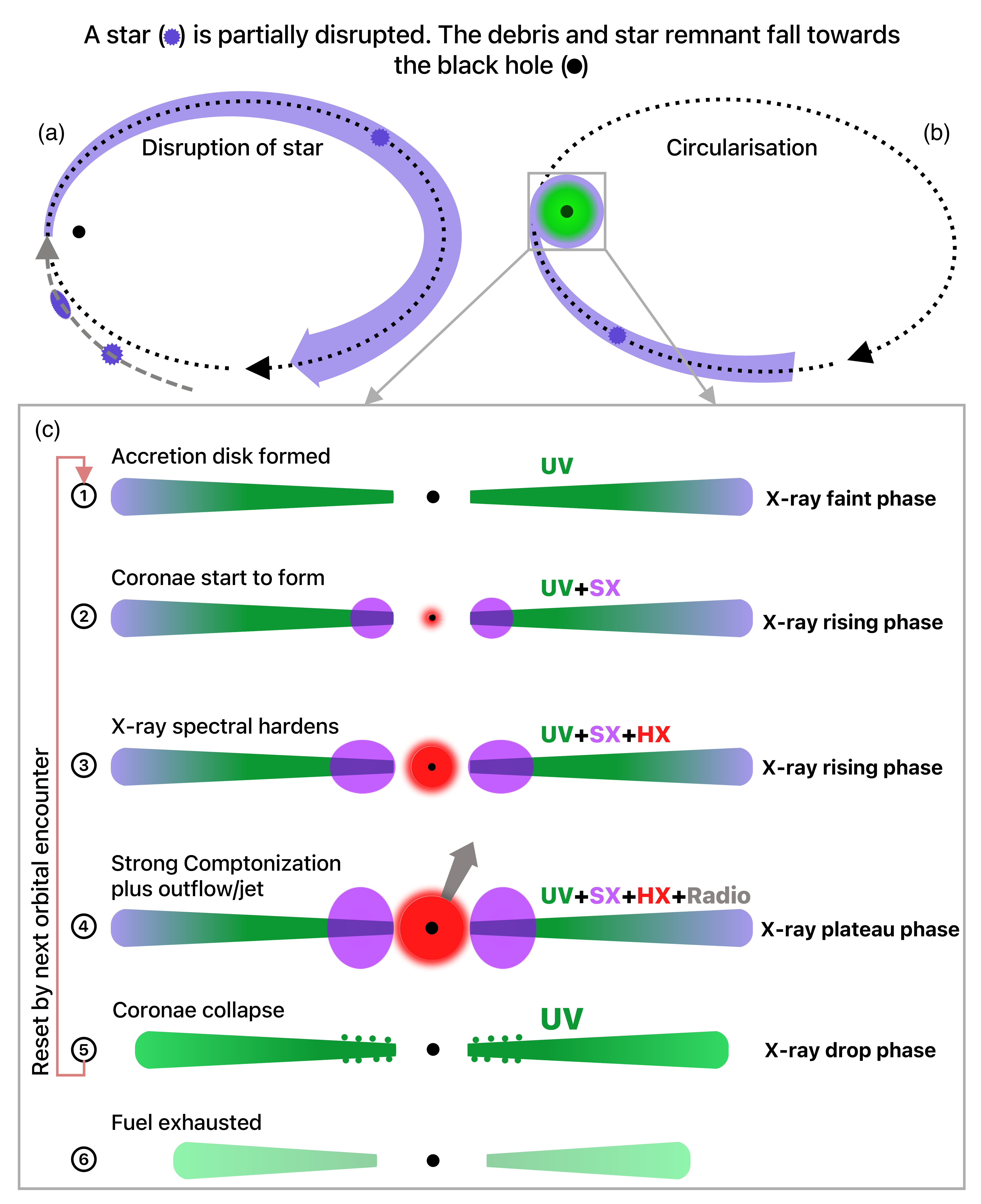}
  \caption{Partially disrupted star upon pericentre passage around a BH (panel a). The initial orbit of the star is represented with a dashed grey curve. The dotted black curve represents the orbit of the stellar debris and remnant (panel a and b). The stellar debris is rapidly circularised, and an accretion disk is formed (panel b). The freed gas (blue) is accreted onto the BH, with the accretion process proceeding in phases \Circled{1}-\Circled{6} (panel c), with a sequential build-up \Circled{1}-\Circled{4} of an accretion disk (UV, green colour), warm (soft X-rays, purple) and hot corona (hard X-rays, red), as well as an outflow or jet (radio, grey). When the gas reservoir depletes, the corona collapses for unknown reasons (\Circled{5}, no X-ray emission), and the accretion of the remaining accretion disk gas is even more rapid, leading to enhanced UV emission. Finally, as the fuel is completely exhausted (6), the UV emission drops as well. However, the subsequent pericentre passage resets the cycle.}
  \label{fig:illustration-evolution}
\end{figure}

The X2 and X3 observations provide high-quality data to understand the physical processes for the X-ray emission during the X-ray bright state (bright end of the \fxrise phase for X2 and \fxplat for X3). As mentioned in \ref{subsec:xray_spec}, the X-ray spectra of X2 and X3 can be best interpreted as inverse Comptonisation of the seed photons from a multi-colour thermal disk by the warm and hot coronae (sequence 4 in Fig.\,\ref{fig:illustration-evolution}). The best-fitting temperature of the inner region $T_\text{in}$ of the multi-colour disk is $\sim60\,\text{eV}$, suggesting that a thin thermal disk exists in the X-ray bright state. This thermal disk causes the UV emission and also contributes substantially to the very soft X-ray emission ($<0.5\,$keV band), while the $0.5-2.0\,\text{keV}$ X-ray emission is dominated by X-ray photons produced from the inverse Componisation by the two coronae.

The results of the X-ray spectral fitting are less certain in the faint state because the S/N is low. However, all three X-ray spectra observed during this state, that is, eRASS2, X1 (in the \fxdrop phase), and Swift5 (at the beginning of the \fxrise phase), are very soft, suggesting that the X-ray emission may come from a thermal accretion disk with no indication of strong X-ray emission from the corona. The value of $T_\text{in}$ is $67^{+29}_{-17}\,\text{eV}$ for eRASS2, $\sim78\,\text{eV}$ for Swift5, and $99^{+120}_{-40}~\text{eV}$ for X1 if fitted with a multi-colour accretion disk (the \mmcd model, see Sect\,\ref{subsubsec:xray_spec_long}). We tested the emergence and potential evolution of a thermal accretion disk by modelling the broad-band UVOT/UVM2 and XRT/X-ray data of the five Swift observations (ObsIDs: 00014135005--00014135009), starting with Swift5, during the first \fxrise phase (e.g. the dashed black line in the upper panel of Fig.\ref{fig:multi_lc}). To increase the S/N,we stacked the observations 00014135006 and 00014135007, as well as 00014135008 and 00014135009, to two combined \mission{Swift} observations (hereafter Swift6-7 and Swift 8-9). We then obtained the X-ray spectra and the UVM2 flux densities for each of the two combined \mission{Swift} observations.

We first modelled the broad band UV and X-ray data with the \mmcd model for Swift5, Swift6-7, and Swift8-9. We again used $E_{\text{B-V}}=0.03$\,mag to correct the Galactic reddening. We note that due to the low photon counts, the XRT spectra follow Poisson distribution (fewer than ten counts per bin). While the UVOT/UVM2 data follow Gaussian statistics. Thus, the results from this spectral modelling should be treated with caution and should not be considered as the best-fitting model. Nevertheless, we found that the broad-band UV/X-ray data can be described with a single multi-colour disk model with a temperature of $\sim45\,\text{eV}$ and $L_\text{bol}$ of $9.0\times10^{43}\,\unitlumi$ ($6.0\times10^{-12}\,\unitflux$, $\lambda_\text{Edd}\approx0.07$) for Swift5. The unfolded spectrum, \mmcd model, and absorption corrected intrinsic model are shown in Fig.\ref{fig:uvxray_sed}. However, a simple \mmcd model cannot describe the X-ray data of Swift6-7 and Swift8-9 well, with clear excess above $\sim0.7\,$keV. Therefore, we added a Comptonisation component to the \mmcd model, that is, TBabs*zashift*(cflux*thcomp*diskbb) in \textsc{Xspec}, to fit the Swift6-7 and Swift8-9 data. The electron temperature $kT_\text{e}$ and the covering fraction cannot be constrained. We thus fixed $kT_\text{e}$ at 10\,keV and covering fraction at 0.01. As shown in Fig.\,\ref{fig:uvxray_sed}, this model can indeed describe the broad band UV and X-ray data of both Swift6-7 and Swift8-9. The values of $T_\text{in}$ are $\sim44\,$eV for Swift6-7 and $\sim48\,$eV for Swift8-9. For comparison, we also show the unfolded X2 data using the best-fitting \mcomp model found in Sect.\,\ref{subsubsec:xmm_ave_spec}. We conclude that the UV and X-ray emission mainly originates from the thin thermal accretion disk in the faint state (illustrated in Fig.\,\ref{fig:illustration-evolution}, sequence \Circled{2}). As the \sflux increases during the \fxrise phase, the warm and hot coronae start to form (sequence \Circled{3}--\Circled{4} in Fig.\,\ref{fig:illustration-evolution}). X-ray photons produced from the inverse Comptonisation of the two coronae start to dominate the X-ray emission.

Our results also indicate a rapid formation of the accretion disk, with a timescale comparable to the duration of \fxfaint phase ($\lesssim3\,$months). For repeating \ptde, the stellar debris should be on an elliptical orbit. \citet{hayasaki_etal2013} showed in their simulations that an accretion disk can be formed rapidly for a star on an elliptical orbit with moderate eccentricity ($e=0.8$) tidally disrupted by a $10^{6}\msun$ SMBH with a deep encounter ($r_\text{t}/r_\text{p}=5$, where $r_\text{p}$ is the pericentre distance). The rapid circularisation is due to stellar stream collisions induced by relativistic apsidal precession. Relativistic precession has also been proposed to explain the rapid disk formation in TDE AT2018fyk \citep{wevers_etal2019}. For \ptde like \jsrc, a very deep encounter is unlikely. However, a moderate encounter with $r_\text{t}/r_\text{p}<2$ is still possible for more centrally concentrated stars \citep{guillochon_ramirez2013, ryu_etal2020c}. Combined with a higher BH mass of $10^{7}\mbh$, rapid disk formation caused by relativistic precession is also possible for \jsrc.

\subsection{X-ray and UV variability: Rapid formation and destruction of the corona and/or accretion-state transition?}\label{subsec:acc_trans}

As argued above, a warm and a hot corona cause the X-ray emission in the \fxplat phase and the bright end of the \fxrise phase. We find no strong evidence for such coronae at the faint end of the \fxrise and \fxdrop phases, suggesting that the X-ray and UV variability could be caused by a change in the structure of the accretion flow at different accretion rates. Theoretical calculation suggests that a transition in accretion mode occurs when the accretion rate reaches critical values \citep{meyer_etal2000}. Strong evidence for accretion-state transitions within single AGNs remains elusive, although there are indications of multiple accretion modes in AGN samples \citep[e.g.][]{gu_cao2009}. Transitions like this have been frequently observed in BHXRBs, which typically show three states based largely on the X-ray spectral and temporal properties \citep{remillard_mcclintock2006}. The \textup{hard} state is characterised by a strong non-thermal emission in the form of a power-law photon spectrum $N(E)\propto E^{-\Gamma}$ with a photon index $\Gamma$ of $\sim1.4-2.1$ and a weak thermal disk emission. In contrast, the X-ray emission in the \textup{soft} state is dominated by thermal disk emission with weak non-thermal emission. The \textup{steep power-law} (SPL) state (or the very high state) is characterised by strong non-thermal power-law component with $\Gamma>2.4$ and also substantial thermal disk emission.

The accretion rate can change by orders of magnitude within a timescale of months to years in nuclear transients, making them the ideal objects for studying accretion-state transitions in SMBH--accreting systems. An accretion-state transition has been reported in a few TDEs. For instance, the detection of radio emission and change in X-ray spectral properties in ASASSN-14li are suggested as evidence of an accretion-state transition from hard to thermal states \citep{vanvelzen_etal2016}. \citet{wevers_etal2021} suggested that the observed evolution of the X-ray and UV emission of AT2018fyk can be explained by a rapid accretion-state transition between the thermal and hard states. In the following, we discuss whether the observed X-ray, UV and radio variability can be explained with accretion-state transitions in the framework of a coupling of the corona with the accretion disk.

\begin{figure}
  \centering
  \includegraphics[width=\columnwidth]{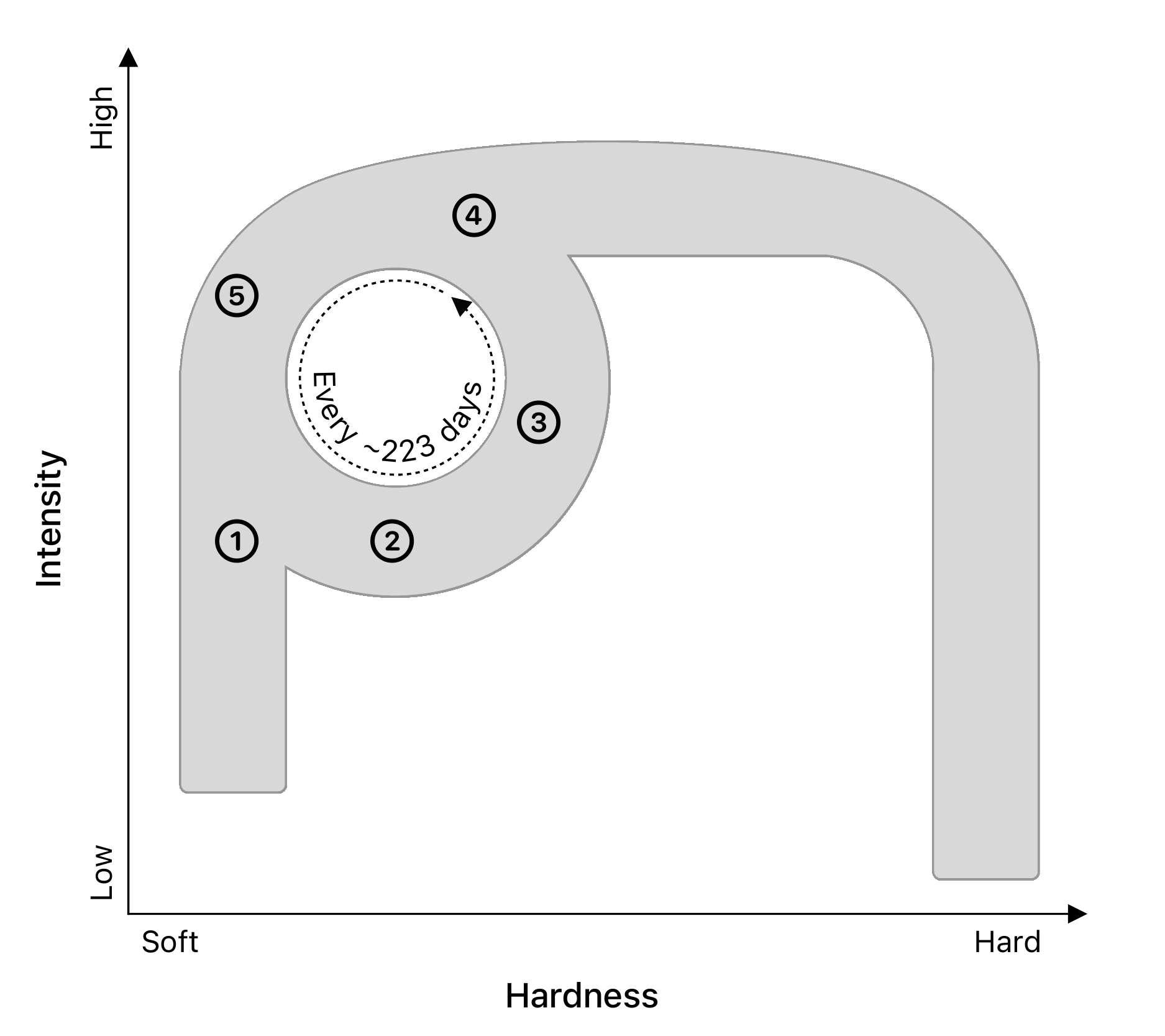}
  \caption{States of \jsrc{} overlaid on the BH X-ray binary state diagram. The numbers are as labelled in Fig.~\ref{fig:illustration-evolution}. The source cycles between a low soft state dominated by thermal emission ({\Circled{1}-\Circled{2}}), ignition of Comptonised emission (\Circled{3}-\Circled{4}), and returns to a thermally dominated state \Circled{5}. The phase \Circled{6} in Fig.~\ref{fig:illustration-evolution} corresponds to the quiescent state of BHXRBs and is not shown in this plot.}
  \label{fig:TDEBHXB}
\end{figure}

\subsubsection{Analogy to BHXRBs: Soft state when the X-ray emission is faint and SPL state when the X-ray emission is bright}
As mentioned above, the X-ray spectra of \jsrc can best be described by a multi-colour disk with $T_\text{in}$ of $\sim70\,$eV at the faint state during the X-ray rise phase. The X-ray spectrum from X1, taken soon after the rapid X-ray flux drop, is also very soft, with $99^{+120}_{-40}~\text{eV}$. These results suggest that the X-ray emission of \jsrc is dominated by thermal emission from a thin disk when the X-ray flux is faint (\Circled{1} and \Circled{5} in Fig.\,\ref{fig:illustration-evolution} panel c), which is reminiscent of the soft state in BHXRBs. In Fig.\,\ref{fig:TDEBHXB} we show the location of \jsrc during the X-ray faint state on the BHXRB hardness-intensity diagram \citep{fender_etal2004}.

The top panel of Fig.\,\ref{fig:xuv_char} shows that the X-ray spectra during the \fxplat phase and the bright end of the \fxrise phase can be described by a power law with a photon index in the range of $\sim2.5-3.0$. A thermal component, which contributes $\sim26\%$ of the total $0.2-2.0\,\text{keV}$ band, is also detected in the high-quality X2 and X3 data. These characteristics suggest that \jsrc is in the SPL state when the source is bright in the X-ray band (\Circled{4} in Fig.\,\ref{fig:illustration-evolution} and \ref{fig:TDEBHXB}). In addition, the X-ray spectra from X2 and X3 require two Comptonisation components, a requirement that was reported in some BHXRBs in the SPL state \citep{gierlinski_done2003}. Perhaps the strongest support for the SPL analogy comes from the detection of radio emission in \jsrc. Unlike the compact steady jet observed in the hard state, transient radio emission from an optically thin jet is often detected at the point of accretion transition to the SPL state in BHXRBs \citep{fender_etal2004}. Transient radio emission like this has also been detected in \jsrc. We detected radio emission during the third plateau phase, but no radio emission was detected during the \fxfaint, \fxrise, and \fxdrop phases. Furthermore, the radio flux decreased by a factor of $\sim3(4)$ within $\sim2$ weeks in the 9(5.5)~GHz band (see Fig.\,\ref{fig:atca_obs}). This is also consistent with the transient radio emission observed in the SPL state of BHXRBs, although the current radio data suggest a flat radio spectra for \jsrc, which is in contrast to the steep radio spectra observed in the SPL state in BHXRBs. Future radio observations at different flux levels are required to understand the radio emission mechanism. It also worth noting that strong high-frequency QPOs (HFQPOs) with frequencies in the range $100-450$\,Hz observed in BHXRBs are all detected in the SPL state \citep{remillard_mcclintock2006}. When we assume that the frequency of HFQPOs scales inversely with $\mbh$, then the expected frequency is around $1-4.5\times10^{-4}\,$Hz for $\mbh$ of $10^7\msun$, which is consistent with the tentative signal at a frequency of $2.5\times10^{-4}$\,Hz found in X2.

\subsubsection{Accretion-state transition between SPL and soft states?}
In the second panel of Fig.\,\ref{fig:xuv_char}, we show the evolution the ratio of \sflux to the $2246\,\AA$ monochromatic flux ($f_\text{0.2-2.0keV}/\nu f_{\nu(2246\AA)}$), which represents the relative contribution from the corona to that of the disk (similar to the X-ray band hardness ratio used in BHXRBs) for the three cycles. It is clear from Fig.\,\ref{fig:xuv_char} that the X-ray spectrum of \jsrc hardens (upper panel) and $f_\text{0.2-2.0keV}/\nu f_{\nu(2246\AA)}$ can increase by a factor of $\sim10$ during the \fxrise phase, suggesting that a transition from the soft state to the SPL state is taking place during the \fxrise phase (\Circled{2} and \Circled{3} in Fig.\,\ref{fig:illustration-evolution} and \ref{fig:TDEBHXB}). This accretion-state transition is accompanied by a rapid formation of the warm and hot coronae, perhaps through the disk evaporation mechanism \citep{liu_etal2002, liu_etal2003, qiao_liu2015}. The exact timescale of this soft-to-SPL state transition and formation of the coronae is challenging to calculate with the current data set. However, we can roughly estimate this timescale to be shorter than two months, assuming that the coronae were totally destroyed during the \fxdrop phase and started to re-form at the beginning of the \fxrise phase. Accretion-state transitions have also been invoked to explain the X-ray variability in several TDEs. For instance, similar to \jsrc, the X-ray spectra of AT2018fyk \citep{wevers_etal2021} and AT2021ehb \citep{yao_etal2022} are very soft and likely dominated by the emission from a thermal accretion disk in the early stages. As the X-ray flux increases, the X-ray spectra of both sources gradually become harder. Interestingly, both sources also show a plateau phase in their X-ray light curves. However, the photon index of the power-law component during the X-ray plateau phase for both sources are in the range of $\sim2.0-2.2$, which is much harder than that in \jsrc and is consistent with the X-ray properties of BHXRBs in the hard state (thus a soft-to-hard state transition) rather than the SPL state. No radio emission is detected in either of these two sources.

The X-ray spectrum of \jsrc becomes softer and the $f_\text{0.2-2.0keV}/\nu f_{\nu(2246\AA)}$ decreases by a factor of $\gtrsim10$ during the \fxdrop phase. We explain the \fxdrop phase in \jsrc by the transition from the SPL state to the soft state, which may result from the collapse or destruction of the coronae (sequence 5--6 in Fig.\ref{fig:illustration-evolution}). A rapid X-ray flux drop has also been reported in a few other TDEs. For instance, a sharp flux drop (a factor of $>50$ in four~days) has been found in the late-time X-ray light curve of the jetted TDE Swift~1644+57 \citep{levan_etal2016}, and is explained as the shutdown of the jet as the accretion disk transitions from a thick disk to a thin disk \citep[e.g.][]{mangano_etal2016}. We do not find strong evidence for powerful jets in \jsrc, and the \fxdrop occurred several weeks after the radio non-detection. Thus it is unlikely the \fxdrop in \jsrc is related to the shutdown of a jet. AT2021ehb also shows a rapid X-ray flux drop by a factor of 10 within three~days, during which the X-ray spectrum becomes significantly softer. \citet{yao_etal2022} proposed that this rapid drop is likely due to a hard-to-soft state transition. Unlike \jsrc, the power-law component still contributed substantially to the X-ray emission of AT2021ehb after the X-ray flux drop, indicating that the corona did not change drastically. Corona destruction has been proposed to explain the X-ray flux variability in the AGN 1ES\,1927+654 \citep{ricci_etal2020, masterson_etal2022}. The timescale of the SPL-to-soft transition or corona destruction in \jsrc is shorter than a few weeks, which is similar to that found in 1ES\,1927+654 ($\text{about one}\,$month). It is clear from Fig.\,\ref{fig:multi_lc} that the \fxdrop phase is always followed by a sudden increase in the UV brightness, which may be attributed to a temporary increase of the accretion rate caused by the collapse of the coronae at the inner region of the disk. The physical process that triggers the SPL-to-soft transition, thus the \fxdrop phase, is unknown. There is an indication that the \fxdrop phase occurs at a UV faint state for cycle2 and cycle3 (see Fig.\,\ref{fig:multi_lc}). However, the UV brightness showed very different evolution before the \fxdrop between cycle2 (decline on a timescale of weeks) and cycle3 (almost constant, but also showed variability within a week). Thus it is currently unclear whether the \fxdrop and the UV faint state are connected. %

By studying the X-ray and UV properties of a sample of seven X-ray bright TDEs, \citet{wevers2020} found that the X-ray emission of TDEs with $\lambda_\text{Edd}\gtrsim0.03$ is dominated by the thermal emission, whereas the power-law component dominates the X-ray emission for TDEs with $\lambda_\text{Edd}\lesssim0.03$. They also found that, similar to BHXRBs and AGNs, the X-ray spectral state transition occurs around $\lambda_\text{Edd}\approx0.03$. For \jsrc, the X-ray spectrum from Swift5, during which \jsrc has a $\lambda_\text{Edd}$ of 0.07, is indeed dominated by thermal emission. However, as $\lambda_\text{Edd}$ increases from 0.07 (Swift5) to a peak value of $\sim0.23$ in cycle2, the contribution from the power-law component increases, and the power-law component dominates the X-ray emission (see the second panel of Fig.~\ref{fig:xuv_char} and also Fig.~\ref{fig:uvxray_sed}). Although contrary to the general trend found in \citet{wevers2020}, our results are consistent with that found in AT2021ehb during its X-ray rising phase \citep{yao_etal2022}, suggesting that the dependence of X-ray spectral state on $\lambda_\text{Edd}$ may vary from source to source. In addition, accretion-state transitions between the soft and hard states in BHXRBs can happen in a wide range of $\lambda_\text{Edd}$. For instance, \citet{tetarenko_etal2016} found that the hard-to-soft transition of a sample of BHXRBs occurs at around $\log (\lambda_\text{Edd})=-0.94$ with a standard deviation of 0.41. The hard-to-soft transition can happen in a wide range of $\lambda_\text{Edd}$ even for individual BHXRBs (e.g. H1743$-$322, \citealt{coriat_etal2011}). The soft-to-hard state transition normally happens at a lower $\lambda_\text{Edd}$ with $\log (\lambda_\text{Edd})=-1.5$ (standard deviation of 0.37, \citealt{tetarenko_etal2016}). The peak $\lambda_\text{Edd}$ of \jsrc monotonically decreases from $\sim0.37$ in cycle1 to $\sim0.07$ in cycle3, which indicates that the accretion-state transitions between soft and SPL states may also occur at very different $\lambda_\text{Edd}$ in accreting SMBH systems. Finally, the similarity between \jsrc and the X-ray light curves of some BHXRBs is worth noting (e.g. \gro, \citealt{sobczak_etal1999}, and 4U\,1630$-$47, \citealt{abe_etal2005}). In particular, the evolution of the X-ray light curve of \jsrc shows remarkable similarity to \gro (e.g. Figure\,1 in \citealt{sobczak_etal1999} and Figure\,4a in \citealt{remillard_mcclintock2006}), which was in an outburst in 1996-1997. After the outburst, \gro initially brightened in the soft state for 50\,days until it entered a plateau-like phase. The source then stayed in this phase for about five months, during which the spectrum was reminiscent of the SPL state (SPL, or very high state). It then quickly went into a faint state within two weeks. The resemblance provides further evidence that similar accretion processes are at work for SMBH and stellar mass BH accreting systems.

\section{Summary\label{sec:summary}}
We presented the detailed multi-wavelength analysis of the \ptde candidate \erasst, which is located in the nucleus of a quiescent galaxy at $z=0.077$. The main observed features are listed below.

\begin{enumerate}
\item \jsrc likely is a repeating nuclear transient that shows four distinctive X-ray phases. We observed three cycles so far. The tentatively estimated recurrence time is $\sim\trecur\,$days.
\item The most prominent feature of \jsrc is the drastic X-ray flux decline by a factor of $\gtrsim100$ within one week during the \textup{X-ray flux drop} phase. Before this rapid X-ray flux drop phase, the X-ray flux of \jsrc is in the \textup{X-ray plateau} phase, which can last for $\text{about two}$~months. After the rapid X-ray flux drop phase, \jsrc goes into the \textup{X-ray faint} state for about $\text{two to three}$~months before the \textup{X-ray rising} phase starts again.
\item The X-ray spectra are generally very soft in the X-ray rising phase, with a photon index $\gtrsim3.0$ and evidence of a multi-colour disk with a temperature of $\sim70\,\text{eV}$ at the faint end of this phase. The X-ray spectra become slightly harder during the X-ray plateau phase, with a photon index $\lesssim3.0$.
\item Non-thermal components are clearly seen in the high-quality \mission{XMM-Newton} data during the plateau phase and also at the bright end of the rising phase. These non-thermal components are likely produced via inverse Comptonisation of soft seed photons by a warm and hot corona.
\item \jsrc showed only moderate UV variability and no significant optical variability. The optical spectra taken at different X-ray phases are always consistent with a typical quiescent galaxy with no indication of emission lines.
\item Radio emission, which decreases rapidly on a timescale of two weeks, has been detected only in the X-ray plateau phase (as of now).
\end{enumerate}

\jsrc is an exceptional repeating nuclear transient discovered with eROSITA. We discussed several scenarios to explain the cause of the repeating X-ray flares in \jsrc: a repeating partial TDE, a pair of EMRIs, and radiation pressure instability. We favour a repeating \ptde as the most plausible scenario. The mass loss is estimated to be $0.05\msun$, $0.015\msun$, and $0.005\msun$ for the first, second, and third cycle, respectively. These values are low enough for a \ptde. The observed peak rest-frame $0.2-2.0$~keV luminosity decreases monotonically from the first cycle ($\sim1.6\times10^{44}\,\unitlumi$) to the third cycle ($\sim3.0\times10^{43}\,\unitlumi$). This also agrees with the expectation from a \ptde. Long-term multi-wavelength monitoring in the future is required to further support the \ptde scenario and to accurately measure the recurrence time. Our analysis shows that the X-ray/UV emission is dominated by thermal emission from a thin accretion disk when the X-ray flux is faint, which is reminiscent of the soft state in BHXRBs. The thermal disk still causes the UV emission in the X-ray bright state. The X-ray emission, however, is dominated by X-ray photons from the inverse Comptonisation of the soft photons by a warm and a hot corona. In addition, the X-ray spectral property and the detection of transient radio emission suggest that \jsrc is in the SPL state when the source is bright in the X-ray band. We propose that a transition from the soft state to the SPL state, accompanied by the formation of the warm and hot coronae, can explain the X-ray rising phase, while the rapid X-ray flux drop is caused by a transition from the SPL state to the soft state, during which the coronae are destroyed. High-cadence X-ray and UV observations, particularly during the rapid X-ray flux drop phase, are needed to further understand the evolution of the thermal and non-thermal emission in \jsrc. This may provide new clues for the accretion-state transition in SMBH--accreting systems and for the formation and destruction of the corona. \jsrc shows transient radio emission with rapid variability on a timescale of a few weeks, which is rare in nuclear transients. Future radio observations at multiple frequencies are crucial for understanding the mechanism for the radio emission in \jsrc. 

\begin{acknowledgements}
ZL is grateful to the \mission{XMM-Newton}, \mission{Swift}, and \mission{NICER} teams for approving the ToO/DDT requests and arranging the follow-up observations. ZL thanks Dr. Taeho Ryu and Dr. Erlin Qiao for helpful discussion.
AM acknowledges support by DLR under the grant 50 QR 2110 (XMM\_NuTra, PI: ZL). MK acknowledges support by DFG grant KR 3338/4-1.
GEA is the recipient of an Australian Research Council Discovery Early Career Researcher Award (project number DE180100346). AGM acknowledges partial support from Narodowe Centrum Nauki (NCN) grants 2016/23/B/ST9/03123, 2018/31/G/ST9/03224, and 2019/35/B/ST9/03944. MG is supported by the EU Horizon 2020 research and innovation programme under grant agreement No 101004719. DAHB acknowledges research support from the National Research Foundation. CJ acknowledges the National Natural Science Foundation of China through grant 11873054, and the support by the Strategic Pioneer Program on Space Science, Chinese Academy of Sciences through grant XDA15052100.
This work is based on data from eROSITA, the soft X-ray instrument aboard SRG, a joint Russian-German science mission supported by the Russian Space Agency (Roskosmos), in the interests of the Russian Academy of Sciences represented by its Space Research Institute (IKI), and the Deutsches Zentrum für Luft- und Raumfahrt (DLR). The SRG spacecraft was built by Lavochkin Association (NPOL) and its subcontractors, and is operated by NPOL with support from the Max Planck Institute for Extraterrestrial Physics (MPE). The development and construction of the eROSITA X-ray instrument was led by MPE, with contributions from the Dr. Karl Remeis Observatory Bamberg \& ECAP (FAU Erlangen-Nuernberg), the University of Hamburg Observatory, the Leibniz Institute for Astrophysics Potsdam (AIP), and the Institute for Astronomy and Astrophysics of the University of Tübingen, with the support of DLR and the Max Planck Society. The Argelander Institute for Astronomy of the University of Bonn and the Ludwig Maximilians Universität Munich also participated in the science preparation for eROSITA.
This work was supported by the Australian government through the Australian Research Council’s Discovery Projects funding scheme (DP200102471).
This paper made use of data based on observations obtained with \mission{XMM-Newton}, an ESA science mission with instruments and contributions directly funded by ESA Member States and NASA.
This work made use of data supplied by the UK Swift Science Data Centre at the University of Leicester.
This paper includes data gathered with the 6.5 meter Magellan Telescopes located at Las Campanas Observatory, Chile. Based on observations made with ESO Telescopes at the La Silla Paranal Observatory under ESO programme 106.21RU.
The Australia Telescope Compact Array (ATCA) is part of the Australia Telescope National Facility, which is funded by the Australian Government for operation as a National Facility managed by CSIRO. \textcolor{black}{We acknowledge the Gomeroi people as the traditional owners of the Observatory site.}
The SALT observations were obtained under the SALT Large Science Programme on transients (2018-2-LSP-001; PI: DAHB) which is also supported by Poland under grant no. MEiN 2021/WK/01.
This research has made use of data obtained through the High Energy Astrophysics Science Archive Research Center Online Service, provided by the NASA/Goddard Space Flight Center.
\end{acknowledgements}

\bibliographystyle{aa}
\bibliography{references}

\begin{appendix} 
\section{False-colour image of the host galaxy and astrometric correction}\label{sec:fc_img}
\begin{figure}
  \centering
  \includegraphics[width=\columnwidth]{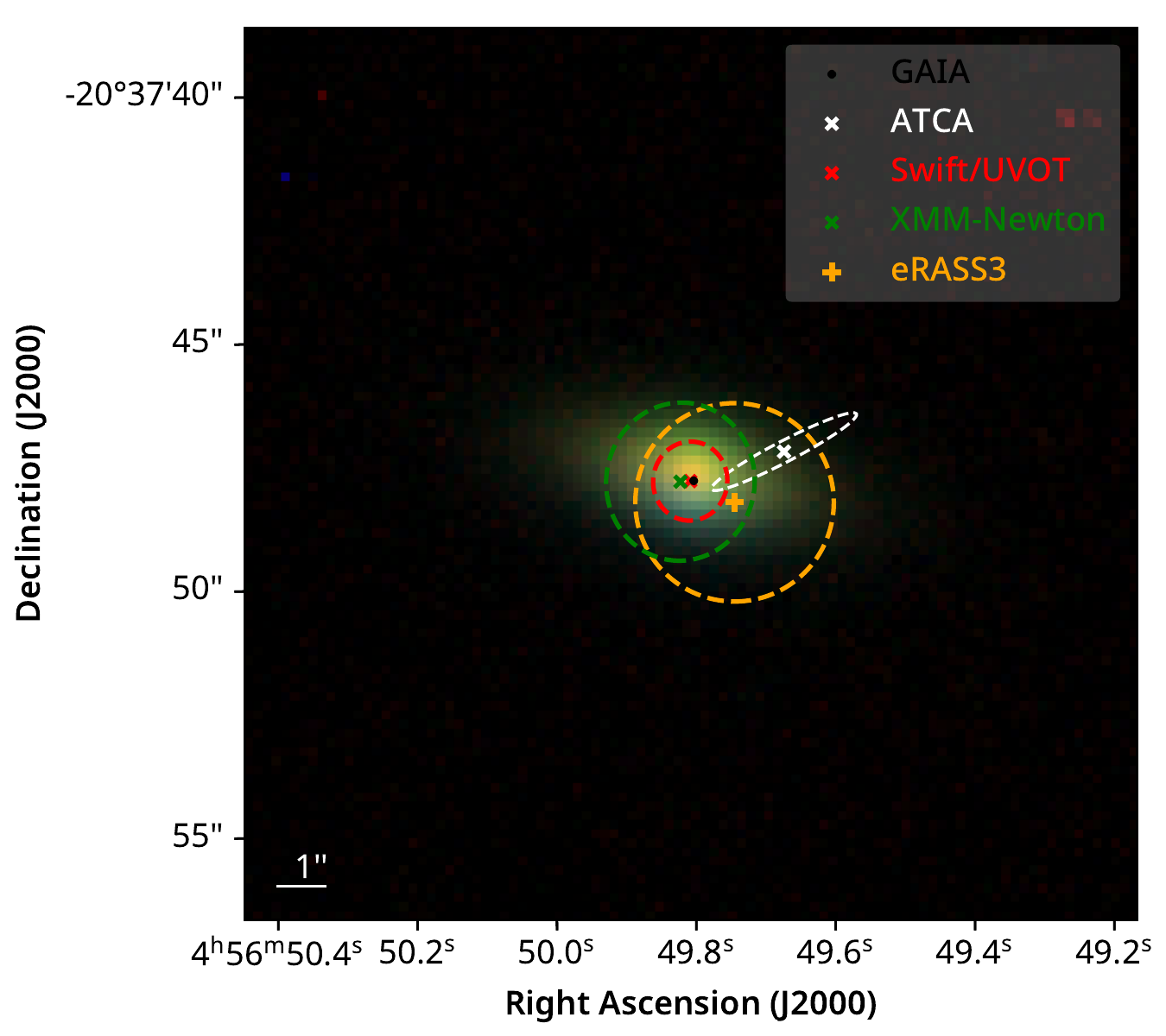}
  \caption{False-colour image of the host galaxy, composited from the CFHT MegaCam $gri$ images. The black dot marks the position given in GAIA EDR3. The crosses represent the positions measured from different instruments, and the circles indicate the $2\sigma$ positional uncertainties: \mission{Swift}/UVOT (red, $2\sigma=0.8\arcsec$), ATCA (white, see Sect.\,\ref{subsec:atca}), \mission{XMM-Newton} (green, $2\sigma=1.6\arcsec$), and eRASS3 (orange, $2\sigma=2.0\arcsec$).}
  \label{fig:fc_img}
\end{figure}

Fig.\,\ref{fig:fc_img} shows the false-colour image of the host galaxy. The image is composited based on the Canada–France–Hawaii Telescope (CFHT) MegaCam $gri$ images. For the CFHT images, astrometric correction was performed using the \texttt{SCAMP} software \citep{bertin2006}. The PanSTARRS1-DR1 catalogue was used as the reference catalogue. The \mission{XMM-Newton} coordinates were estimated from the EPIC/pn data of X2. The Swift5 UVOT observation, which has a long exposure time, was used to determine the location of the UV flare. The UVOT image was also corrected for astrometry using the PanSTARRS1-DR1 catalogue as reference. The position of the UV flare is $(\text{RA, Dec})=(\text{04h56m49.81s}, -20\degr37\arcmin47.99\arcsec),$ with a $1\sigma$ uncertainty of $0.4\arcsec$. The UV position is consistent with the X-ray positions measured from eRASS3 and X2.

\section{Estimating the upper limit of the X-ray flux}\label{sec:cal_ul}
To calculate the upper limit of \jsrc flux in eRASS1, we first adopted the Bayesian approach presented in \citet{kraft_etal1991} to estimate the photon count upper limit at a given confidence level. Following the suggestion in \citet{ruiz_etal2022}, we chose a single-side confidence level of 0.9987 (this corresponds to a double-side confidence level of 0.997) to estimate the $3\sigma$ upper limit. Using the source and background regions defined for eRASS2, we measured total and background photon counts of 2 and 0.54, respectively, in the $0.2-2.3\,\text{keV}$ energy band. Based on the exposure time of 325~s for eRASS1, this resulted in a $3\sigma$ count-rate upper limit of 0.04\,$\mathrm{s}^{-1}$ after correcting for the encircled energy fraction (EEF), which measures the fraction of total source photons within the source region ($\text{EEF}\sim0.8$ for \jsrc in eRASS1). We then calculated the unabsorbed intrinsic flux upper limit over the $0.2-2.0~\text{keV}$ energy band for eRASS1.
We converted the source photon count rate into flux using the \mul model. We derived a 3$\sigma$ upper limit of $2.2\times10^{-13}\,\unitflux$. We did not include the host galaxy absorption as no evidence for significant host galaxy absorption was found in the high-quality \mission{XMM-Newton} data (see Sect.\,\ref{subsubsec:xmm_ave_spec}).

\section{Observation log and details of the UV and X-ray observation}\label{sec:obs_log}
\begin{table*}
\caption{\mission{Swift}/UVOT and \mission{XMM-Newton}/OM observation log of \jsrc. MJD is the mid-date of the coverage for each observation; $T_\text{exp}$ is the effective exposure time in units of seconds; $m_\text{AB}$ is the apparent AB magnitude with the UVM2 filter; $f_\lambda$ is the UVM2 flux density in units of $10^{-17}\,\unitfluxd$. Quoted uncertainties are at the $1\sigma$ confidence level; $N_\text{sigma}$ is the detection significance. For observations with $N_\text{sigma}<3.0$, the values of the $3\sigma$ upper limits are given for $m_\text{AB}$ and $f_\lambda$; ObsID is the observation ID for each observation.}
\centering
\label{tab:sw_uvot}
\begin{tabular}{lcccccc}
\hline\hline
MJD  & $T_\text{exp}$ & $m_\text{AB}$ & $f_\lambda$ & $N_\text{sigma}$ &  ObsID & Instrument \\\hline
59284.89 & 1659   & $22.16\pm{0.28}$ & $3.0\pm{0.8}$ &  3.8 & 00014135001 & \mission{Swift} \\[0.5mm]
59300.20 & 11800& $21.07\pm{0.18}$ & $7.56\pm1.01$ &  9.7 & 0862770201 & \mission{XMM-Newton} \\[0.5mm]
59320.20 & 3201   & $21.01\pm{0.09}$ & $8.7\pm{0.7}$ & 12.3 & 00014135002 & \mission{Swift} \\[0.5mm]
59329.73 & 3285   & $21.10\pm{0.10}$ & $8.0\pm{0.7}$ & 11.9 & 00014135003 & \mission{Swift} \\[0.5mm]
59340.45 & 1375   & $21.63\pm{0.20}$ & $5.0\pm{0.9}$ &  5.5 & 00014135004 & \mission{Swift} \\[0.5mm]
59391.16 & 7628   & $21.64\pm{0.09}$ & $4.9\pm{0.4}$ & 12.1 & 00014135005 & \mission{Swift} \\[0.5mm]
59399.00 & 6347   & $21.50\pm{0.09}$ & $5.5\pm{0.4}$ & 12.5 & 00014135006 & \mission{Swift} \\[0.5mm]
59405.13 & 1778   & $21.41\pm{0.16}$ & $6.0\pm{0.9}$ &  7.1 & 00014135007 & \mission{Swift} \\[0.5mm]
59413.03 & 1623   & $21.43\pm{0.17}$ & $5.9\pm{0.9}$ &  6.4 & 00014135008 & \mission{Swift} \\[0.5mm]
59427.10 & 1609   & $21.64\pm{0.19}$ & $4.9\pm{0.8}$ &  5.8 & 00014135009 & \mission{Swift} \\[0.5mm]
59434.71 & 1936   & $21.73\pm{0.19}$ & $4.5\pm{0.8}$ &  5.7 & 00014135010 & \mission{Swift} \\[0.5mm]
59447.51 & 51100  & $21.99\pm{0.16}$ & $3.01\pm0.47$ &  8.7 & 0891801101  & \mission{XMM-Newton} \\[0.5mm]
59453.89 & 2914   & $21.82\pm{0.17}$ & $4.1\pm{0.6}$ &  6.5 & 00014135011 & \mission{Swift} \\[0.5mm]
59468.45 &  438   & $     <21.59   $ & $    <5.1   $ &  2.5 & 00014135012 & \mission{Swift} \\[0.5mm]
59472.51 & 2504   & $21.85\pm{0.18}$ & $4.0\pm{0.7}$ &  6.0 & 00014135013 & \mission{Swift} \\[0.5mm]
59481.30 & 127500 & $22.91\pm{0.19}$ & $1.27\pm0.24$ &  7.4 & 0891801701  & \mission{XMM-Newton} \\[0.5mm]
59481.49 & 2458   & $22.33\pm{0.25}$ & $2.6\pm{0.6}$ &  4.5 & 00014135014 & \mission{Swift} \\[0.5mm]
59496.67 & 3034   & $22.66\pm{0.29}$ & $1.9\pm{0.5}$ &  3.7 & 00014135016 & \mission{Swift} \\[0.5mm]
59510.83 & 2738   & $22.49\pm{0.26}$ & $2.2\pm{0.5}$ &  4.1 & 00014135017 & \mission{Swift} \\[0.5mm]
59524.53 & 1349   & $21.74\pm{0.21}$ & $4.5\pm{0.8}$ &  5.3 & 00014135018 & \mission{Swift} \\[0.5mm]
59527.63 & 1401   & $21.88\pm{0.24}$ & $3.9\pm{0.8}$ &  4.6 & 00014135019 & \mission{Swift} \\[0.5mm]
59529.87 &  440   & $21.63\pm{0.33}$ & $4.9\pm{1.5}$ &  3.3 & 00014135021 & \mission{Swift} \\[0.5mm]
59543.72 & 1045   & $21.32\pm{0.17}$ & $6.6\pm{1.0}$ &  6.3 & 00014135022 & \mission{Swift} \\[0.5mm]
59557.41 & 2288   & $21.64\pm{0.15}$ & $4.9\pm{0.7}$ &  7.2 & 00014135023 & \mission{Swift} \\[0.5mm]
59566.74 &  604   & $21.36\pm{0.26}$ & $6.3\pm{1.5}$ &  4.3 & 00014135025 & \mission{Swift} \\[0.5mm]
59572.58 & 2591   & $21.46\pm{0.13}$ & $5.7\pm{0.7}$ &  8.7 & 00014135026 & \mission{Swift} \\[0.5mm]
59584.90 & 2835   & $21.74\pm{0.15}$ & $4.5\pm{0.6}$ &  7.5 & 00014135027 & \mission{Swift} \\[0.5mm]
59635.12 & 9989   & $21.91\pm{0.10}$ & $3.8\pm{0.3}$ & 11.7 & 00014135999$^a$ & \mission{Swift} \\[0.5mm]
59643.68 & 2452   & $21.96\pm{0.20}$ & $3.6\pm{0.7}$ &  5.4 & 00014135033 & \mission{Swift} \\[0.5mm]
59650.19 & 2935   & $22.61\pm{0.29}$ & $2.0\pm{0.5}$ &  3.8 & 00014135034 & \mission{Swift} \\[0.5mm]
59656.16 & 2640   & $22.11\pm{0.21}$ & $3.1\pm{0.6}$ &  5.3 & 00014135036 & \mission{Swift} \\[0.5mm]
59663.24 & 2734   & $22.28\pm{0.25}$ & $2.7\pm{0.6}$ &  4.5 & 00014135037 & \mission{Swift} \\[0.5mm]
59670.56 & 3088   & $22.01\pm{0.20}$ & $3.5\pm{0.6}$ &  5.5 & 00014135038 & \mission{Swift} \\[0.5mm]
59678.08 & 4241   & $22.25\pm{0.20}$ & $2.7\pm{0.5}$ &  5.6 & 00014135039 & \mission{Swift} \\[0.5mm]
59684.37 & 2894   & $22.16\pm{0.22}$ & $3.0\pm{0.6}$ &  4.9 & 00014135040 & \mission{Swift} \\[0.5mm]
59691.98 & 6307   & $22.05\pm{0.14}$ & $3.4\pm{0.4}$ &  7.8 & 00014135041 & \mission{Swift} \\[0.5mm]
59698.54 & 2592   & $     <22.63   $ & $    <2.0   $ &  2.2 & 00014135042 & \mission{Swift} \\[0.5mm]
59705.34 & 2443   & $22.22\pm{0.26}$ & $2.9\pm{0.7}$ &  4.2 & 00014135043 & \mission{Swift} \\\hline

\end{tabular}
\tablefoot{\tablefoottext{a}{This is a surrogate ObsID that consists of three observations with ObsIDs of  00014135030, 00014135031, and 00014135032.}}
\end{table*}

\clearpage
\onecolumn
\LTcapwidth=\textwidth
\begin{longtable}{cccccccc}
\caption{Log of X-ray observations of \jsrc. MJD is the mid-date of the coverage for each observation; $T_\text{exp}$ is the effective exposure time in units of seconds. Exposure for \mission{XMM-Newton} ObsID 0862770201 is calculated from combined MOS data, while the pn exposures are given for the other \mission{XMM-Newton} observations; $\text{CR}_\text{total}$ and $\text{CR}_\text{bkg}$ are respectively the total and background photon counts rate calculated over the $0.3-5.0\,$keV band for \mission{Swift}, $0.2-2.3\,$keV band for eROSITA, $0.3-1.0\,$keV band for \mission{NICER} detections and $0.4-1.0\,$keV for non-detections (avoid potential contribution from optical loading), and $0.2-6.0\,$keV for \mission{XMM-Newton}. For \mission{XMM-Newton} ObsID 0862770201, count rates from the combined MOS data are given, while the count rates from the pn data are given for the other \mission{XMM-Newton} observations; $T_\text{in}/\Gamma$: The inner temperature of a multi-color disk is given for eRASS2, \mission{Swift} observation 00014135005 (Swift5), and \mission{XMM-Newton} observation 0862770201. The photon index of the power-law is given for the other observations. A value without an uncertainty means that the parameter is fixed at the given value during spectral fitting; \sflux is the rest frame intrinsic $0.2-2.0\,$keV flux in units of $10^{-12}\,\unitflux$. The $3\sigma$ upper limits are given for observations in which the \jsrc is not detected; ObsID is the observation ID for each observation. Quoted uncertainties are at the 90\% confidence level.}
\label{tab:logxray}\\
\hline\hline
MJD & $T_\text{exp}$ & $\text{CR}_\text{total}$ & $\text{CR}_\text{bkg}$ & $T_\text{in}/\Gamma$ & \sflux & ObsID & Instrument \\ 
 & (s) & ($\text{cts}\,\text{s}^{-1}$) & ($\text{cts}\,\text{s}^{-1}$) & (eV/-) & $10^{-12}\,\unitflux$ &   &   \\
\hline
\endfirsthead
\caption{continued.}\\
\hline\hline
MJD & $T_\text{exp}$ & $\text{CR}_\text{total}$ & $\text{CR}_\text{bkg}$ & $T_\text{in}/\Gamma$ & \sflux & ObsID & Instrument \\[0.5mm]
 & (s) & ($\text{cts}\,\text{s}^{-1}$) & ($\text{cts}\,\text{s}^{-1}$) & (eV/-) & $10^{-12}\,\unitflux$ &   &   \\
 \hline
\endhead
\hline
\endfoot
58919.15 &   325 & 0.00616 & 0.00167 & $3.0              $ & $< 0.22              $ & eRASS1      & eROSITA\\[0.5mm]
59101.73 &   340 & 0.03534 & 0.00107 & $64._{-18.}^{+29.}$ & $0.51_{-0.26}^{+0.51}$ & eRASS2      & eROSITA\\[0.5mm]
59272.63 &   194 & 2.22111 & 0.01939 & $2.5_{-0.2}^{+0.2}$ & $7.76_{-0.76}^{+0.84}$ & eRASS3-1    & eROSITA\\[0.5mm]
59281.86 &   123 & 2.35918 & 0.02659 & $2.8_{-0.2}^{+0.2}$ & $11.3_{-1.40}^{+1.61}$ & eRASS3-2    & eROSITA\\[0.5mm]
59284.89 &  1683 & 0.21745 & 0.00110 & $2.5_{-0.1}^{+0.1}$ & $11.6_{-1.45}^{+1.66}$ & 00014135001 &   \mission{Swift}\\[0.5mm]
59290.94 &  1664 & 0.21575 & 0.20971 & $3.0              $ & $< 0.19              $ & 4604010101  &   \mission{NICER}\\[0.5mm]
59291.53 &  9941 & 0.17916 & 0.17999 & $3.0              $ & $< 0.13              $ & 4604010102  &   \mission{NICER}\\[0.5mm]
59292.21 &  6394 & 0.19706 & 0.19072 & $3.0              $ & $< 0.14              $ & 4604010103  &   \mission{NICER}\\[0.5mm]
59298.15 &  4387 & 0.29770 & 0.25157 & $3.0              $ & $< 0.18              $ & 4604010201  &   \mission{NICER}\\[0.5mm]
59300.20 & 10040 & 0.00159 & 0.00047 & $99^{+120}_{-40}$   & $0.03^{+0.06}_{-0.02}$ & 0862770201  &   \mission{XMM-Newton}\\[0.5mm]
59305.09 &  2735 & 0.22157 & 0.25379 & $3.0              $ & $< 0.20              $ & 4604010301  &   \mission{NICER}\\[0.5mm]
59320.20 &  3254 & 0.00000 & 0.00009 & $3.0              $ & $< 0.28              $ & 00014135002 &   \mission{Swift}\\[0.5mm]
59322.59 &  1808 & 0.31029 & 0.36862 & $3.0              $ & $< 0.30              $ & 4604010501  &   \mission{NICER}\\[0.5mm]
59323.49 &  4525 & 0.24840 & 0.28309 & $3.0              $ & $< 0.21              $ & 4604010502  &   \mission{NICER}\\[0.5mm]
59325.27 &   230 & 0.14348 & 0.14935 & $3.0              $ & $< 0.35              $ & 4604010504  &   \mission{NICER}\\[0.5mm]
59328.31 &   420 & 0.15476 & 0.12096 & $3.0              $ & $< 0.24              $ & 4604010507  &   \mission{NICER}\\[0.5mm]
59329.73 &  3342 & 0.00000 & 0.00007 & $3.0              $ & $< 0.24              $ & 00014135003 &   \mission{Swift}\\[0.5mm]
59340.45 &  1402 & 0.00000 & 0.00007 & $3.0              $ & $< 0.54              $ & 00014135004 &   \mission{Swift}\\[0.5mm]
59391.16 &  7669 & 0.00065 & 0.00004 & $78$                & $0.18_{-0.11}^{+0.18}$ & 00014135005 &   \mission{Swift}\\[0.5mm]
59399.00 &  6406 & 0.00250 & 0.00015 & $4.3_{-1.1}^{+1.3}$ & $0.47_{-0.27}^{+0.70}$ & 00014135006 &   \mission{Swift}\\[0.5mm]
59405.13 &  1793 & 0.00558 & 0.00014 & $5.1_{-1.6}^{+1.9}$ & $1.65_{-1.14}^{+4.34}$ & 00014135007 &   \mission{Swift}\\[0.5mm]
59413.03 &  1641 & 0.01402 & 0.00037 & $2.6_{-0.7}^{+0.7}$ & $0.68_{-0.29}^{+0.53}$ & 00014135008 &   \mission{Swift}\\[0.5mm]
59422.51 &  3282 & 0.99116 & 0.32306 & $3.7_{-0.3}^{+0.3}$ & $2.77_{-0.42}^{+0.58}$ & 4595020102  &   \mission{NICER}\\[0.5mm]
59423.35 &  1597 & 0.99249 & 0.22507 & $3.7_{-0.3}^{+0.3}$ & $3.47_{-0.60}^{+0.86}$ & 4595020103  &   \mission{NICER}\\[0.5mm]
59426.30 &  2093 & 1.21739 & 0.32895 & $3.9_{-0.3}^{+0.3}$ & $3.97_{-0.66}^{+0.93}$ & 4595020104  &   \mission{NICER}\\[0.5mm]
59427.10 &  1631 & 0.02024 & 0.00022 & $2.9_{-0.5}^{+0.6}$ & $1.21_{-0.45}^{+0.74}$ & 00014135009 &   \mission{Swift}\\[0.5mm]
59428.95 &  2879 & 1.22612 & 0.38511 & $3.3_{-0.3}^{+0.3}$ & $3.17_{-0.43}^{+0.60}$ & 4595020105  &   \mission{NICER}\\[0.5mm]
59431.03 &  2131 & 1.15767 & 0.21195 & $4.0_{-0.3}^{+0.3}$ & $4.82_{-0.83}^{+1.17}$ & 4595020106  &   \mission{NICER}\\[0.5mm]
59432.96 &   764 & 1.01047 & 0.17102 & $4.0_{-0.4}^{+0.4}$ & $4.20_{-0.91}^{+1.40}$ & 4595020107  &   \mission{NICER}\\[0.5mm]
59433.12 &   999 & 1.04304 & 0.19543 & $3.7_{-0.4}^{+0.4}$ & $3.73_{-0.72}^{+1.08}$ & 4595020108  &   \mission{NICER}\\[0.5mm]
59434.71 &  1968 & 0.03100 & 0.00022 & $3.5_{-0.5}^{+0.5}$ & $3.07_{-0.96}^{+1.45}$ & 00014135010 &   \mission{Swift}\\[0.5mm]
59441.95 &   104 & 1.82692 & 0.35394 & $3.8_{-0.7}^{+0.7}$ & $5.96_{-1.63}^{+2.81}$ & 4595020109  &   \mission{NICER}\\[0.5mm]
59445.50 &  1760 & 2.15909 & 0.24284 & $3.3_{-0.3}^{+0.3}$ & $6.83_{-0.96}^{+1.36}$ & 4595020111  &   \mission{NICER}\\[0.5mm]
59447.51 & 40200 & 1.46844 & 0.01909 & ---                 & $4.79_{-0.90}^{+1.10}$ & 0891801101  &   \mission{XMM-Newton}\\[0.5mm]
59448.38 &  1169 & 1.92729 & 0.17791 & $3.0_{-0.3}^{+0.3}$ & $5.10_{-0.56}^{+0.76}$ & 4595020112  &   \mission{NICER}\\[0.5mm]
59453.89 &  2954 & 0.05822 & 0.00064 & $2.8_{-0.2}^{+0.2}$ & $3.20_{-0.59}^{+0.72}$ & 00014135011 &   \mission{Swift}\\[0.5mm]
59460.31 &   345 & 0.93832 & 0.01655 & $3.4_{-0.2}^{+0.2}$ & $5.26_{-0.67}^{+0.76}$ & eRASS4      &   eROSITA\\[0.5mm]
59467.45 &  1568 & 0.99490 & 0.20285 & $2.8_{-0.3}^{+0.3}$ & $2.39_{-0.29}^{+0.42}$ & 4604010901  &   \mission{NICER}\\[0.5mm]
59468.45 &   445 & 0.02924 & 0.00037 & $2.9_{-0.9}^{+0.9}$ & $1.73_{-0.91}^{+2.05}$ & 00014135012 &   \mission{Swift}\\[0.5mm]
59470.50 &   341 & 1.07625 & 0.70582 & $4.1_{-0.9}^{+1.1}$ & $1.67_{-0.61}^{+1.21}$ & 4604010902  &   \mission{NICER}\\[0.5mm]
59472.51 &  2542 & 0.03029 & 0.00038 & $2.6_{-0.4}^{+0.4}$ & $1.71_{-0.44}^{+0.61}$ & 00014135013 &   \mission{Swift}\\[0.5mm]
59476.22 &  1560 & 0.69231 & 0.26452 & $2.9_{-0.4}^{+0.4}$ & $1.23_{-0.17}^{+0.24}$ & 4604010903  &   \mission{NICER}\\[0.5mm]
59479.33 &   929 & 1.16039 & 0.18996 & $3.0_{-0.3}^{+0.3}$ & $2.81_{-0.35}^{+0.49}$ & 4604010904  &   \mission{NICER}\\[0.5mm]
59481.30 & 85960 & 0.78391 & 0.01568 & ---                 & $2.40_{-0.36}^{+0.76}$ & 0891801701  &   \mission{XMM-Newton}\\[0.5mm]
59481.49 &  2490 & 0.02611 & 0.00026 & $2.7_{-0.4}^{+0.4}$ & $1.48_{-0.41}^{+0.58}$ & 00014135014 &   \mission{Swift}\\[0.5mm]
59482.46 &  4637 & 0.88851 & 0.15370 & $2.9_{-0.3}^{+0.3}$ & $2.33_{-0.26}^{+0.37}$ & 4604010905  &   \mission{NICER}\\[0.5mm]
59485.04 &  1183 & 0.96196 & 0.22790 & $2.7_{-0.3}^{+0.3}$ & $2.03_{-0.23}^{+0.31}$ & 4604010906  &   \mission{NICER}\\[0.5mm]
59488.08 &  1532 & 1.19648 & 0.32618 & $3.0_{-0.3}^{+0.3}$ & $2.78_{-0.38}^{+0.57}$ & 4604010907  &   \mission{NICER}\\[0.5mm]
59491.05 &  1410 & 1.34965 & 0.35106 & $2.7_{-0.3}^{+0.3}$ & $2.94_{-0.34}^{+0.49}$ & 4604010908  &   \mission{NICER}\\[0.5mm]
59494.25 &   811 & 1.13933 & 0.21393 & $2.5_{-0.3}^{+0.3}$ & $2.34_{-0.24}^{+0.31}$ & 4604010909  &   \mission{NICER}\\[0.5mm]
59496.67 &  3079 & 0.03085 & 0.00032 & $2.7_{-0.3}^{+0.3}$ & $2.05_{-0.49}^{+0.64}$ & 00014135016 &   \mission{Swift}\\[0.5mm]
59497.25 &  1133 & 1.13945 & 0.20779 & $2.9_{-0.3}^{+0.3}$ & $2.63_{-0.31}^{+0.44}$ & 4604010910  &   \mission{NICER}\\[0.5mm]
59499.80 &  1703 & 0.69466 & 0.22953 & $2.5_{-0.3}^{+0.3}$ & $2.19_{-0.24}^{+0.31}$ & 4604010911  &   \mission{NICER}\\[0.5mm]
59500.64 &  3647 & 1.16781 & 0.35207 & $2.9_{-0.3}^{+0.3}$ & $2.47_{-0.28}^{+0.41}$ & 4604010912  &   \mission{NICER}\\[0.5mm]
59501.35 & 10186 & 0.91645 & 0.23872 & $2.9_{-0.2}^{+0.2}$ & $2.13_{-0.22}^{+0.31}$ & 4604010913  &   \mission{NICER}\\[0.5mm]
59502.49 &  7885 & 0.90298 & 0.23229 & $2.7_{-0.3}^{+0.3}$ & $1.96_{-0.18}^{+0.24}$ & 4604010914  &   \mission{NICER}\\[0.5mm]
59503.48 &  7906 & 1.08753 & 0.24026 & $2.9_{-0.2}^{+0.2}$ & $2.48_{-0.24}^{+0.34}$ & 4604010915  &   \mission{NICER}\\[0.5mm]
59504.51 &  4370 & 0.94073 & 0.18709 & $2.8_{-0.3}^{+0.3}$ & $2.23_{-0.24}^{+0.35}$ & 4604010916  &   \mission{NICER}\\[0.5mm]
59505.52 &  3600 & 1.07750 & 0.28458 & $2.7_{-0.3}^{+0.3}$ & $2.36_{-0.25}^{+0.35}$ & 4604010917  &   \mission{NICER}\\[0.5mm]
59506.22 &   826 & 1.14407 & 0.25471 & $2.8_{-0.3}^{+0.3}$ & $2.41_{-0.29}^{+0.40}$ & 4604010918  &   \mission{NICER}\\[0.5mm]
59508.15 &   553 & 0.95841 & 0.35484 & $3.3_{-0.5}^{+0.5}$ & $1.95_{-0.39}^{+0.60}$ & 4604010920  &   \mission{NICER}\\[0.5mm]
59510.21 &  3079 & 0.90257 & 0.14294 & $2.8_{-0.3}^{+0.3}$ & $2.31_{-0.26}^{+0.38}$ & 4604010921  &   \mission{NICER}\\[0.5mm]
59510.83 &  2777 & 0.03421 & 0.00044 & $2.3_{-0.3}^{+0.3}$ & $1.71_{-0.39}^{+0.51}$ & 00014135017 &   \mission{Swift}\\[0.5mm]
59511.91 &  2492 & 0.79374 & 0.14859 & $2.9_{-0.3}^{+0.3}$ & $1.91_{-0.20}^{+0.27}$ & 4604010922  &   \mission{NICER}\\[0.5mm]
59517.60 &  2745 & 0.22077 & 0.19720 & $3.0              $ & $< 0.15              $ & 4604011001  &   \mission{NICER}\\[0.5mm]
59520.47 &  3000 & 0.31967 & 0.30448 & $3.0              $ & $< 0.22              $ & 4604011002  &   \mission{NICER}\\[0.5mm]
59523.54 &  1200 & 0.27083 & 0.31064 & $3.0              $ & $< 0.27              $ & 4604011003  &   \mission{NICER}\\[0.5mm]
59524.53 &  1371 & 0.00000 & 0.00011 & $3.0              $ & $< 1.50              $ & 00014135018 &   \mission{Swift}\\[0.5mm]
59526.63 &   936 & 0.24679 & 0.31657 & $3.0              $ & $< 0.29              $ & 4604011004  &   \mission{NICER}\\[0.5mm]
59527.63 &  3924 & 0.00000 & 0.00005 & $3.0              $ & $< 0.24              $ & 00014135019 &   \mission{Swift}\\[0.5mm]
59529.83 &   636 & 0.24528 & 0.28465 & $3.0              $ & $< 0.30              $ & 4604011005  &   \mission{NICER}\\[0.5mm]
59529.87 &  1402 & 0.00000 & 0.00004 & $3.0              $ & $< 0.69              $ & 00014135021 &   \mission{Swift}\\[0.5mm]
59532.54 &  1143 & 0.25984 & 0.34356 & $3.0              $ & $< 0.29              $ & 4604011006  &   \mission{NICER}\\[0.5mm]
59535.32 &  1817 & 0.33352 & 0.37695 & $3.0              $ & $< 0.28              $ & 4604011007  &   \mission{NICER}\\[0.5mm]
59538.23 &  2041 & 0.20137 & 0.22903 & $3.0              $ & $< 0.19              $ & 4604011008  &   \mission{NICER}\\[0.5mm]
59541.33 &  1884 & 0.16030 & 0.14718 & $3.0              $ & $< 0.14              $ & 4604011009  &   \mission{NICER}\\[0.5mm]
59543.75 &  1996 & 0.00000 & 0.00009 & $3.0              $ & $< 0.39              $ & 00014135022 &   \mission{Swift}\\[0.5mm]
59557.41 &  2372 & 0.00042 & 0.00010 & $3.0              $ & $< 0.41              $ & 00014135023 &   \mission{Swift}\\[0.5mm]
59566.74 &   614 & 0.00000 & 0.00004 & $3.0              $ & $< 1.60              $ & 00014135025 &   \mission{Swift}\\[0.5mm]
59572.58 &   193 & 0.00000 & 0.00049 & $3.0              $ & $< 3.60              $ & 00014135024 &   \mission{Swift}\\[0.5mm]
59572.58 &  2635 & 0.00076 & 0.00034 & $3.0              $ & $< 0.38              $ & 00014135026 &   \mission{Swift}\\[0.5mm]
59584.90 &  2886 & 0.00000 & 0.00006 & $3.0              $ & $< 0.39              $ & 00014135027 &   \mission{Swift}\\[0.5mm]
59601.40 &   530 & 0.55849 & 0.40031 & $3.0              $ & $< 0.40              $ & 4595020114  &   \mission{NICER}\\[0.5mm]
59602.56 &  3203 & 0.33406 & 0.38072 & $3.0              $ & $< 0.26              $ & 4595020115  &   \mission{NICER}\\[0.5mm]
59603.21 &  1953 & 0.36457 & 0.39628 & $3.0              $ & $< 0.29              $ & 4595020116  &   \mission{NICER}\\[0.5mm]
59605.47 &   456 & 0.32456 & 0.28567 & $3.0              $ & $< 0.35              $ & 4595020118  &   \mission{NICER}\\[0.5mm]
59607.27 &   916 & 0.25218 & 0.26658 & $3.0              $ & $< 0.26              $ & 4595020119  &   \mission{NICER}\\[0.5mm]
59608.60 &  3275 & 0.25893 & 0.26392 & $3.0              $ & $< 0.20              $ & 4595020120  &   \mission{NICER}\\[0.5mm]
59609.72 &  1664 & 0.25661 & 0.24056 & $3.0              $ & $< 0.21              $ & 4595020121  &   \mission{NICER}\\[0.5mm]
59610.47 &  2922 & 0.21253 & 0.17810 & $3.0              $ & $< 0.15              $ & 4595020122  &   \mission{NICER}\\[0.5mm]
59611.79 &   997 & 0.17352 & 0.16539 & $3.0              $ & $< 0.18              $ & 4595020123  &   \mission{NICER}\\[0.5mm]
59612.47 &  1252 & 0.19968 & 0.18684 & $3.0              $ & $< 0.18              $ & 4595020124  &   \mission{NICER}\\[0.5mm]
59613.37 &  1469 & 0.29340 & 0.25956 & $3.0              $ & $< 0.22              $ & 4595020125  &   \mission{NICER}\\[0.5mm]
59619.34 &   859 & 0.27125 & 0.13602 & $3.5_{-0.8}^{+0.8}$ & $0.48_{-0.15}^{+0.25}$ & 4595020126  &   \mission{NICER}\\[0.5mm]
59619.40 &   872 & 0.23969 & 0.13188 & $4.4_{-1.0}^{+1.3}$ & $0.57_{-0.23}^{+0.49}$ & 4595020126  &   \mission{NICER}\\[0.5mm]
59619.72 &   905 & 0.25535 & 0.12572 & $4.6_{-0.9}^{+1.1}$ & $0.77_{-0.30}^{+0.61}$ & 4595020126  &   \mission{NICER}\\[0.5mm]
59625.99 &   787 & 0.42948 & 0.19563 & $3.4_{-0.6}^{+0.7}$ & $0.77_{-0.19}^{+0.30}$ & 4595020127  &   \mission{NICER}\\[0.5mm]
59626.06 &  1560 & 0.40192 & 0.18540 & $3.7_{-0.5}^{+0.5}$ & $0.81_{-0.18}^{+0.27}$ & 4595020128  &   \mission{NICER}\\[0.5mm]
59633.19 &  3138 & 0.56724 & 0.26562 & $3.6_{-0.4}^{+0.4}$ & $1.21_{-0.23}^{+0.35}$ & 4595020129  &   \mission{NICER}\\[0.5mm]
59635.13 & 10129 & 0.00938 & 0.00033 & $3.2_{-0.3}^{+0.3}$ & $0.77_{-0.20}^{+0.27}$ & 00014135999 &   \mission{Swift}\\[0.5mm]
59639.90 &   149 & 0.69128 & 0.11484 & $2.3_{-0.9}^{+0.9}$ & $1.68_{-0.36}^{+0.51}$ & 4604010104  &   \mission{NICER}\\[0.5mm]
59640.10 &  3423 & 0.76833 & 0.22940 & $3.2_{-0.3}^{+0.3}$ & $1.84_{-0.26}^{+0.38}$ & 4604010105  &   \mission{NICER}\\[0.5mm]
59642.13 &  3267 & 0.78206 & 0.27016 & $3.2_{-0.3}^{+0.3}$ & $1.75_{-0.24}^{+0.36}$ & 4604010106  &   \mission{NICER}\\[0.5mm]
59643.68 &  2485 & 0.02173 & 0.00044 & $3.2_{-0.5}^{+0.5}$ & $1.78_{-0.58}^{+0.88}$ & 00014135033 &   \mission{Swift}\\[0.5mm]
59645.48 &  2460 & 0.67195 & 0.18330 & $3.2_{-0.3}^{+0.3}$ & $1.71_{-0.26}^{+0.38}$ & 4604010107  &   \mission{NICER}\\[0.5mm]
59648.20 &  1991 & 0.59769 & 0.16090 & $3.2_{-0.4}^{+0.4}$ & $1.54_{-0.24}^{+0.36}$ & 4604010108  &   \mission{NICER}\\[0.5mm]
59650.19 &  2977 & 0.01445 & 0.00032 & $3.2_{-0.5}^{+0.5}$ & $1.58_{-0.56}^{+0.88}$ & 00014135034 &   \mission{Swift}\\[0.5mm]
59651.90 &   668 & 0.95060 & 0.47320 & $2.6_{-0.6}^{+0.6}$ & $1.39_{-0.24}^{+0.35}$ & 4604010109  &   \mission{NICER}\\[0.5mm]
59652.06 &   720 & 0.87778 & 0.38022 & $3.6_{-0.5}^{+0.6}$ & $2.12_{-0.52}^{+0.91}$ & 4604010110  &   \mission{NICER}\\[0.5mm]
59654.83 &  1787 & 0.68942 & 0.27128 & $3.7_{-0.4}^{+0.4}$ & $1.83_{-0.37}^{+0.59}$ & 4604010111  &   \mission{NICER}\\[0.5mm]
59656.16 &  2665 & 0.02177 & 0.00106 & $3.0_{-0.4}^{+0.4}$ & $1.48_{-0.47}^{+0.68}$ & 00014135036 &   \mission{Swift}\\[0.5mm]
59657.15 & 18390 & 0.55037 & 0.01046 & ---                 & $2.44_{-0.21}^{+0.24}$ & 0891801101  &   \mission{XMM-Newton}\\[0.5mm]
59663.82 &  3172 & 0.02176 & 0.00032 & $2.8_{-0.4}^{+0.4}$ & $1.52_{-0.42}^{+0.58}$ & 00014135037 &   \mission{Swift}\\[0.5mm]
59670.56 &  3127 & 0.02303 & 0.00049 & $2.5_{-0.4}^{+0.4}$ & $1.22_{-0.32}^{+0.44}$ & 00014135038 &   \mission{Swift}\\[0.5mm]
59678.08 &  4305 & 0.01115 & 0.00032 & $2.8_{-0.4}^{+0.5}$ & $0.68_{-0.22}^{+0.34}$ & 00014135039 &   \mission{Swift}\\[0.5mm]
59684.37 &  2922 & 0.01643 & 0.00087 & $2.6_{-0.5}^{+0.5}$ & $0.88_{-0.29}^{+0.44}$ & 00014135040 &   \mission{Swift}\\[0.5mm]
59691.98 &  6391 & 0.01377 & 0.00041 & $3.1_{-0.3}^{+0.4}$ & $1.08_{-0.28}^{+0.39}$ & 00014135041 &   \mission{Swift}\\[0.5mm]
59698.54 &  2627 & 0.01218 & 0.00135 & $3.5_{-0.8}^{+0.8}$ & $1.18_{-0.56}^{+1.06}$ & 00014135042 &   \mission{Swift}\\[0.5mm]
59705.30 &  3014 & 0.00199 & 0.00033 & $3.0_{-0.0}^{+0.0}$ & $0.12_{-0.08}^{+0.13}$ & 00014135043 &   \mission{Swift}\\[0.5mm]
\hline
\end{longtable}
\end{appendix}

\end{document}